\documentclass[useAMS,usegraphicx,usenatbib]{mn2e}

\usepackage{ulem}
\usepackage{color}
\usepackage{graphics,graphicx}
\usepackage{times} 
\usepackage{amssymb}
\usepackage{amsmath}
\usepackage{lscape}
\usepackage{url}
\usepackage{multirow}
\usepackage{ctable}
\usepackage{threeparttable}
\newif\ifAMStwofonts
\AMStwofontstrue
\def\xmm{{\it XMM-Newton}}

\def\suzaku{{\it Suzaku}}

\def\epicmos1{{\it EPIC}{\rm-MOS1~\/}}
\def\epicmos2{{\it EPIC}{\rm-MOS2 ~\/}}
\def\epicmos{{\it EPIC}{\rm-MOS}}
\def\xis{{\rm XIS}}
\def\pin{{\rm PIN}}
\def\hxd{{\rm HXD}}
\def\nustar{{\it NuStar}}

\def\deg{$^{\circ}$}

\def\kmps{\hbox{$\rm\thinspace km~s^{-1}$}}

\def\H0{{\rm ~km~s^{-1}~Mpc^{-1}}}

\def\kev{\hbox{\rm keV}}

\def\atpcm{{\rm atom~cm$^{-2}$}}

\def\ergpcmsqps{\hbox{$\rm\thinspace erg~cm^{-2}~s^{-1}$}}

\def\ergcmps{\hbox{\rm erg~cm~s$^{-1}$}}

\def\chisq{{$\chi^{2}$}}
\def\rchi{{$\chi^{2}_{\nu}$~\/}}

\def\xspec{\hbox{\small XSPEC}}
\def\xspecv{\hbox{\small XSPEC}\, v12.6.0f}

\def\heasoftv{\hbox{\rm{\small HEASOFT~\/}}}
\def\xselect{\hbox{\rm{\small XSELECT~\/}}}

\def\ftool{\hbox{\rm{\small FTOOL}}}

\def\addascaspec{\hbox{\rm{\small ADDASCASPEC~\/}}}

\def\addascaspec{\hbox{\rm{\small ADDASCASPEC}}}

\def\xstar{\hbox{\rm{\small XSTAR}}}
\def\grid25{\hbox{\rm{\small GRID25}}}

\def\tbabs{\rm{\small TBABS}}

\def\reflionx{\rm{\small REFLIONX}}

\def\relconv{\rm{\small RELCONV}}

\def\pexriv{\rm{\small PEXRIV}}

\def\eg{{\it e.g.~\/}}

\def\ie{{\it i.e.~\/}}

\def\la{\mathrel{\hbox{\rlap{\hbox{\lower4pt\hbox{$\sim$}}}{\raise2pt\hbox{$<$}}}}}
\def\ga{\mathrel{\hbox{\rlap{\hbox{\lower4pt\hbox{$\sim$}}}{\raise2pt\hbox{$>$}}}}}

\def\d25{D$_{25}$}
\def\nh{{$N_{\rm H}$}}

\def\.25{0.25 keV\thinspace}

\def\rg{$R_{\rm G}$}

\title[Suzaku Observations of `Bare' AGN]{Suzaku Observations of `Bare' Active Galactic Nuclei}

\author[D.\,J. Walton, E. Nardini, A.\,C. Fabian, L.\,C. Gallo, R.\,C. Reis]
{\parbox{7.in}{D.\,J. Walton$^{1,2}$ \thanks{E-mail: dwalton@ast.cam.ac.uk},
E. Nardini$^{3}$,
A.\,C. Fabian$^{1}$,
L.\,C. Gallo$^{4}$ and
R.\,C. Reis$^{5}$ \\
\footnotesize
$^{1}$ \it{Institute of Astronomy, Cambridge University, Madingley Road, Cambridge, CB3 0HA, UK} \\
$^{2}$ \it{Space Radiation Laboratory, California Institute of Technology, Pasadena, CA 91125, USA} \\
$^{3}$ \it{Harvard-Smithsonian Center for Astrophysics, 60 Garden St., Cambridge, MA 02138, USA} \\
$^{4}$ \it{Department of Astronomy and Physics, Saint Mary’s University, 923 Robie Street, Halifax, NS B3H 3C3, Canada}\\
$^{5}$ \it{Department of Astronomy, University of Michigan, 500 Church Street, Ann Arbor, MI 48109, USA}}}
\date{}

\begin{document}
\pagerange{\pageref{firstpage}--\pageref{lastpage}}
\maketitle
\label{firstpage}

\begin{abstract}We present a X-ray spectral analysis of a large sample of 25
`bare' active galactic nuclei, sources with little or no complicating intrinsic
absorption, observed with \suzaku. Our work focuses on studying the potential
contribution from relativistic disc reflection, and examining the implications
of this interpretation for the intrinsic spectral complexities frequently
displayed by AGN in the X-ray bandpass. During the analysis, we take the unique
approach of attempting to simultaneously undertake a systematic analysis of the
whole sample, as well as a detailed treatment of each individual source, and
find that disc reflection has the required flexibility to successfully reproduce
the broadband spectrum observed for all of the sources considered. Where
possible, we use the reflected emission to place constraints on the black hole
spin for this sample of sources. Our analysis suggests a general preference for
rapidly rotating black holes, which if taken at face value is most consistent
with the scenario in which SMBH growth is dominated by prolonged, ordered
accretion. However, there may be observational biases towards AGN with high spin
in the compiled sample, limiting our ability to draw strong conclusions for the
general population at this stage. Finally, contrary to popular belief, our
analysis also implies that the dichotomy between radio loud/radio quiet AGN is
not solely related to black hole spin.
\end{abstract}


\section{Introduction}

Knowledge of the spin distribution for the supermassive black holes powering
active galactic nuclei (AGN) provides important information on the growth
history of these black holes, which would otherwise be difficult to
observationally constrain. The accretion and merger events experienced by
these nuclear black holes over their lifetimes can impart enough angular
momentum to substantially modify the black hole spins from their natal values
(\citealt{Moderski96,Volonteri05}). Recent simulations by \cite{Berti08}
demonstrate that the exact spin distribution expected depends strongly on the
nature of these events, and in particular how the angular momentum of the
accreted material/coalesced black holes relate to one another. If the accretion
is primarily `ordered', \ie the accreted material has angular momentum in the
same sense as that of the black hole, the black holes are spun up and will
preferentially display high spins. If instead the accretion is chaotic, \ie the
angular momenta of the accreted material and the AGN are not typically aligned,
the opposite will be true, leading to a strong preference for non-rotating
black holes.

Measuring black hole spin requires knowledge of the radius of the innermost
stable circular orbit (ISCO). Under the assumption that the inner accretion
disc extends in to this radius, the ISCO can be measured through study of the
reflected emission produced as the optically thick disc is irradiated by the
hard X-rays generated in the corona, most likely via Compton up-scattering.
This emission contains a combination of backscattered continuum emission and
atomic features lines, the most prominent of which is often the iron K$\alpha$
emission line (6.4--6.97\,\kev, depending on ionisation state) owing to its
high cosmic abundance and fluorescent yield (see \eg \citealt{George91}). The
reflected emission from the inner disc is blurred and broadened by the strong
Doppler and relativistic effects inherent to material in close orbit around a
black hole (\citealt{Fabian89, kdblur}). This can results in broad and skewed
observed emission lines profiles, even though the atomic emission is
intrinsically narrow, and if the amount of relativistic blurring can be
constrained, the inner radius of the disc can be quantified. Active galaxies
do often display broad emission features that can be associated with
relativistically broadened iron lines (see \eg \citealt{Tanaka95},
\citealt{Nandra99}, \citealt{Fabian00}, \citealt{Nandra07}, \citealt{FabZog09}),
and although questions have been raised over whether this is the correct
interpretation for these features, with absorption based models proposed as an
alternative (\citealt{Miller08, LMiller09}), in \cite{Walton12xrbAGN} we
present a strong argument in favour of the relativistic disc reflection origin
through comparison with the similar features frequently observed in Galactic
black hole binaries.

Although the iron K$\alpha$ line is frequently the most prominent reflected
feature, for a broad range of ionisation states the full reprocessed emission
spectrum also contains a host of other emission lines from lighter elements at
lower energies. In addition, reflection spectra also display a broad emission
feature peaking at $\sim$30\,\kev, referred to as the Compton hump, due to the
combined effects photoelectric absorption of photons at lower energies and
Compton down-scattering of photons at higher energies. In order to self
consistently treat the reflected emission arising from the inner regions of
the accretion disc, the same processes that broaden and skew the iron line must
be applied to all of these features. These effects can blend the soft complex
of reflected emission lines together into a smooth emission feature, offering a
natural explanation for the soft excess frequently observed in the spectra of
AGN (\citealt{Crummy06}). Therefore, where possible, the broadband spectra
covering all the key reflected emission features should be considered when
attempting to constrain the spin of AGN, and in principle it should be possible
to constrain the spin from the broadband spectrum even for sources that lack
prominent broad iron lines, assuming the soft excess is indeed associated with
reflection.

The best sources with which to perform such analysis and rigorously test the
reflection interpretation are therefore `bare' active galaxies, sources with
little or no intrinsic absorption to complicate the determination of the
reflected emission. Notable examples of such sources are Fairall\,9
(\citealt{Grondoin01, Schmoll09, Emman11a}), Ark\,120 (\citealt{Vaughan04,
Nardini11}), Ton\,S180 (\citealt{Vaughan02, Nardini12}), and the extreme narrow
line Seyfert 1 galaxy 1H\,0707-495 (\citealt{FabZog09}).

However, even for apparently bare AGN, interpretations other than reflection
have been proposed to explain the soft excess. Although bare AGN show little
or no absorption by partially ionised material when studied by high resolution
grating spectrometers, it was suggested that the broad, overall curvature
observed at low energies could still be caused by such absorption in an
extremely turbulent outflow, in which large velocity broadening smears out the
discrete atomic features (\citealt{Gierlin04, Middleton07}). However, more
detailed simulations show that outflows driven from the disc are not actually
able to provide the velocity broadening required (\citealt{Schurch07,
Schurch09}). More recently, complex hybrid Comptonisation interpretations have
successfully been applied to the data, in which the soft excess arises due to
Comptonisation by thermal electrons within the inner disc, possibly existing
as a skin-like layer on the surface of the disc, while the higher energy
continuum arises in a separate, more traditional corona (\citealt{Done12,
Jin12}).

Nevertheless, there is very good evidence to suggest that the soft excess does
indeed arise due to reprocessing of the primary continuum. First of all, the
energy of the soft excess is observed to be remarkably constant over many
orders of magnitude in black hole mass and also over a fair range in Eddington
ratio (\citealt{Gierlin04, Miniutti09}), which would require a roughly constant
electron temperature in the Comptonisation interpretation. Indeed, roughly
constant electron temperatures were obtained by \cite{Done12} and \cite{Jin12}.
The temperature of the accretion disc is dependent on both black hole mass and
Eddington ratio, and should vary substantially over the range of black hole
masses and luminosities observed. It is probably not unreasonable to expect
that some variation should also be seen in the temperature of the disc
electrons over these ranges. The constant energy of this feature therefore
strongly hints at an atomic origin which, given the inability of absorption to
reproduce the soft excess highlighted previously, supports a reflection
interpretation.

In addition, one of the key observational discoveries over the last couple of
years is the detection of lags between the soft and hard X-ray emission, which
demonstrate that, on short timescales, the soft excess responds to changes in the
high energy continuum (\citealt{Zoghbi10, Zoghbi11b, Emman11b, deMarco11}), as
predicted by the reflection based interpretation. Although there have been
claims that absorption based interpretations can also reproduce these lags via
reflection from the more distant obscuring clouds in the case of 1H\,0707-495
(\citealt{LMiller10}), this requires both a special source geometry and viewing
alignment, while evidence for such lags have now been seen in a large number of
sources (\citealt{deMarco12}), arguing strongly against this interpretation.
Crucially, a similar lag has also now been detected between the high energy
continuum and the broad component of the iron line profile in NGC\,4151
(\citealt{Zoghbi12}), strongly suggesting that such lags are indeed
characteristic for reflected emission. These lags also appear to argue strongly
against the Comptonisation interpretation, as fluctuations in the corona should
not precede fluctuations in the inner disc.

In this paper, we present a reflection based analysis of a sample of bare AGN,
with the purpose of both testing the robustness of the reflection interpretation,
and where possible providing initial spin constraints for the AGN considered. We
specifically focus on \suzaku\ observations in this work in order to utilise its
combination of broadband spectral coverage and large collecting area. The paper
is structured as follows: section \ref{sec_c3_red} describes our data reduction
and sample selection procedure, and section \ref{sec_c3_spec} details our spectral
analysis. We discuss our results in section \ref{sec_c3_disc} and finally summaries
our conclusions in section \ref{sec_c3_conc}.

\section{Observations and Data Reduction}
\label{sec_c3_red}

\begin{table*}
  \caption[Observational details for the sample of `bare' AGN compiled.]
{Observational details for the sample of `bare' AGN compiled in this work, ordered
in terms of the total XIS counts recorded. The uncertainties quoted on the XIS
countrates are the 1$\sigma$ uncertainties, while the PIN source rates are quoted
as a percentage of the total rate recorded by the detector. Galactic column
densities are taken from \cite{NH}, and positions and redshifts are compiled from
the NASA Extragalactic Database (NED).}
\begin{center}
\begin{tabular}{c c c c c c c c c c c}
\hline
\hline
\\[-0.3cm]
Source & RA & DEC & $z$ & $N_{\rm H, Gal}$ & OBSID & Obs. Date & Exposure & FI XIS Rate & PIN \% \\
& (h:m:s) & (d:m:s) & & ($10^{20}$ cm$^{-2}$) & & & (XIS/PIN; ks) & (ct/s) & \\
\\[-0.3cm]
\hline
\hline
\\[-0.25cm]
Mrk\,509 & 20:44:09.7 & -10:43:25 & 0.0344 & 4.25 & 701093010 & 25/04/2006 & 22/15 & $3.446\pm0.008$ & 30 \\
& & & & & 701093020 & 14/10/2006 & 24/22 & $4.416\pm0.008$ & 25 \\
& & & & & 701093030 & 15/11/2006 & 18/17 & $4.42\pm0.01$ & 24 \\
& & & & & 701093040 & 27/11/2006 & 28/28 & $3.821\pm0.008$ & 23 \\
\\[-0.25cm]
3C\,382 & 18:35:03.4 & +32:41:47 & 0.05787 & 6.98 & 702125010 & 01/11/2007 & 120/114 & $3.075\pm0.004$ & 22 \\
\\[-0.25cm]
Mrk\,335 & 00:06:19.5 & +20:12:10 & 0.0258 & 3.56 & 701031010 & 21/06/2006 & 151/132 & $1.569\pm0.002$ & 7 \\
\\[-0.25cm]
Fairall 9 & 01:23:45.8 & -58:48:20 & 0.0470 & 3.16 & 702043010 & 07/06/2007 & 145/127 & $1.963\pm0.003$ & 15 \\
\\[-0.25cm]
1H\,0419-577 & 04:26:00.8 & -57:12:00 & 0.1040 & 1.26 & 702041010 & 25/07/2007 & 179/161 & $1.453\pm0.002$ & 14 \\
& & & & & 704064010 & 16/01/2010 & 119/107 & $0.953\pm0.002$ & 7 \\
\\[-0.25cm]
Ark\,564 & 22:42:39.3 & +29:43:31 & 0.0247 & 5.34 & 702117010 & 26/06/2007 & 89/110 & $2.678\pm0.004$ & 7 \\
\\[-0.25cm]
Ark\,120 & 05:16:11.4 & -00:08:59 & 0.0327 & 9.78 & 702014010 & 01/04/2007 & 91/104 & $2.240\pm0.004$ & 16 \\
\\[-0.25cm]
3C\,390.3 & 18:42:09.0 & +79:46:17 & 0.0561 & 3.47 & 701060010 & 14/12/2006 & 91/92 & $2.083\pm0.004$ & 22 \\
\\[-0.25cm]
PKS\,0558-504 & 05:59:47.4 & -50:26:52 & 0.1372 & 3.36 & 701011010 & 17/01/2007 & 21/18 & $1.415\pm0.006$ & 8 \\
& & & & & 701011020 & 18/01/2007 & 19/17 & $2.001\pm0.007$ & 6 \\
& & & & & 701011030 & 19/01/2007 & 21/18 & $1.251\pm0.006$ & 4 \\
& & & & & 701011040 & 20/01/2007 & 20/16 & $2.049\pm0.007$ & 7 \\
& & & & & 701011050 & 21/01/2007 & 20/15 & $2.078\pm0.007$ & 5 \\
\\[-0.25cm]
NGC\,7469 & 23:03:15.6 & +08:52:26 & 0.0163 & 4.45 & 703028010 & 24/06/2008 & 92/85 & $1.258\pm0.003$ & 19 \\
\\[-0.25cm]
Mrk\,110 & 09:25:12.9 & +52:17:11 & 0.0353 & 1.30 & 702124010 & 02/11/2007 & 91/80 & $1.414\pm0.003$ & 13 \\
\\[-0.25cm]
Swift\,J0501.9-3239 & 05:19:35.8 & -32:39:28 & 0.0124 & 1.75 & 703014010 & 11/04/2008 & 36/34 & $2.362\pm0.006$ & 22 \\
\\[-0.25cm]
Mrk\,841 & 15:04:01.2 & +10:26:16 & 0.0364 & 2.22 & 701084010 & 22/01/2007 & 45/41 & $0.950\pm0.003$ & 12 \\
& & & & & 701084020 & 23/07/2007 & 45/60 & $0.961\pm0.003$ & 12 \\
\\[-0.25cm]
Ton\,S180 & 00:57:19.9 & -22:22:59 & 0.0620 & 1.36 & 701021010 & 09/12/2006 & 121/102 & $0.820\pm0.002$ & 4 \\
\\[-0.25cm]
PDS\,456 & 17:28:19.8 & -14:15:56 & 0.1840 & 19.6 & 701056010 & 24/02/2007 & 174/209 & $0.381\pm0.001$ & 2 \\
\\[-0.25cm]
1H\,0323+342 & 03:24:41.1 & +34:10:46 & 0.0610 & 12.7 & 704034010 & 26/07/2009 & 76/69 & $0.559\pm0.002$ & 11 \\
\\[-0.25cm]
UGC\,6728 & 11:45:16.0 & +79:40:53 & 0.0065 & 4.42 & 704029010 & 06/06/2009 & 46/39 & $0.897\pm0.003$ & 14 \\
\\[-0.25cm]
Mrk\,359 & 01:27:32.5 & +19:10:44 & 0.0174 & 4.26 & 701082010 & 06/02/2007 & 98/96 & $0.307\pm0.001$ & 4 \\
\\[-0.25cm]
MCG--2-14-9 & 05:16:21.2 & -10:33:41 & 0.0285 & 7.13 & 703060010 & 28/08/2008 & 131/120 & $0.217\pm0.001$ & 5 \\
\\[-0.25cm]
ESO\,548-G081 & 03:42:03.7 & -21:14:40 & 0.0145 & 2.29 & 704026010 & 03/08/2009 & 39/33 & $0.661\pm0.003$ & 15 \\
\\[-0.25cm]
Mrk\,1018 & 02:06:16.0 & -00:17:29 & 0.0424 & 2.43 & 704044010 & 03/07/2009 & 37/35 & $0.548\pm0.003$ & 13 \\ 
\\[-0.25cm]
RBS\,1124 & 12:31:36.4 & +70:44:14 & 0.2080 & 1.52 & 702114010 & 14/04/2007 & 79/83 & $0.241\pm0.001$ & 10 \\
\\[-0.25cm]
IRAS\,13224-3809 & 13:12:19.4 & -38:24:53 & 0.0658 & 5.34 & 701003010 & 26/01/2007 & 166/159 & $0.102\pm0.001$ & - \\
\\[-0.25cm]
1H\,0707-495 & 07:08:41.5 & -49:33:06 & 0.0406 & 4.31 & 700008010 & 03/12/2005 & 121/137 & $0.060\pm0.001$ & - \\
\\[-0.25cm]
IRAS\,05262+4432 & 05:29:55.5 & +44:34:39 & 0.0322 & 32.1 & 703019010 & 12/09/2008 & 68/66 & $0.088\pm0.001$ & - \\
\\[-0.3cm]
\hline
\hline
\end{tabular}
\label{tab_c3_obs}
\end{center}
\end{table*}

\subsection{Data Reduction}

We reduced all the \suzaku\ observations of type 1 AGN publicly available prior
to October 2010. Using the latest \heasoftv software package we processed the
unfiltered event files for each of the \xis\ CCDs and editing modes operational
in the respective observations, following the \suzaku\ Data Reduction
Guide\footnote{http://heasarc.gsfc.nasa.gov/docs/suzaku/analysis/}. We started by
creating new cleaned event files for all the XIS detectors operational during each
observation (note that XIS2, one of the front illuminated detectors, experienced a
charge leak in November 2006 and has not been in operation since) by re-running the
\suzaku\ pipeline with the latest calibration, as well as the associated screening
criteria files. In addition to the standard XIS event selection criteria, we also
required that the cut-off rigidity for clean events be greater than 6\,GeV. The
exact size of the circular source regions used varied from source to source,
depending on \eg the presence of other nearby sources, but were typically $\sim$3.5'
in radius, and background regions were selected from the surrounding areas on the
CCD free of any contaminating sources, with care taken to avoid the calibration
sources in the corners. \xselect was used to extract spectral products from these
event files, and responses were generated for each individual spectrum using the
{\small XISRESP} script with a medium resolution. The spectra and response files
for all the front-illuminated instruments, XIS0, XIS2 (where operational) and XIS3
were combined using the \ftool\ \addascaspec. Since we are interested in the
average spectral properties of active galaxies in this work, where a source was
observed on more than one occasion the individual spectra obtained from these
observations were also combined into a single, averaged spectrum; for these sources,
the spectra obtained from each of the individual observations were found to be
broadly similar in each case. Finally, we grouped the spectra to have a minimum
signal-to-noise (S/N) of 5 per energy bin with the {\small SPECGROUP} task (part of
the \xmm\ {\small SAS}), to allow the use of \chisq\ minimization during spectral
fitting. During modelling, we do not consider the XIS spectra below 0.6\,\kev\ and
also exclude the 1.7--2.5\,\kev\ energy range owing to calibration uncertainties,
and we primarily make use of the front illuminated spectra, owing to its superior
performance at high energies, although the back illuminated spectra are routinely
checked for consistency.

For the \hxd\ \pin\ detector we again reprocessed the unfiltered event files for each
of the observations considered following the data reduction guide. Since the \hxd\ is
a collimating instrument rather than an imaging spectrometer, estimating the background
requires individual consideration of the non X-ray instrumental background (NXB) and
cosmic X-ray background (CXB). The response and NXB files were downloaded for each
observation\footnote{http://www.astro.isas.ac.jp/suzaku/analysis/hxd/}, for all of the
observations considered here the higher quality `tuned' (Model D) background was
available. Spectral products were generated using the {\small HXDPINXBPI} script
provided by the \suzaku\ team, which includes a simulated contribution from the CXB
using the form of \cite{Boldt87}. As with the XIS detectors, we combined the individual
spectra of sources with multiple observations into a single, averaged spectrum, and
rebinned the data so that each energy bin had a minimum S/N of 3, lower than the minimum
S/N required in the XIS spectra owing to the much lower PIN source rates, but still
sufficient to allow the use of \chisq\ minimization. Nominally we consider the PIN
spectrum over the full 15--70\,\kev\ energy band, but in reality the exact energy
range considered varies from source to source, depending on the highest energy at which
each source is reliably detected.

\subsection{Sample Selection}
\label{sec_sel}

In this work, we are interested in `bare' AGN, \ie active galaxies with little or no
intrinsic absorption, that display soft excesses. In order to identify such sources,
we modelled the $\sim$2.5--10\,\kev\ energy range with a powerlaw continuum modified
by Galactic absorption (see Table \ref{tab_c3_obs} for the Galactic neutral column
densities adopted, as determined by \citealt{NH}), modelled with the \tbabs\
absorption code (\citealt{tbabs}; the appropriate solar abundances for this model are
used throughout); the iron K band (generally taken here as 4--7\,\kev) was excluded
to prevent any iron emission present in the spectrum from influencing our continuum
determination. We then extrapolated this continuum back through the 0.6--2.0\,\kev\
data and selected sources that displayed a clear, and relatively smooth (\ie not
dominated by diffuse thermal emission, as is the case for \eg\ NGC\,1365;
\citealt{Wang09}) excess of emission over that predicted at lower energies. The
resulting sample contains 25 sources that met our selection criteria, the
observational details of these sources analysed are listed in Table \ref{tab_c3_obs}.
Note that this is largely based on a visual inspection of the spectra, so some of the
sources selected may (and indeed do) display some mild, partially ionised absorption,
but even in this minority of cases this absorption does not have a large effect on
the observed spectrum.

\section{Spectral Analysis}
\label{sec_c3_spec}

Here we detail our spectral analysis of the sample outlined in the previous section
(see Table \ref{tab_c3_obs}). In section \ref{sec_spec_basic} we outline the basic
approach we adopt for all the sources, and in section \ref{sec_spec_details} we
provide additional, detailed information on a source-by-source basis. Spectral
modelling is performed with \xspecv\ (\citealt{XSPEC}), and throughout this work the
uncertainties quoted on model parameters are the 90 per cent confidence limits for
one parameter of interest, unless stated otherwise. We also adopt as standard the
cross-normalisation constant of either $C_{\rm PIN/XIS0}$ = 1.16 or 1.18 between the
XIS0 and PIN spectra (the former for observations with the nominal pointing position
centred on the XIS detectors, and the latter on the HXD detector), as recommended 
by the \hxd\ calibration team\footnote{http://www.astro.isas.jaxa.jp/suzaku/doc/suzakumemo/suzakumemo-2008-06.pdf}.
However, since we are using combined FI XIS spectra in this work (XIS0, XIS2 and XIS3
where available) we take the extra step of comparing this with the XIS0 spectrum on
a source by source basis in order to calculate the correct normalisation constant for
use between the combined XIS and PIN spectra, $C_{\rm PIN/XIS}$, using equation
\ref{eqn_crossnorm}:

\vspace{-0.2cm}
\begin{equation}
C_{\rm PIN/XIS} = C_{\rm XIS0/XIS} \times C_{\rm PIN/XIS0}
\label{eqn_crossnorm}
\end{equation}
\vspace{-0.2cm}

\noindent{We calculate $C_{\rm XIS0/XIS}$ by simultaneously modelling the combined
FI XIS and XIS0 spectra with a phenomenological broken powerlaw model for the broad
0.6--10.0\,\kev\ continuum (excluding again the energy ranges 1.7--2.5\,\kev\ due to
calibration uncertainties and 4.0--7.0\,\kev\ to exclude any iron emission), with all
the physical parameters linked for the two spectra, and only the normalisation constant
allowed to vary. The values of $C_{\rm PIN/XIS}$ we obtain are presented in Table
\ref{tab_c3_refl}. However, while this gives the appropriate normalisation constant for
the situation in which the PIN background is known perfectly, it does not account for
any systematic uncertainties in the modelled background currently available, which are
estimated by the HXD team to be 3 per cent\footnote{http://heasarc.gsfc.nasa.gov/docs/suzaku/analysis/abc/}
therefore we also investigate the consequences of allowing the XIS/PIN cross-normalisation
parameter to vary within the range equivalent to $\pm$3 per cent of the background,
these ranges and the values obtained are quoted in Table \ref{tab_refl2}.}

\subsection{Basic Approach}
\label{sec_spec_basic}

\begin{figure*}
\begin{center}
\rotatebox{0}{
{\includegraphics[width=159pt]{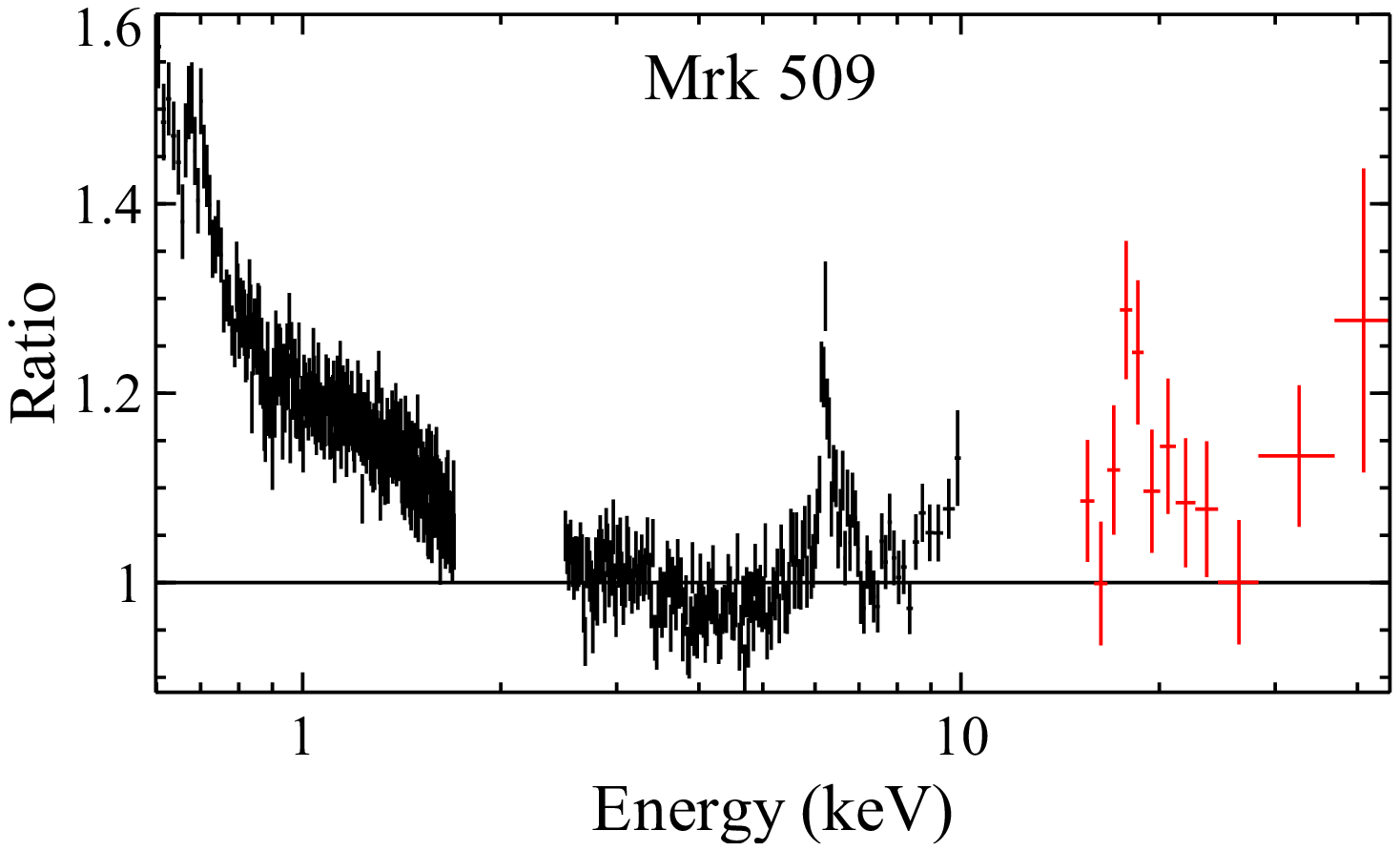}}
}
\rotatebox{0}{
{\includegraphics[width=159pt]{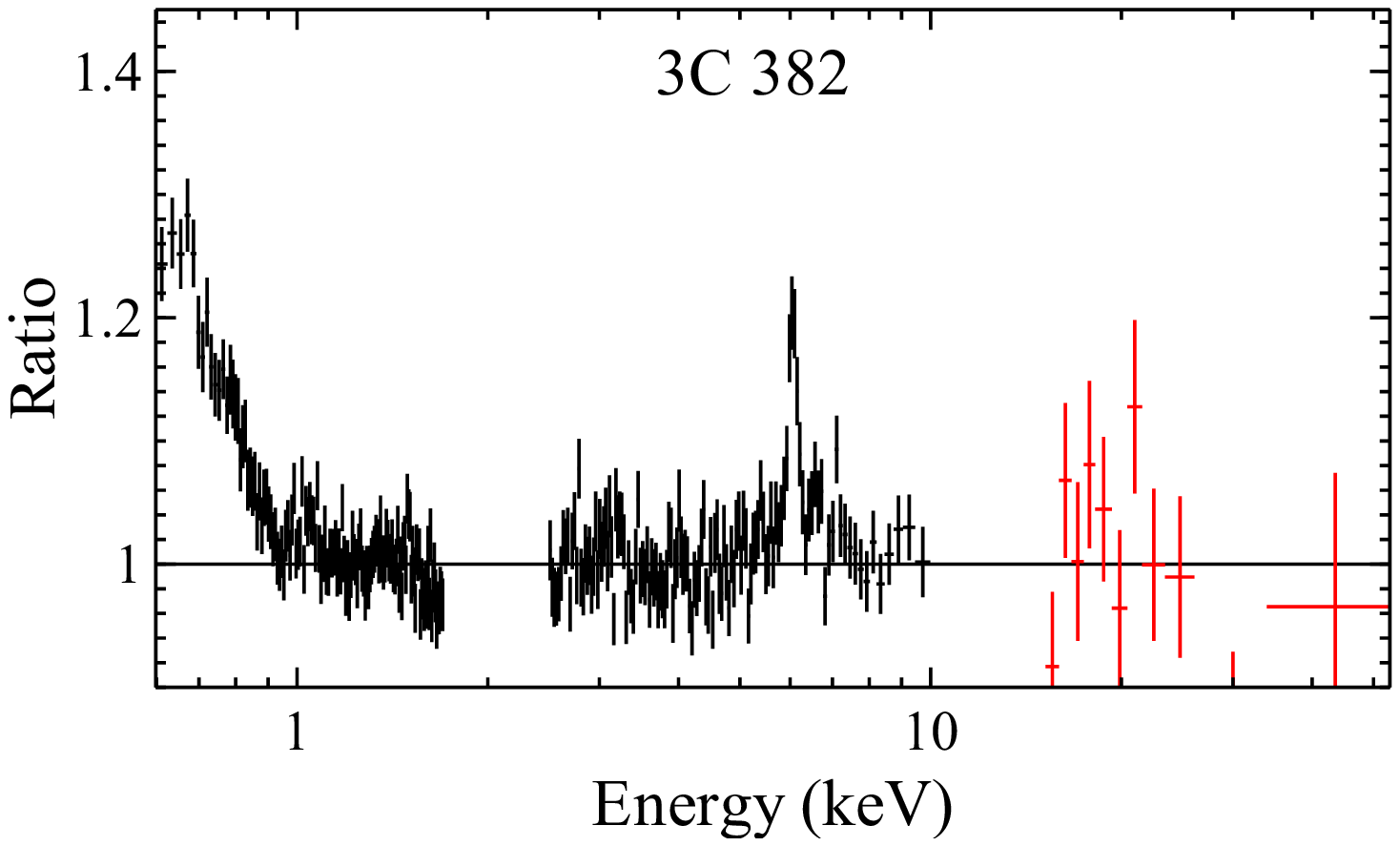}}
}
\rotatebox{0}{
{\includegraphics[width=159pt]{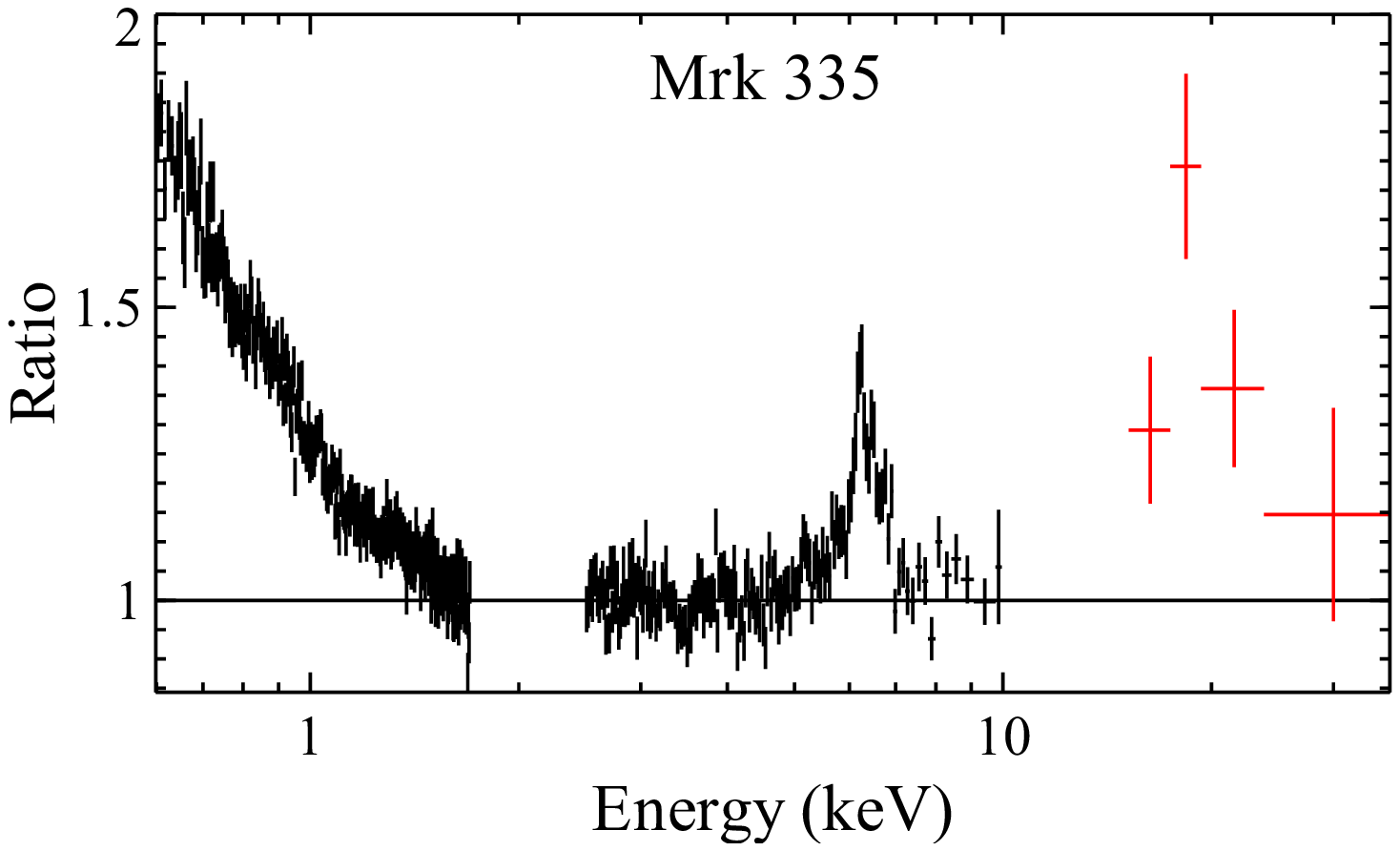}}
}\\
\vspace*{0.4cm}
\rotatebox{0}{
{\includegraphics[width=159pt]{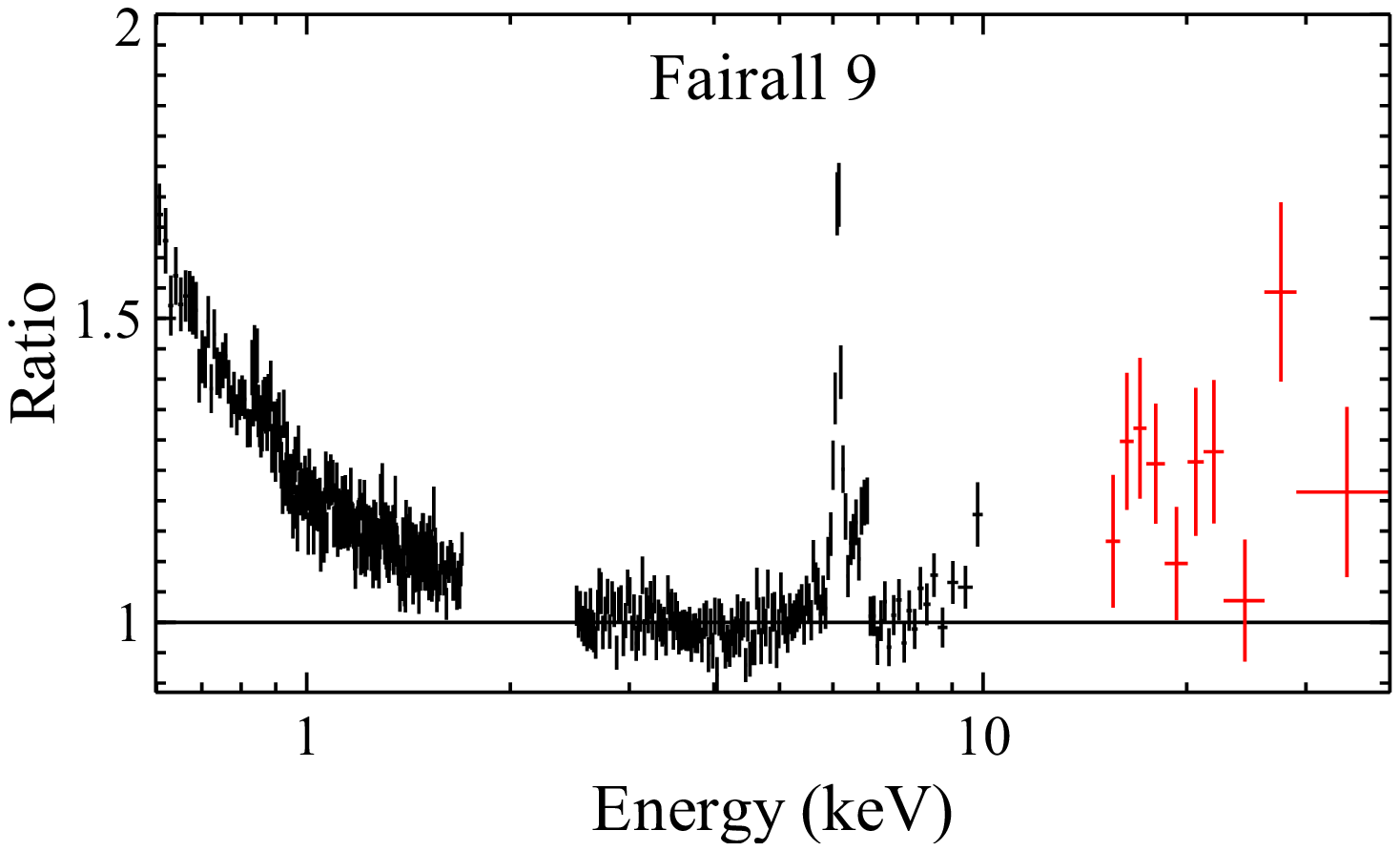}}
}
\rotatebox{0}{
{\includegraphics[width=159pt]{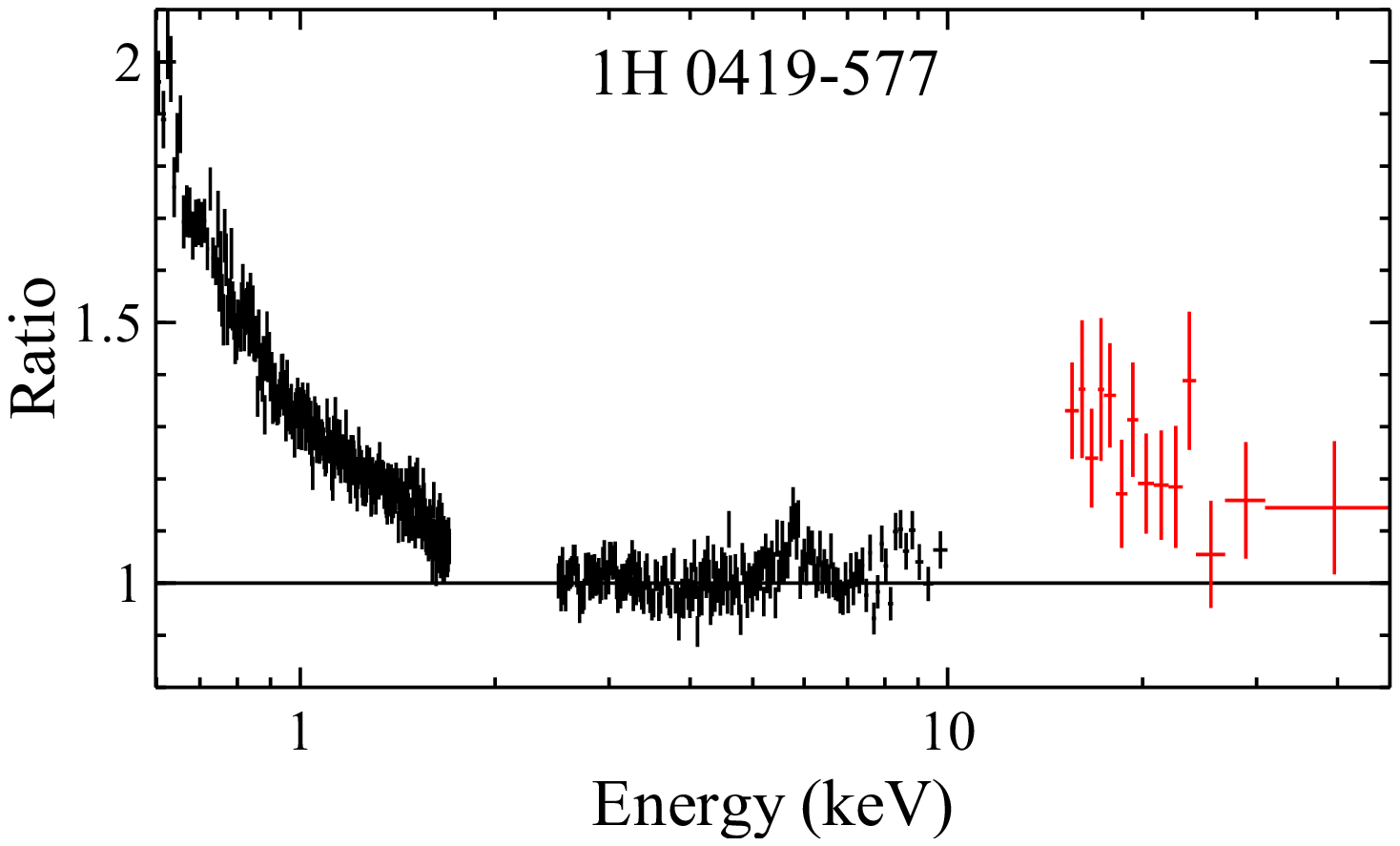}}
}
\rotatebox{0}{
{\includegraphics[width=159pt]{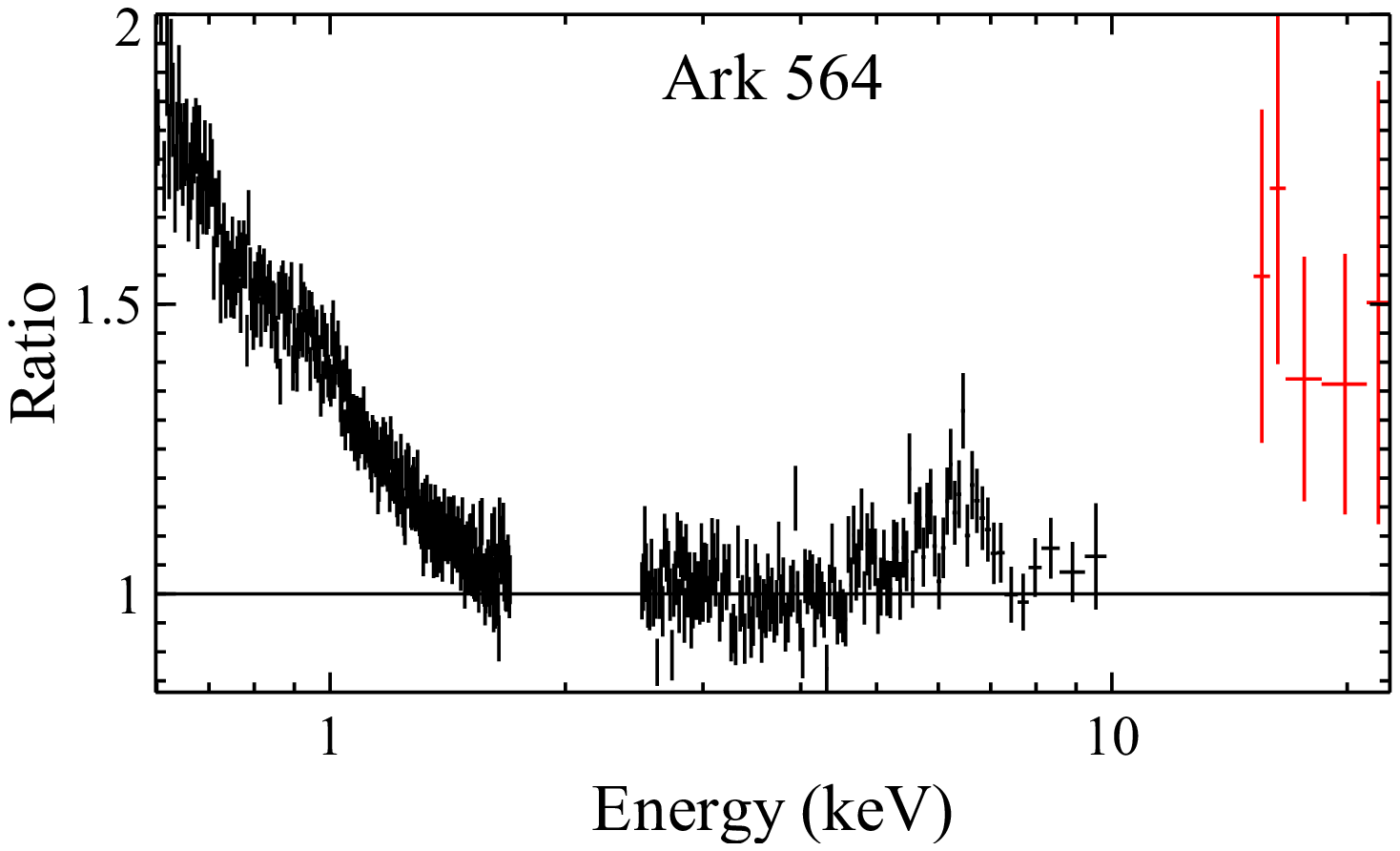}}
}\\
\vspace*{0.4cm}
\rotatebox{0}{
{\includegraphics[width=159pt]{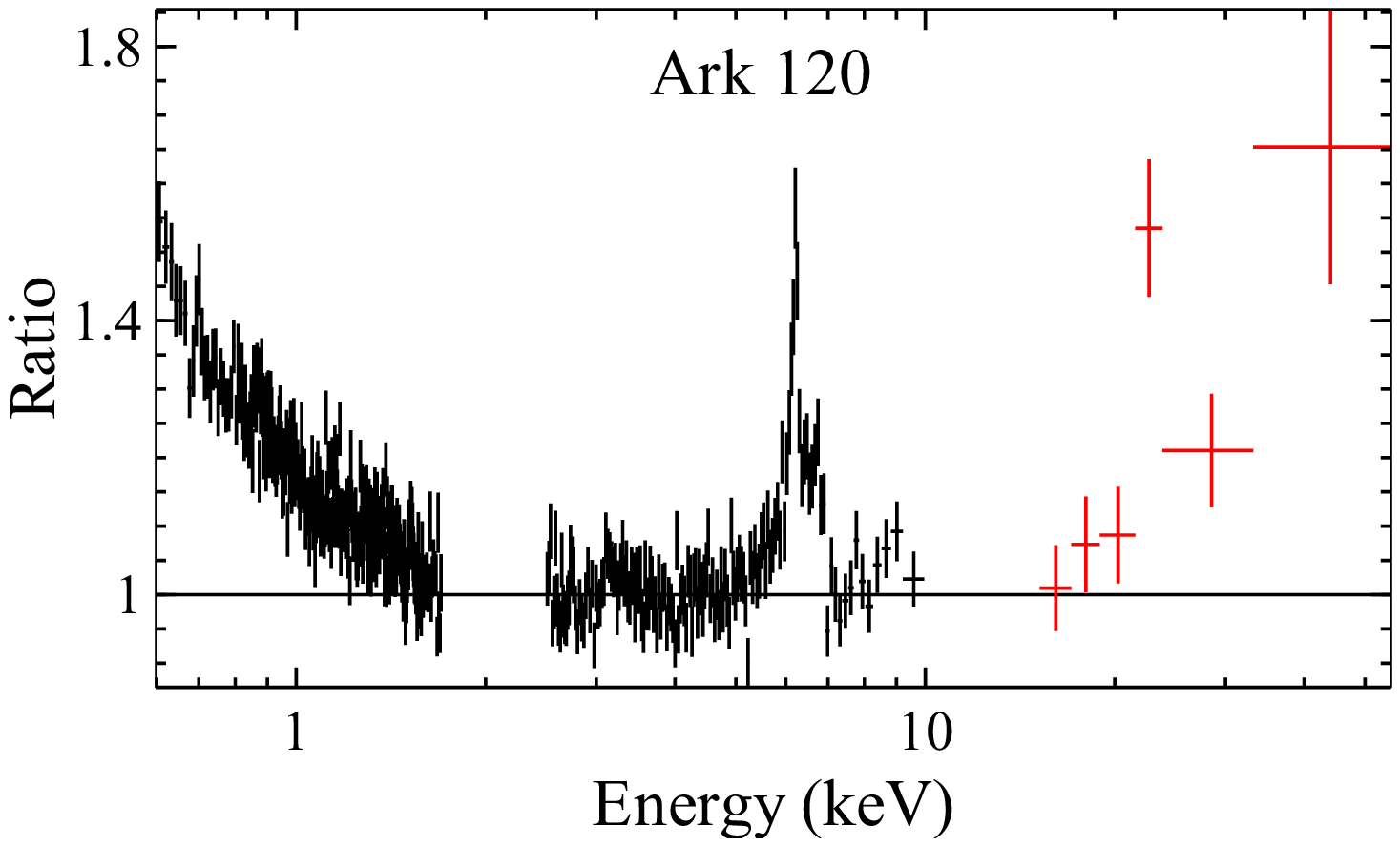}}
}
\rotatebox{0}{
{\includegraphics[width=159pt]{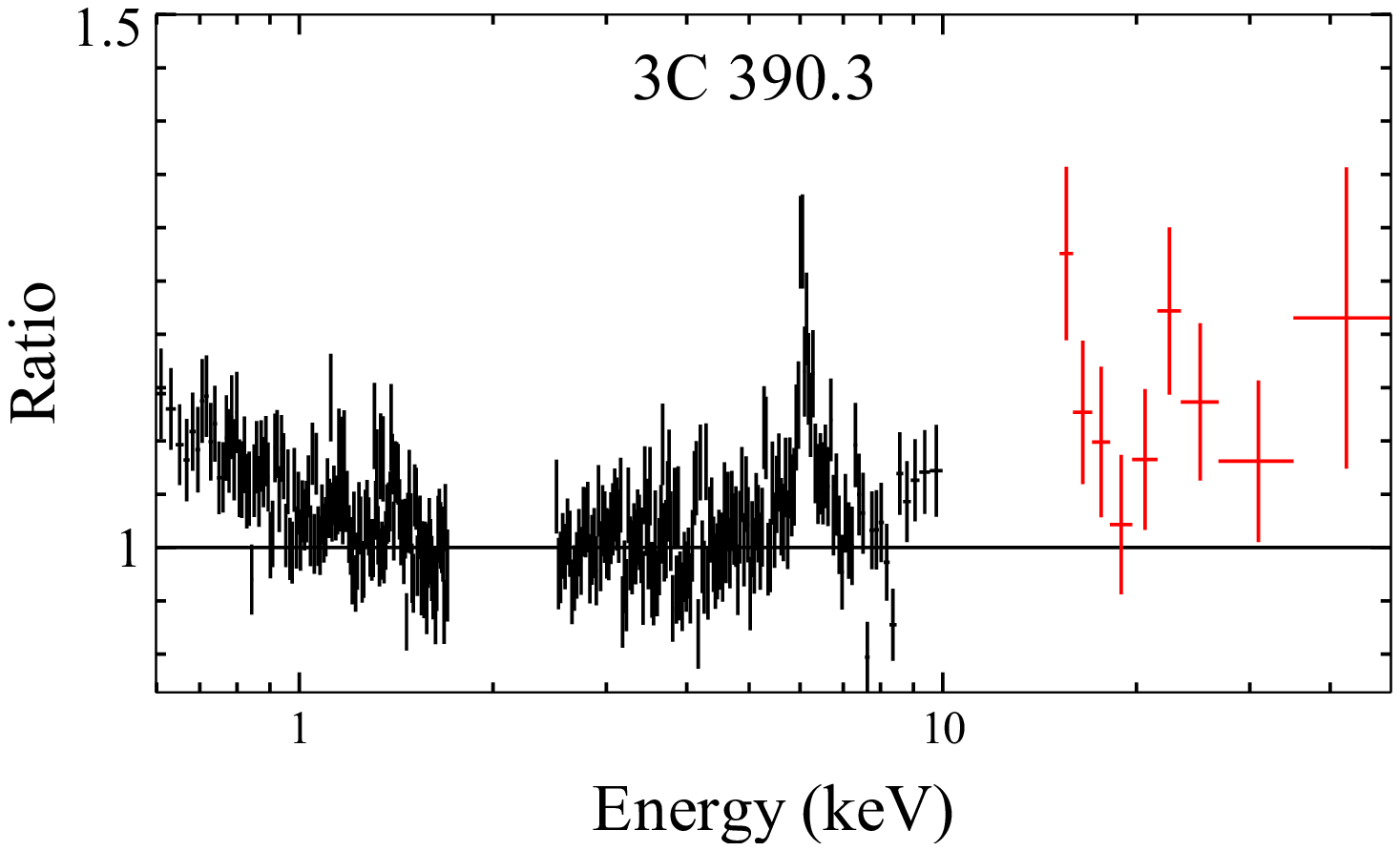}}
}
\rotatebox{0}{
{\includegraphics[width=159pt]{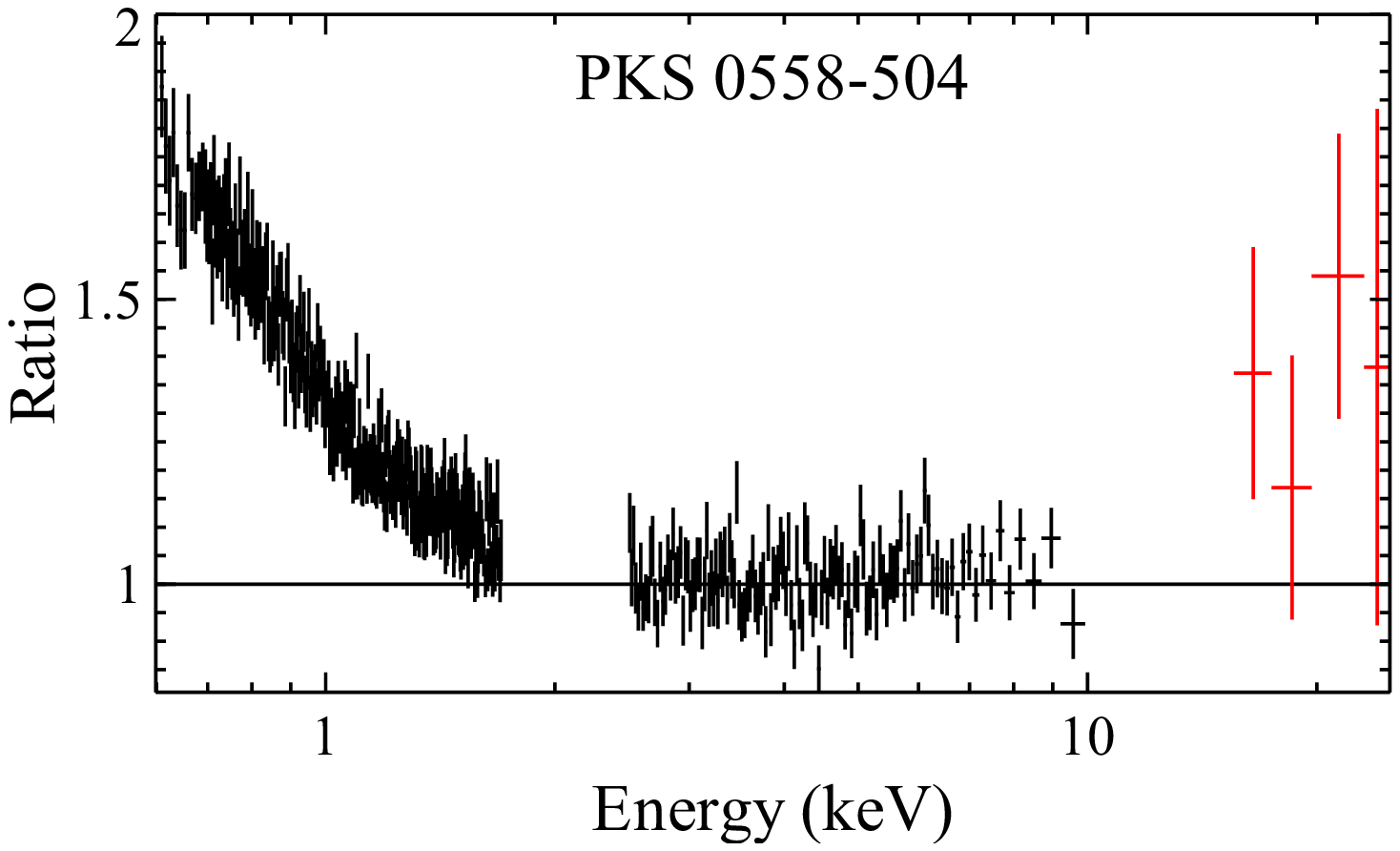}}
}\\
\vspace*{0.4cm}
\rotatebox{0}{
{\includegraphics[width=159pt]{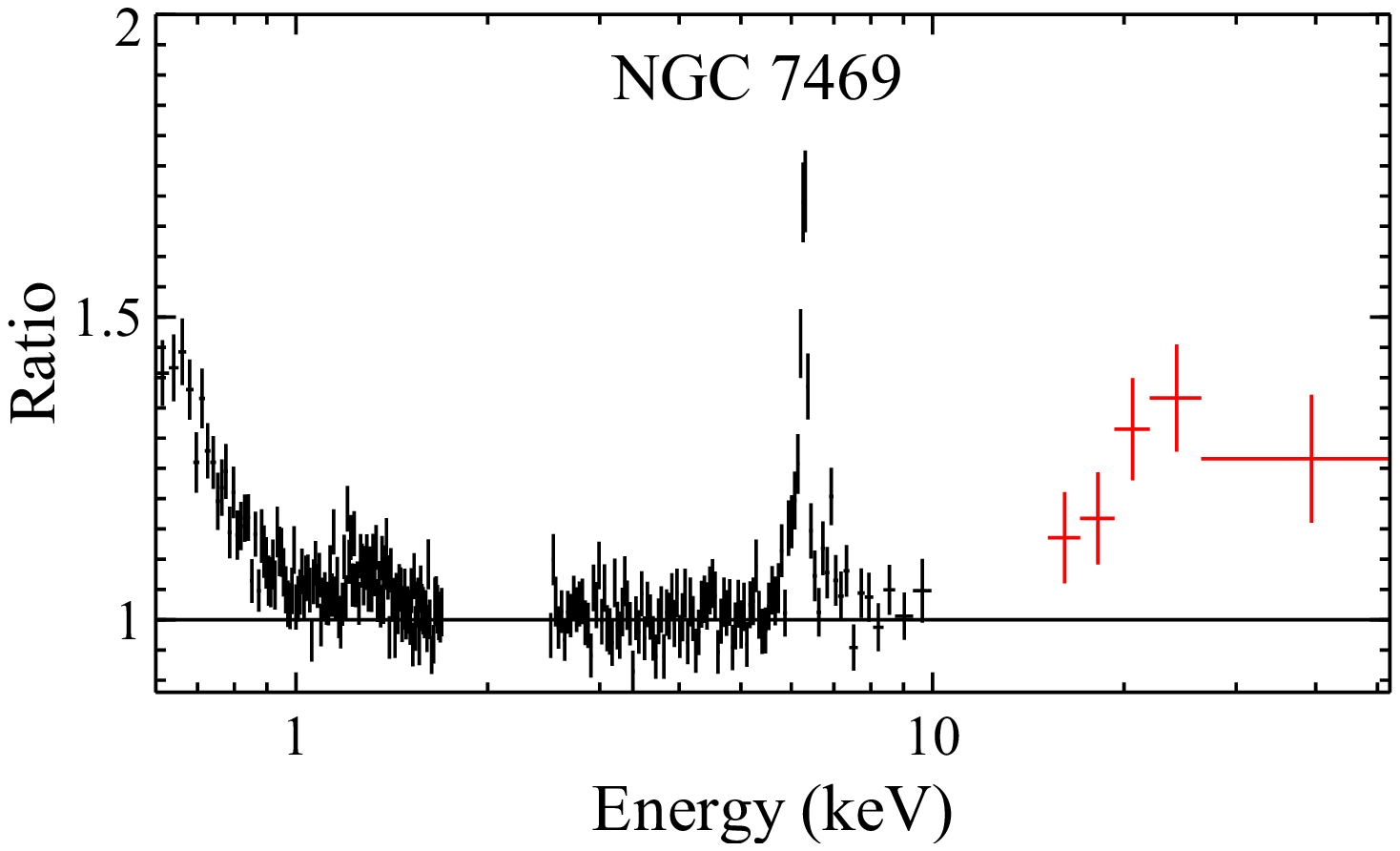}}
}
\rotatebox{0}{
{\includegraphics[width=159pt]{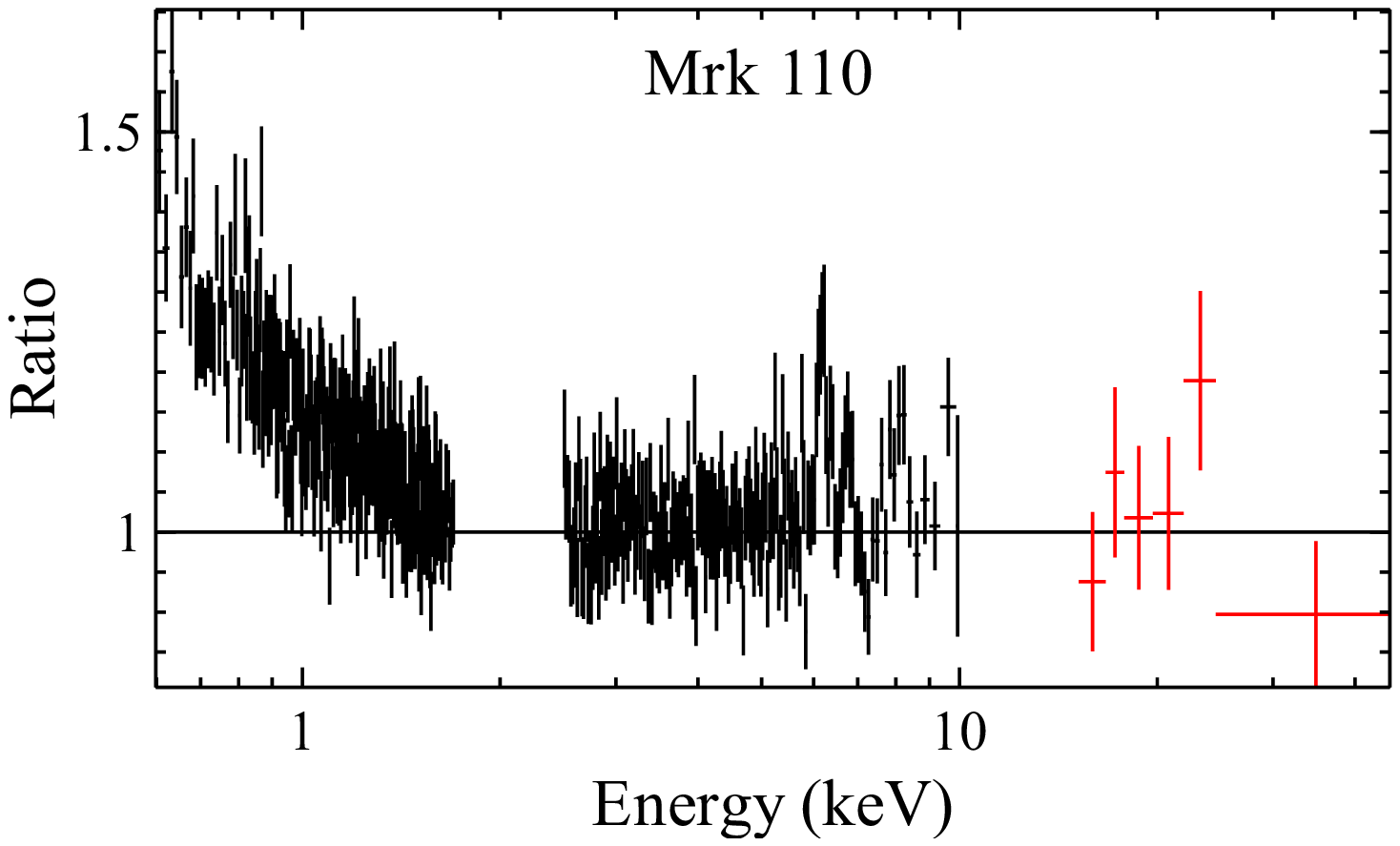}}
}
\rotatebox{0}{
{\includegraphics[width=159pt]{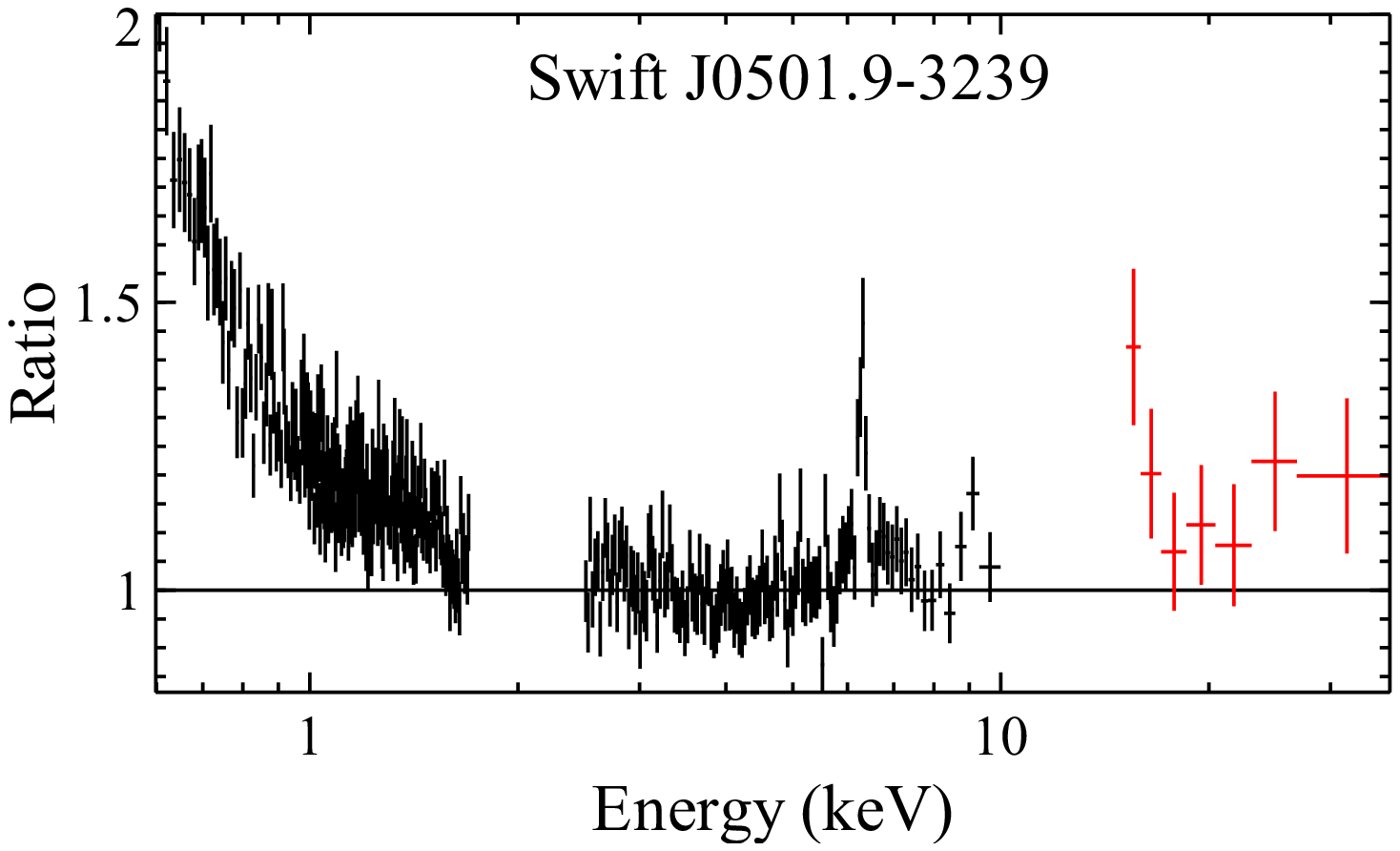}}
}\\
\vspace*{0.4cm}
\rotatebox{0}{
{\includegraphics[width=159pt]{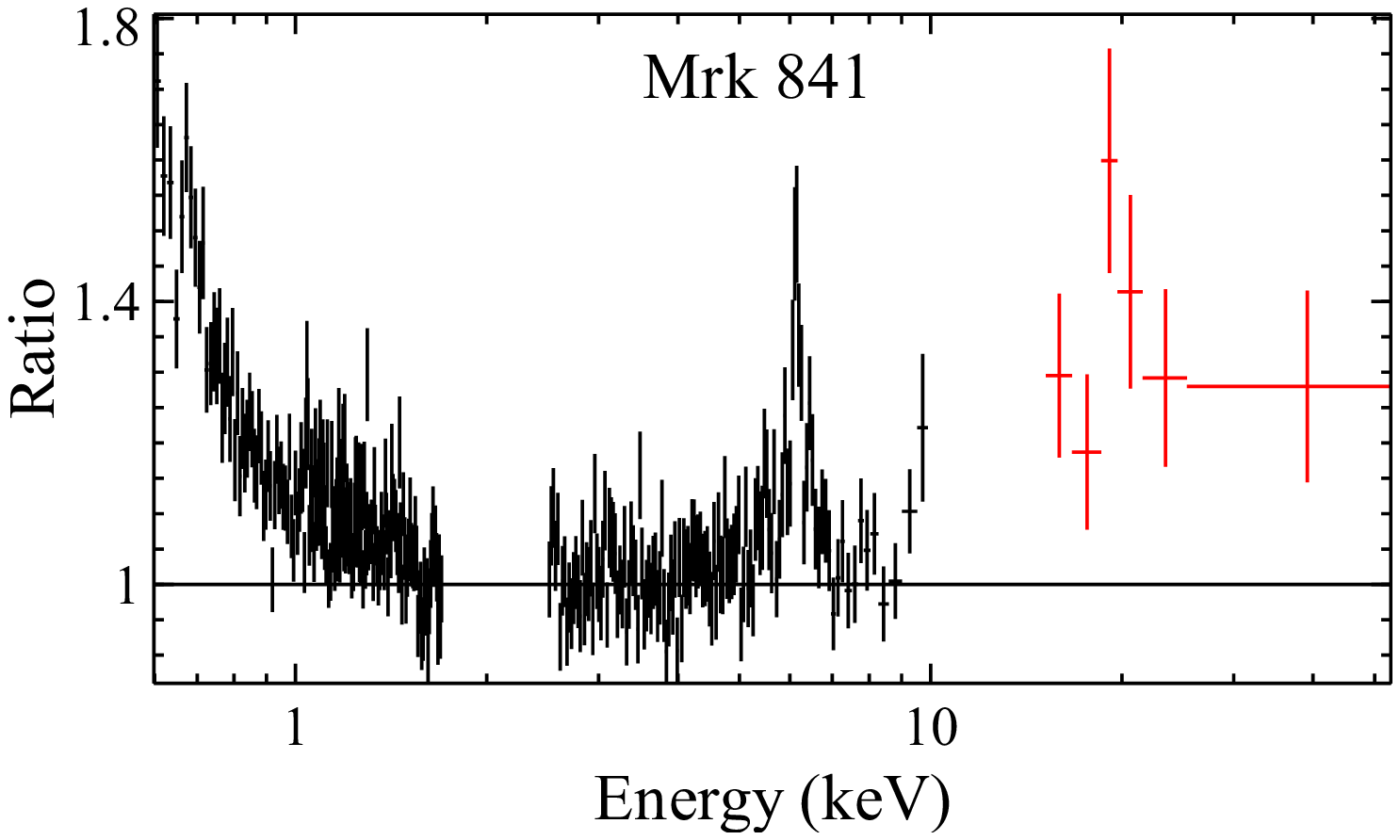}}
}
\rotatebox{0}{
{\includegraphics[width=159pt]{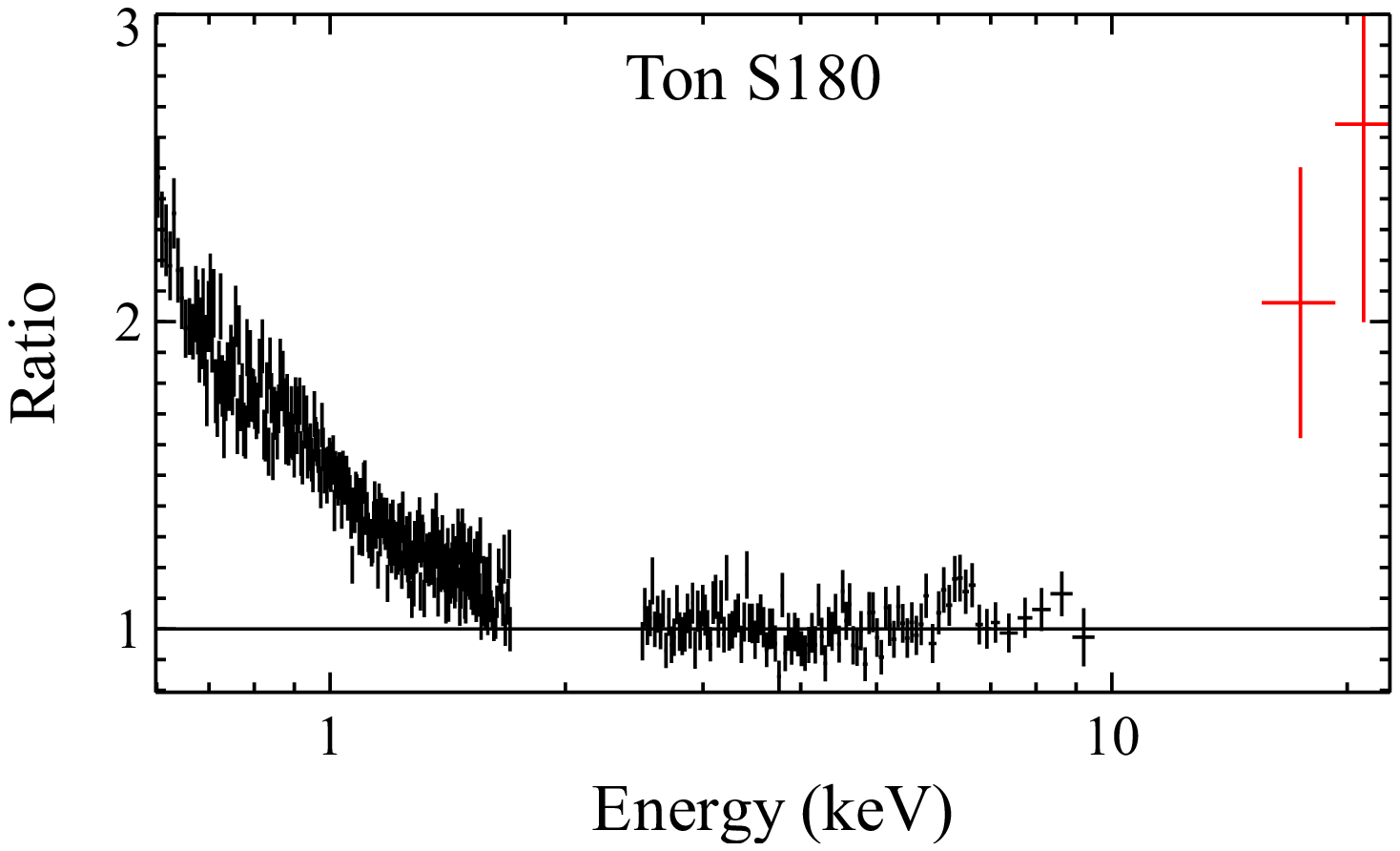}}
}
\rotatebox{0}{
{\includegraphics[width=159pt]{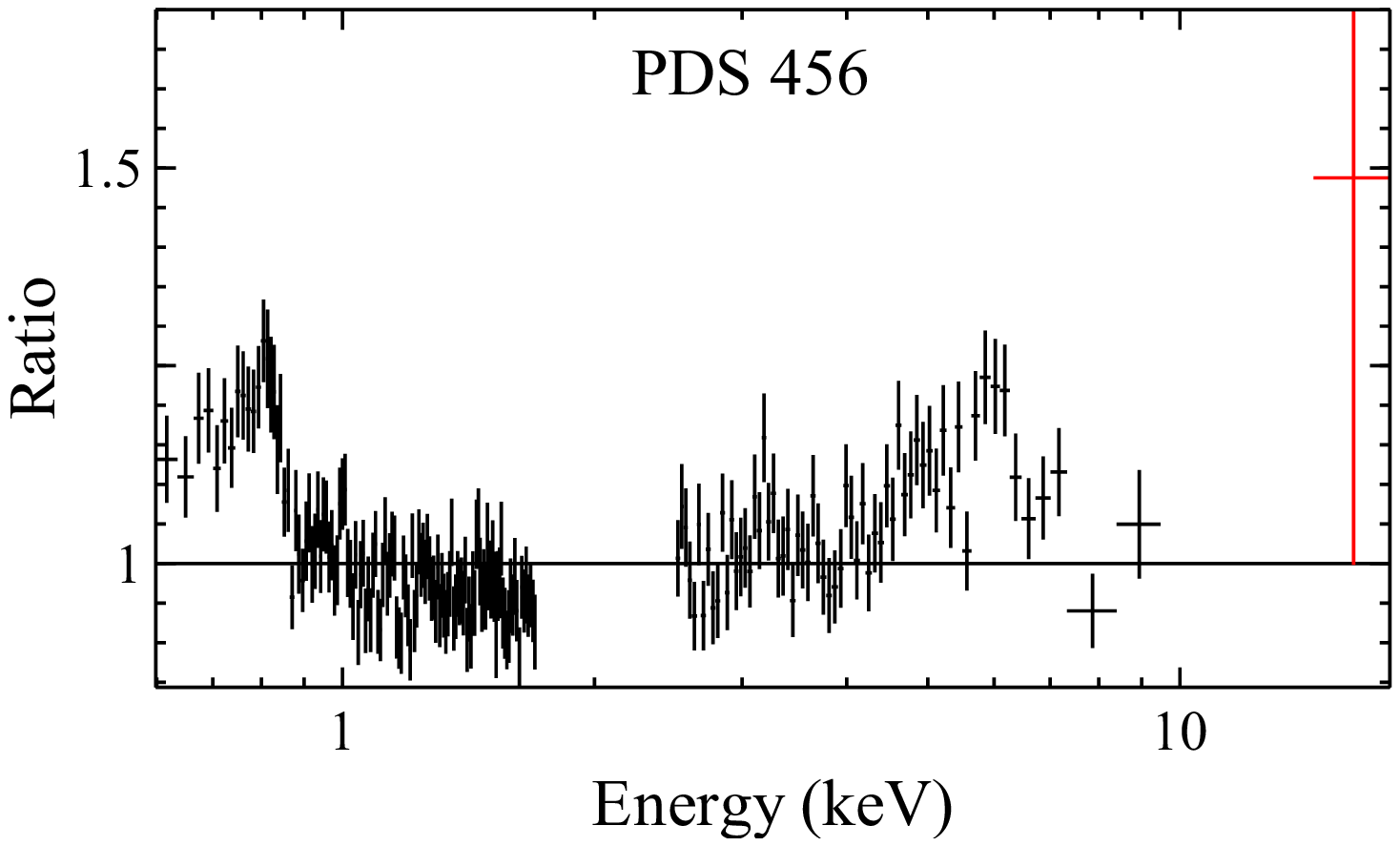}}
}\\
\vspace*{0.4cm}
\rotatebox{0}{
{\includegraphics[width=159pt]{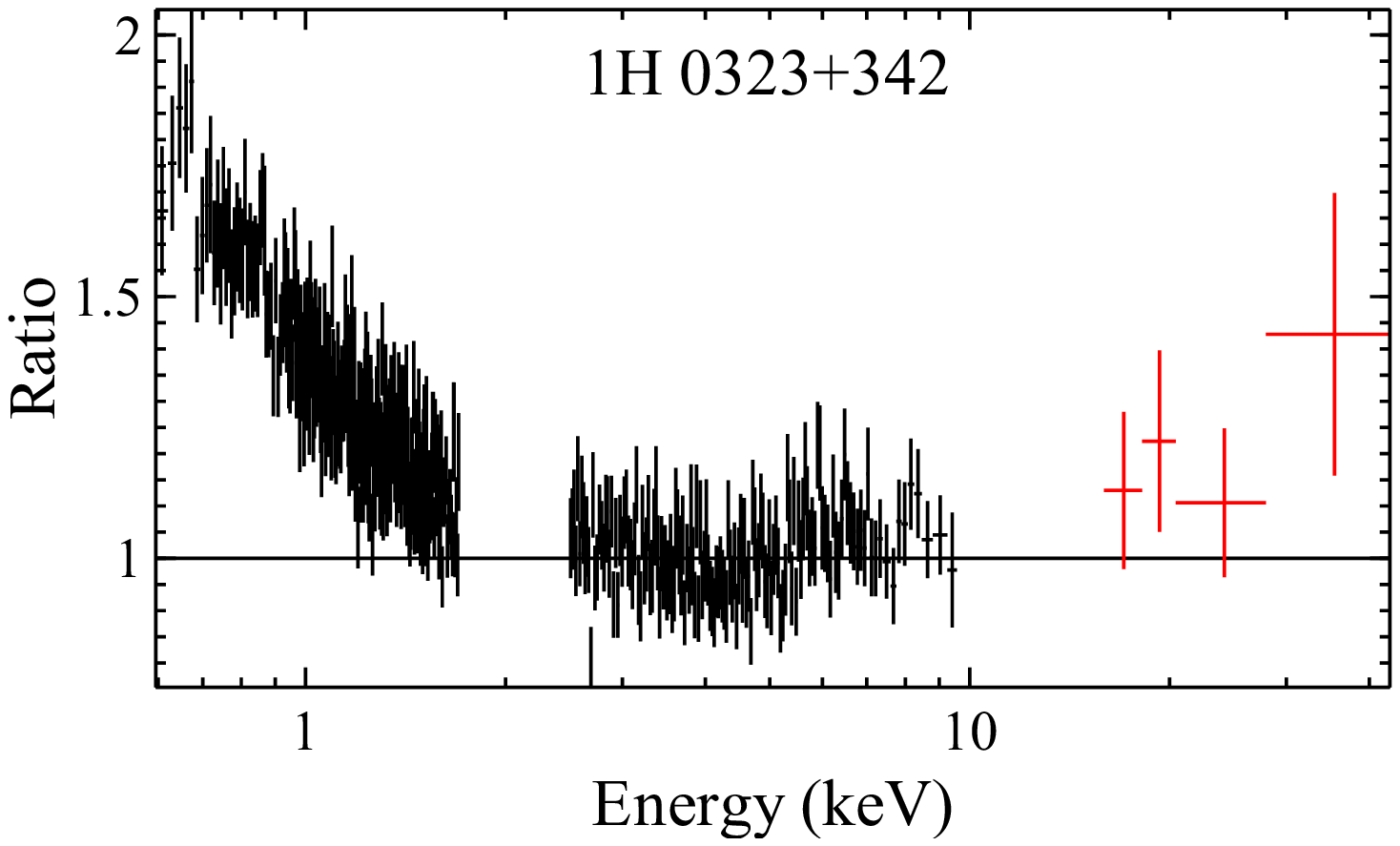}}
}
\rotatebox{0}{
{\includegraphics[width=159pt]{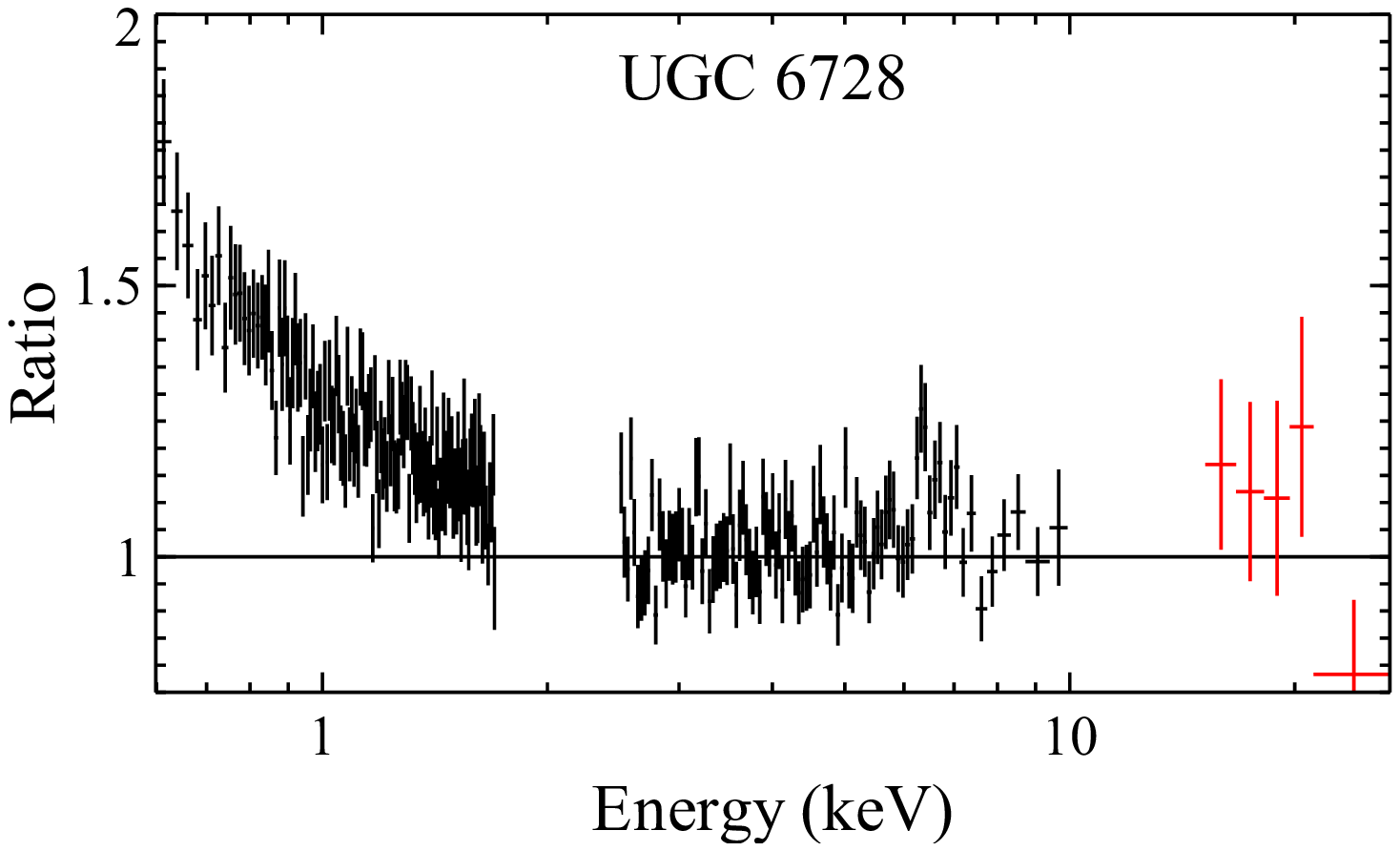}}
}
\rotatebox{0}{
{\includegraphics[width=159pt]{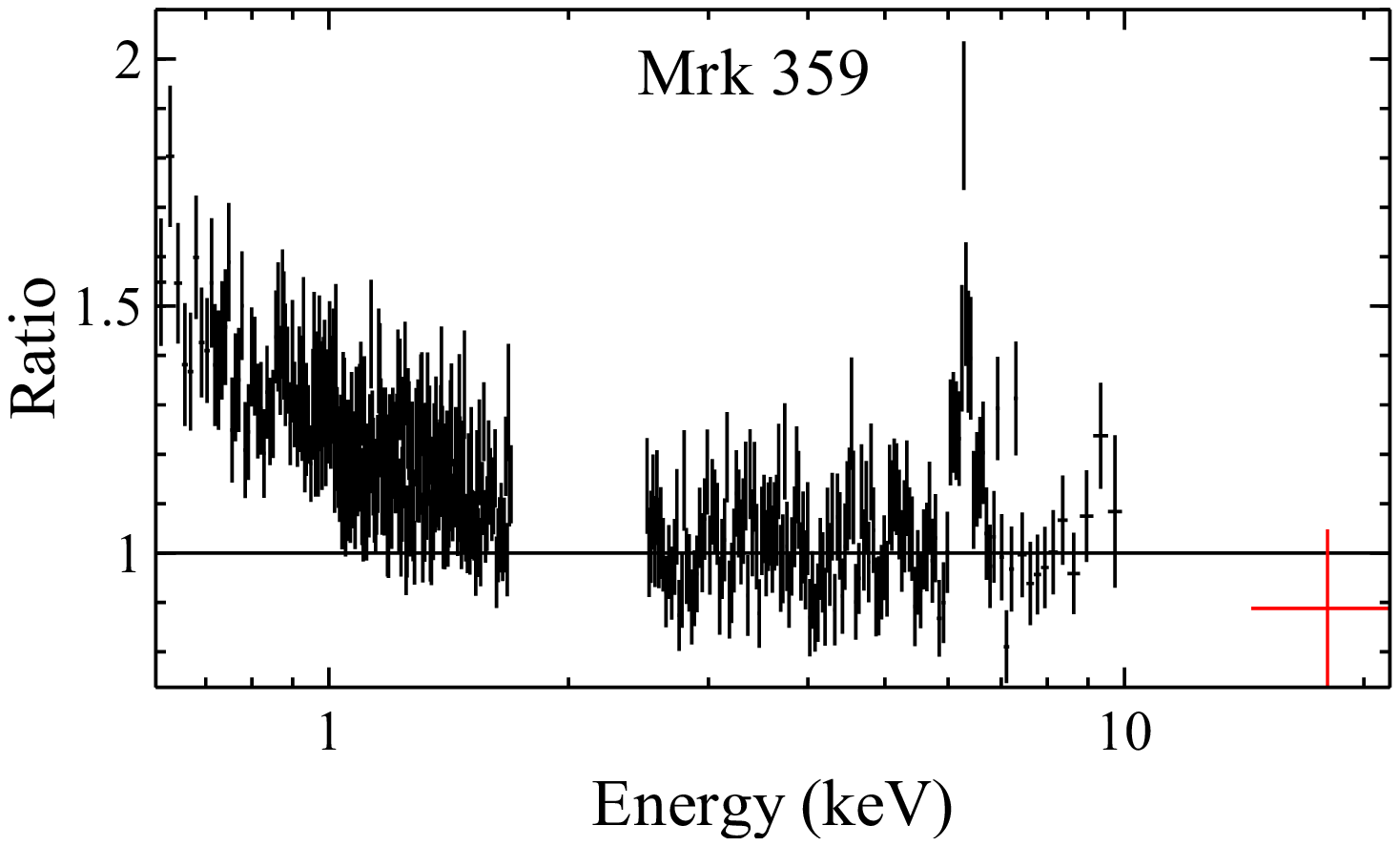}}
}
\end{center}
\caption[Data/model ratio plots of \suzaku\ spectra to a simple powerlaw continuum
for the compiled sample.]
{Data/model ratio plots of \suzaku\ \xis\ (front illuminated; black) and \pin\ (red)
spectra to a powerlaw continuum model, as described in the text (section
\ref{sec_spec_basic}), for the compiled sample. Each source displays a soft excess,
and a number also show evidence for broad iron emission components, and/or for `hard'
excesses above $\sim$10\,\kev, although these features are by no means ubiquitous.
The data shown have been re-binned for plotting purposes.}
\label{fig_ratio_po}
\end{figure*}

\begin{figure*}
\begin{center}
\rotatebox{0}{
{\includegraphics[width=159pt]{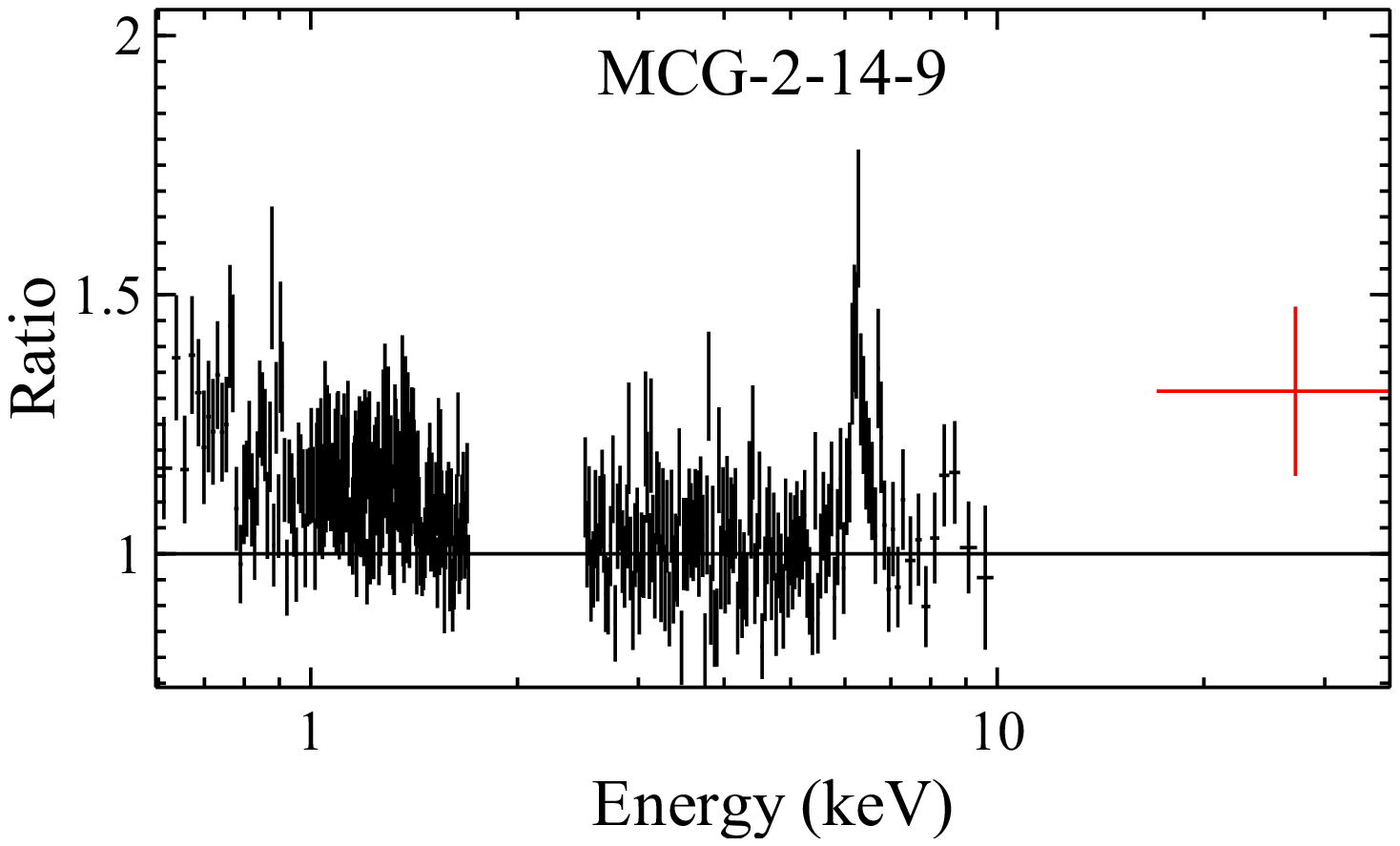}}
}
\rotatebox{0}{
{\includegraphics[width=159pt]{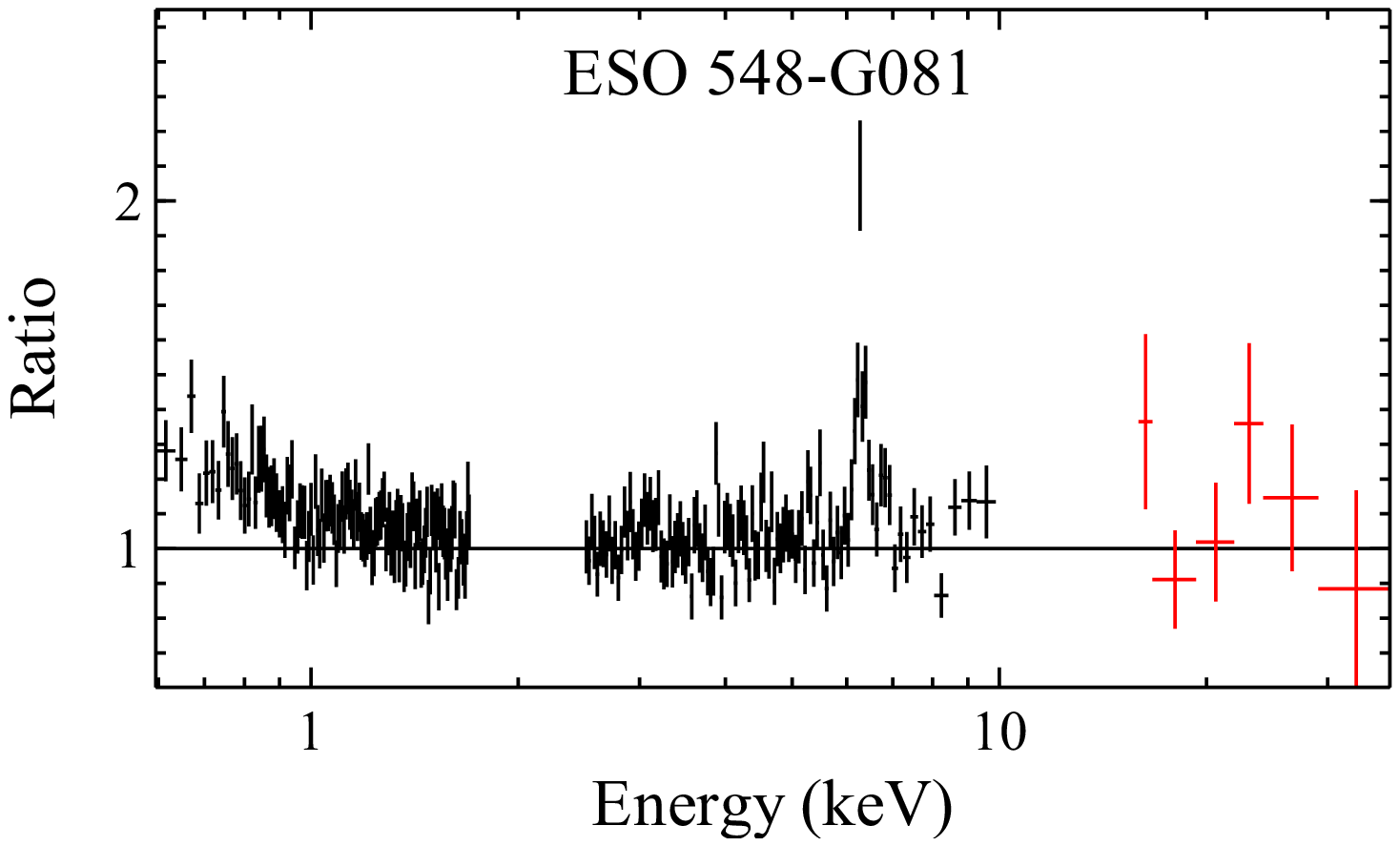}}
}
\rotatebox{0}{
{\includegraphics[width=159pt]{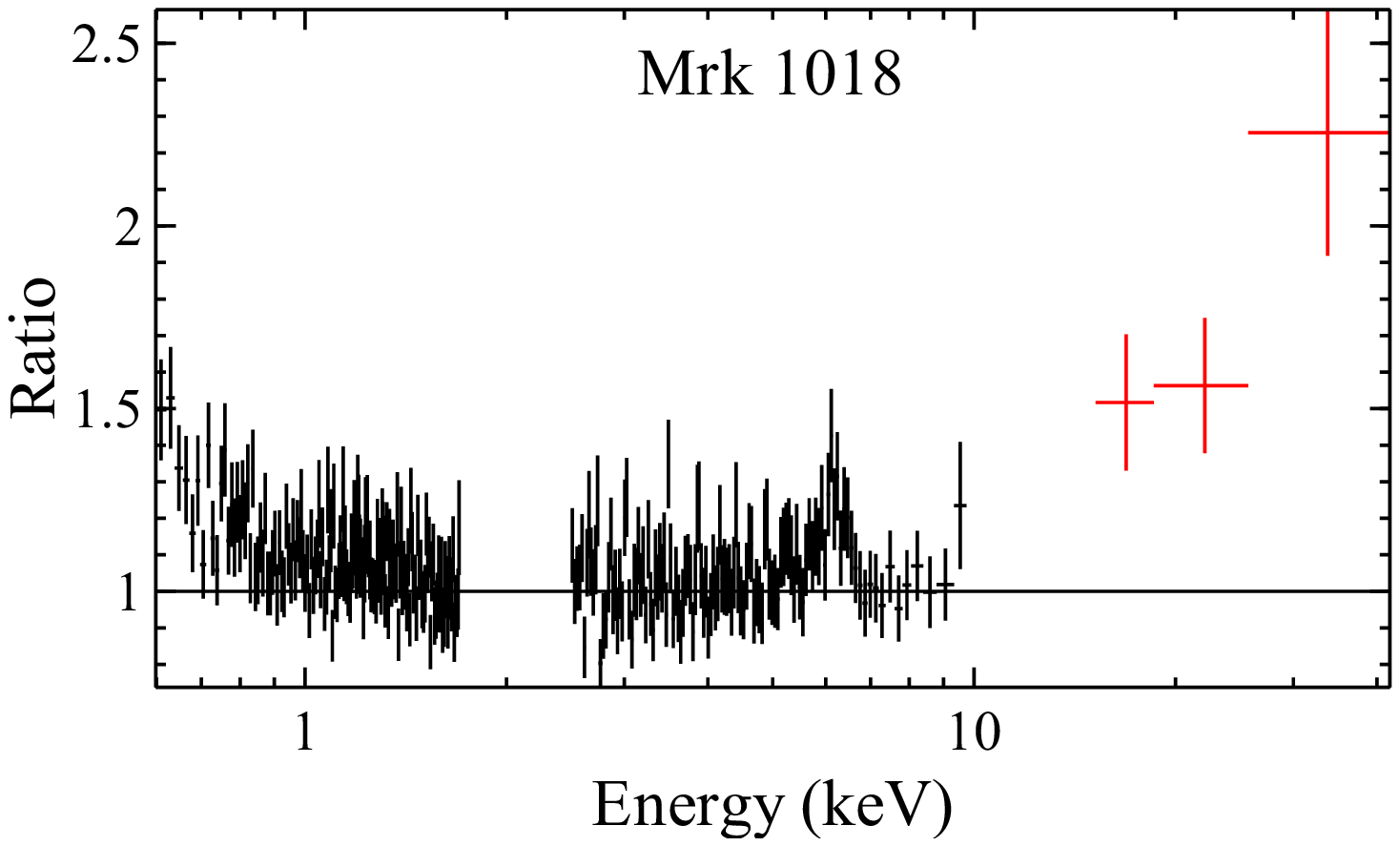}}
}\\
\vspace*{0.4cm}
\rotatebox{0}{
{\includegraphics[width=159pt]{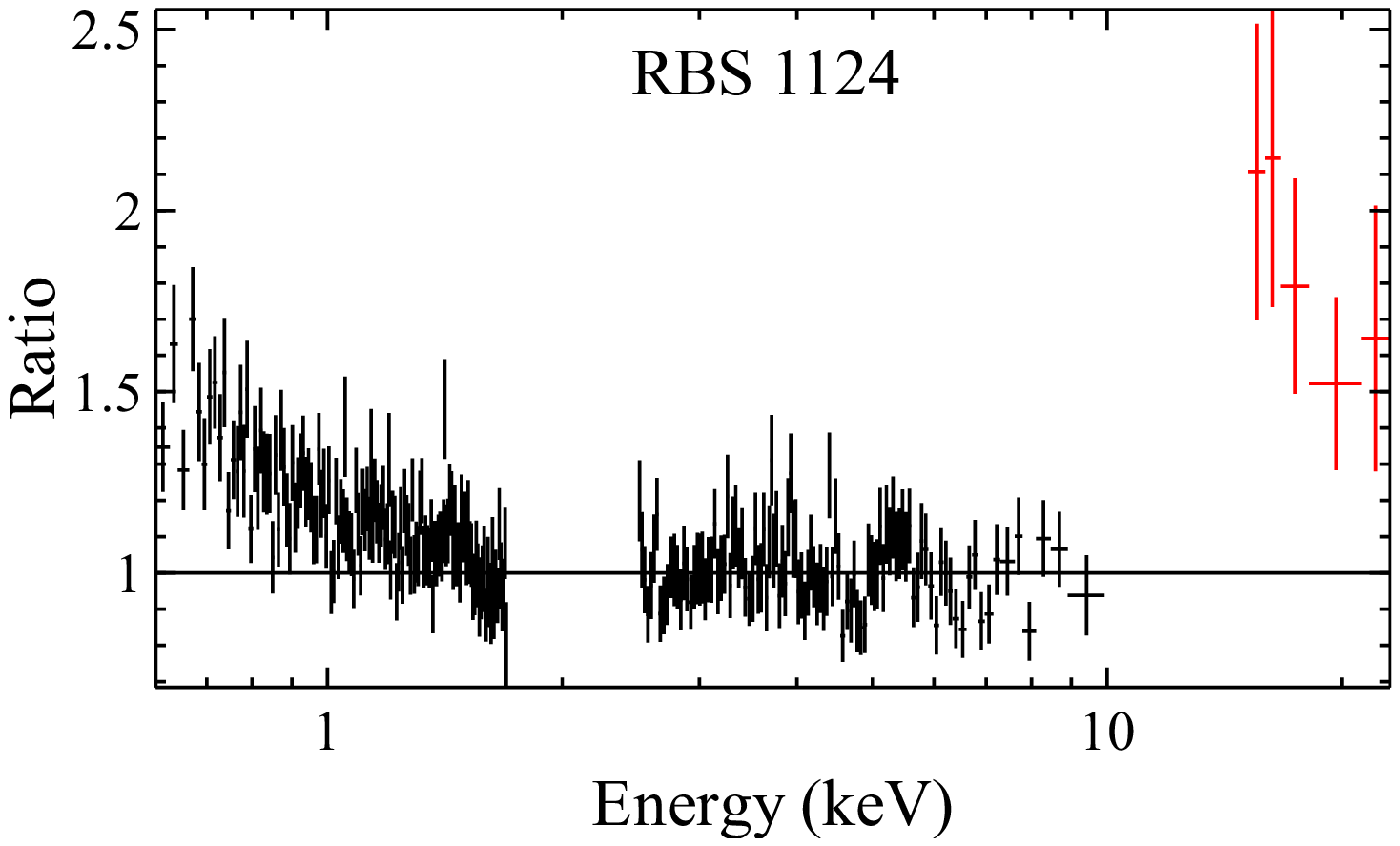}}
}
\rotatebox{0}{
{\includegraphics[width=159pt]{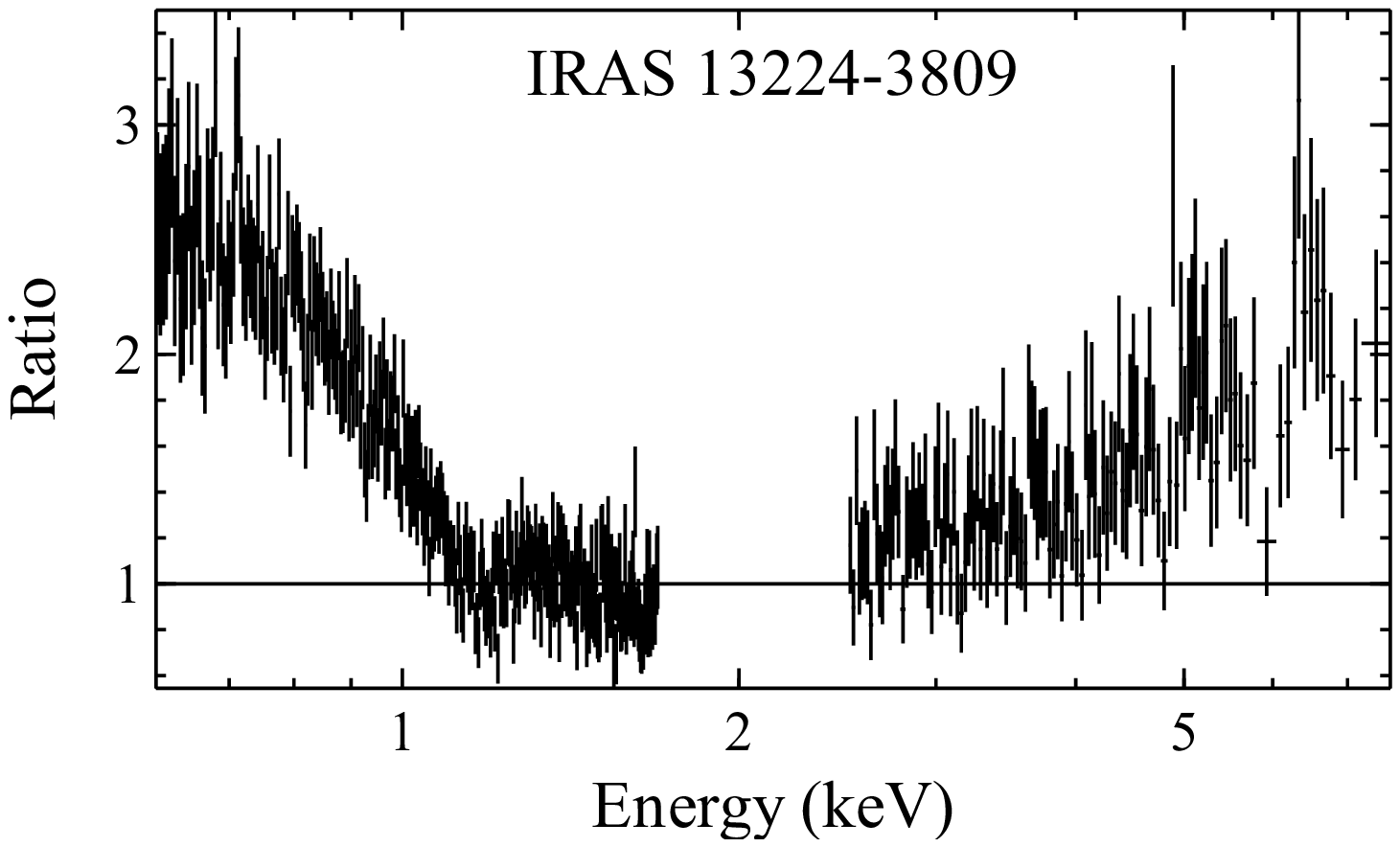}}
}
\rotatebox{0}{
{\includegraphics[width=159pt]{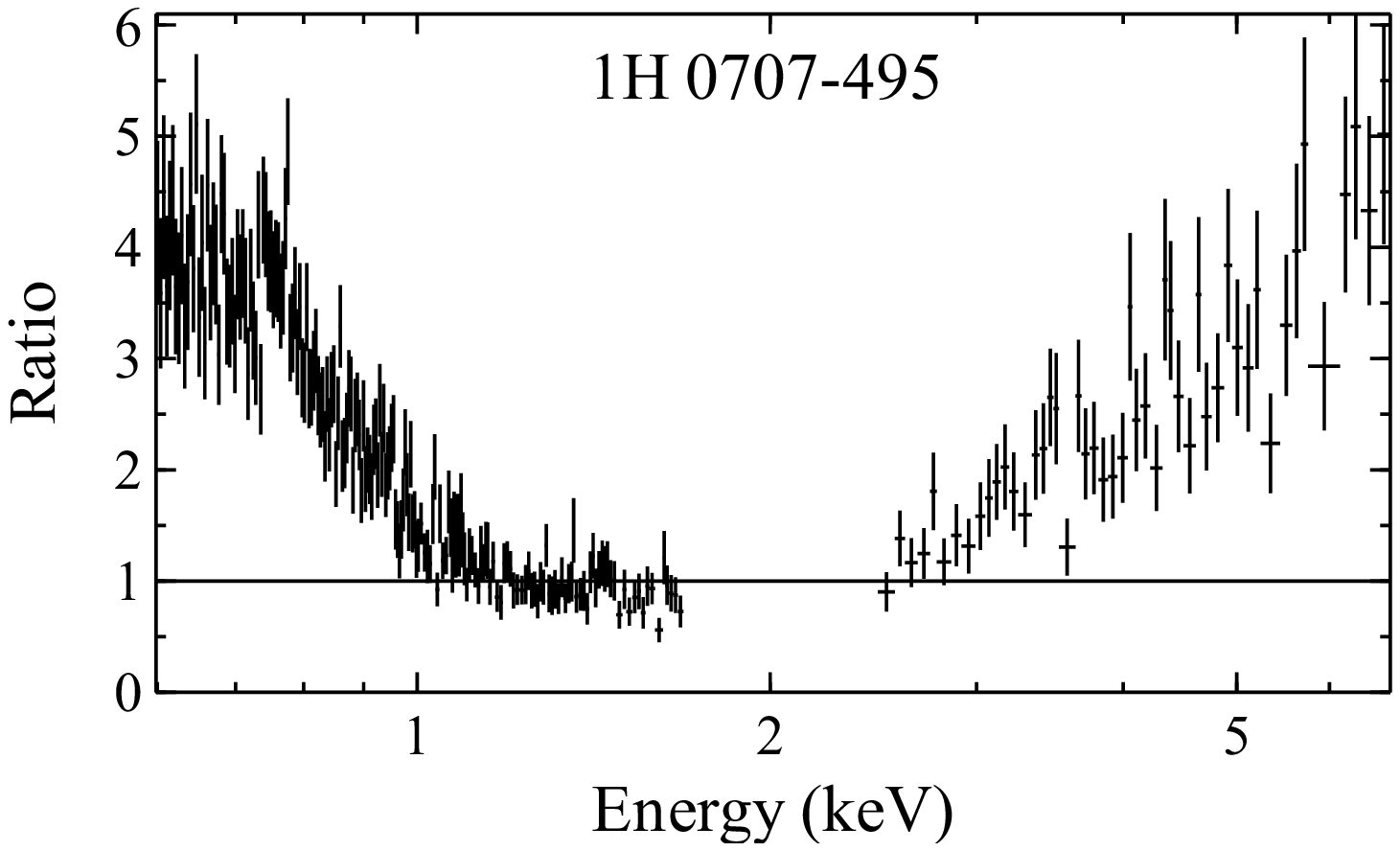}}
}\\
\vspace*{0.4cm}
\rotatebox{0}{
{\includegraphics[width=159pt]{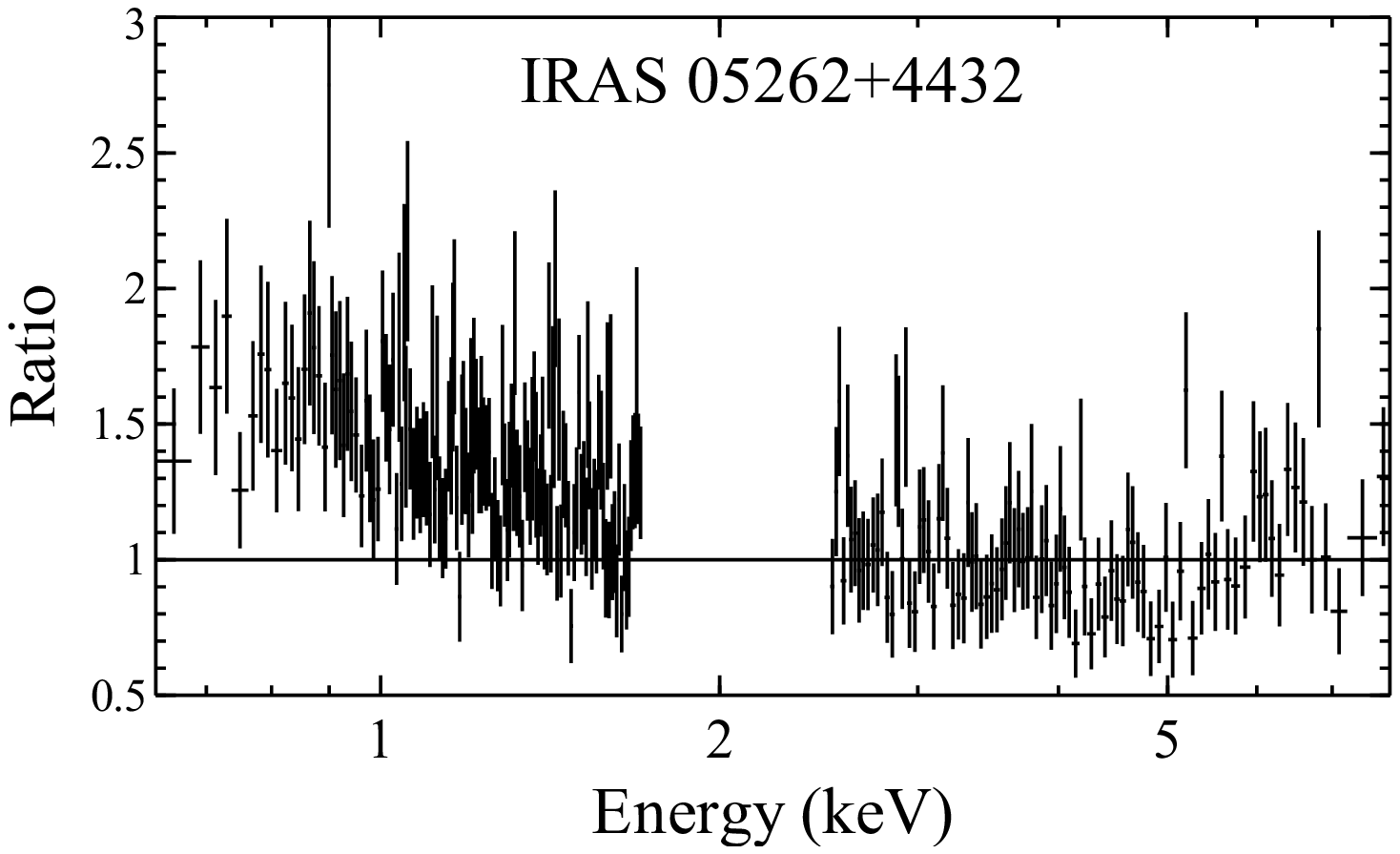}}
}
\end{center}
\vspace*{0.25cm}
\begin{flushleft}
\textbf{Figure~\ref{fig_ratio_po}.} Continued; note that the photon indices of
IRAS\,13224-3809 and 1H\,0707-495 have both been fixed at 2.7 (see section
\ref{sec_spec_details}). Furthermore, there are no robust PIN detections for
either of these sources, or for IRAS\,05262+4432.
\end{flushleft}
\end{figure*}

In this work we are investigating the reflection based interpretation for the broad
band X-ray spectra of active galaxies, \ie that there are two main continuum
components: a powerlaw-like component (PLC) most likely arising via electron
scattering in some kind of corona, and a reflected component (RDC) arising due to
irradiation of the disc by the PLC, which is modified by the strong gravitational
effects present close to the black hole. This interpretation potentially provides a
natural explanation in the form of the RDC for the soft excesses, broad iron lines
and the hard excesses often displayed by AGN, and forms the basis of our spectral
modelling. We try to take a broadly systematic approach, as detailed below, while
also considering the specific details of each source individually, which are
presented in section \ref{sec_spec_details}, although as the data quality declines
so in general does the level of detail to which each source can be considered. The
spectral complexities displayed by each source are highlighted in Figure
\ref{fig_ratio_po}, in which we show the ratio of the XIS (black) and PIN (red)
data to the absorbed powerlaw model discussed previously, using the cross-calibration
normalisations calculated with equation \ref{eqn_crossnorm}.

Naturally the PLC is modelled simply with a powerlaw emission component, with
variable flux and photon index ($\Gamma$). To model the RDC component we make use
of the physically self-consistent reflection code \reflionx\ (\citealt{reflion}).
This model is specifically designed to account for the reprocessed emission from
the cool, partially ionised accretion discs in AGN, intrinsically including both
the reflected continuum and the imprinted atomic features, in particular the iron
K-shell transitions. The key parameters of \reflionx\ are the photon index of the
ionising continuum (assumed to be powerlaw-like), and the iron abundance
($A_{Fe}$) and ionisation state of the reflecting medium. The latter is quantified
as the classic ionisation parameter for a photoionised plasma, $\xi=L/nR^{2}$,
where $L$ is the ionising luminosity, $n$ is the number density of the reflecting
medium and $R$ is its distance from the primary X-ray source. Throughout this
work, we require that the photon index of the ionising continuum be that of the
PLC, while the iron abundance and ionisation parameter are free to vary (unless
stated otherwise, see section \ref{sec_spec_details}).

In order to account for the strong gravitational and Doppler effects that the
emission from material orbiting close to the black hole will experience, we modify
this reflected emission with \relconv\ (\citealt{relconv}), one of the most
sophisticated relativistic convolution kernels. Under the assumption that the inner
radius of the accretion disc is coincident with the ISCO, this kernel allows the
black hole spin to be constrained directly from the RDC, and even allows for the
intriguing possibility of black holes spinning in a retrograde sense with respect
to the material orbiting in their accretion discs. The key free parameters of
\relconv\ are the spin of the black hole, the inclination of the disc ($i$), and
the radial emissivity profile of the disc. We assume a single powerlaw emissivity
profile for simplicity, \ie $\epsilon(r)\propto r^{-q}$, where the emissivity index
$q$ is a free parameter. For a simple Newtonian `lamp post' like geometry, in which
the illuminating source is compact and centrally located, the predicted emissivity
index is $q=3$. However, this does not take into account any of the effects of
general relativistic light bending, which focuses a larger fraction of the emission
onto the inner disc if the X-ray source is also located in a region of extreme
gravity (\citealt{lightbending}), or of relativistic time dilation. Both of these
effects generally serve to steepen the inner emissivity profile (\citealt{Wilkins11}),
therefore throughout this work we require that $q \geq 3$. The inner and outer radii
of the disc are fixed respectively at the (spin dependent) ISCO, and at 400\,\rg\
(\rg\ = $GM/c^{2}$), the maximum value accepted by the model. It is also possible
to include the effects of limb brightening/darkening, and we assume the latter
following the calculations of \cite{kdblur}.

With this RDC we attempt to simultaneously account for the broad curvature across
the 0.5--10.0\,\kev\ energy range that causes the appearance of a soft excess, any
relativistically broadened iron emission detected and also any hard excess displayed
by the AGN. However, this is a complex, multi-parameter model component, and
degeneracies can arise between various parameter combinations, particularly in the
relativistic blurring parameters (see \eg \citealt{Nardini11}); this can be especially
problematic with moderate quality data, or when the features associated with
reflection are weak. One of the main aims of this work is to provide where possible an
initial, albeit model dependent constraint on the black hole spin. Therefore, in cases
where it is not possible to independently constrain all the blurring parameters, we
fix \eg the inclination to 45\deg and/or the emissivity index to $q=3$ in an attempt
to provide at least a loose spin constraint. This was only necessary for the minority
of cases (7 AGN in total) even including our analysis in which $C_{\rm PIN/XIS}$ was
allowed to vary, although in 4 of these sources we were still unable to provide any
constraint on the spin. Furthermore, in a few individual sources certain parameters
might also be fixed for other reasons, \eg\ the inclination being known a priori from
radio constraints. Full details are provided in section \ref{sec_spec_details}.

In addition to their broadband continuum emission, active galaxies also very frequently
display evidence for narrow emission features consistent with neutral iron (6.4\,\kev).
Our general assumption is that these features also arise through reflection of the
primary X-ray emission, but by cold, distant material rather than by the accretion disc,
possibly associated with the torus-like structure that obscures the central X-ray
sources of AGN viewed at high inclination angles. Therefore, where we see evidence for
such a feature, our default procedure is to include a second, cold reflection component.
This was also modelled with \reflionx, with $\xi$ = 1.0 (the lowest ionisation state
accepted by the model), while we required the iron abundance to be the same as that of
the disc reflection component, and we again required that the photon index of the
ionising continuum to be the same as that of the PLC. This second reflector accounts for
the narrow emission from neutral iron, and where present may also contribute some
emission to the hard excess. Less common than narrow emission from neutral iron, which
is almost observed ubiquitously, AGN also sometimes display narrow emission features
consistent with highly ionised iron (Fe {\small XXV} and {\small XXVI}; 6.67 and
6.97\,\kev\ respectively). Where we find evidence for these additional narrow features,
we simply account for them with narrow ($\sigma = 10$\,eV) Gaussian emission lines.

Finally, all emission components are modified by neutral absorption due to the
Galactic ISM. Again, we make use of the \tbabs\ absorption code, and fix the column
density at the Galactic value quoted in Table \ref{tab_c3_obs}. Given that we have
deliberately selected `bare' active galaxies, we do not include any contribution
from neutral absorption intrinsic to the source. However, despite our source
selection method, high resolution grating spectra of some of the sources included
in this work do show evidence for some weak, partially ionised absorption intrinsic
to the source. Where this is the case, we include an ionised absorber using the
\xstar\ photoionisation code (\citealt{xstar}) to account for the main features in
our lower resolution CCD spectra. The free parameters of the absorption model used
here are the ionisation state (again quantified as $\xi$), the column density and
the line-of-sight velocity of the absorbing medium; the abundances of the absorbing
medium are assumed to be solar, and its turbulent velocity is fixed at 200\,\kmps.
In addition, there has been growing evidence for AGN displaying highly blueshifted
absorption features from highly ionised iron (see \eg \citealt{Tombesi10b}),
typically interpreted as evidence for rapid outflows (although see \citealt{Gallo11}
for an alternative explanation). Where we find evidence for similar features, we
also treat these with \xstar, although in these cases we will use grids specifically
designed to treat absorption by highly ionised material, as detailed below.

\subsection{Individual Source Details}
\label{sec_spec_details}

\begin{table*}
  \caption[Key Parameters obtained for the reflection based models constructed for
the compiled sample.]
{Key Parameters obtained for the reflection based models constructed for the compiled
sample (see section \ref{sec_spec_details} for details). Parameters in parentheses have
not been allowed to vary, and where we were unable to constrain the black hole spin
this is indicated with a `U'.}
\begin{center}
\begin{tabular}{c c c c c c c c c c}
\hline
\hline
\\[-0.3cm]
Source & En. Range & $C_{\rm PIN/XIS}$ & $\Gamma$ & $A_{\rm Fe}$ & $\xi$ & $q$ & $i$ & $a^{*}$ & \rchi \\
\\[-0.25cm]
& (keV) & & & (solar) & ($\ergcmps$) & & (\deg) & &  ($\chi^{2}$/d.o.f.)\\
\\[-0.3cm]
\hline
\hline
\\[-0.2cm]
Mrk\,509 & 0.6--44.0 & 1.17 & $2.04\pm0.01$ & $0.5\pm0.1$ & $170^{+30}_{-80}$ & $>7.4$ & $<18$ & $0.86^{+0.02}_{-0.01}$ & 1314/1333 \\
\\[-0.25cm]
3C\,382 & 0.6--53.0 & 1.14 & $1.81\pm0.01$ & $>5.1$ & $500^{+60}_{-240}$ & $>6.1$ & (40) & $0.75^{+0.07}_{-0.04}$ & 1366/1317 \\
\\[-0.25cm]
Mrk\,335 & 0.6--36.0 & 1.17 & $2.16\pm0.01$ & $1.4\pm0.2$ & $220\pm10$ & $>4.9$ & $50^{+8}_{-7}$ & $0.83^{+0.10}_{-0.13}$ & 1233/1152 \\
\\[-0.25cm]
Fairall 9 & 0.6--39.0 & 1.16 & $1.99\pm0.01$ & $1.1\pm0.2$ & $140^{+60}_{-30}$ & $>3.5$ & $45^{+13}_{-9}$ & $>0.64$ & 1276/1253 \\
\\[-0.25cm]
1H\,0419-577 & 0.6--48.0 & 1.19 & $1.98^{+0.01}_{-0.02}$ & $0.9\pm0.1$ & $104^{+4}_{-26}$ & $5.4^{+0.2}_{-1.0}$ & $51^{+4}_{-6}$ & $>0.88$ & 1384/1294 \\
\\[-0.25cm]
Ark\,564 & 0.6--22.0 & 1.20 & $2.52\pm0.01$ & $1.0^{+0.2}_{-0.1}$ & $540^{+40}_{-20}$ & $>6.2$ & $64^{+1}_{-11}$ & $0.96^{+0.01}_{-0.06}$ & 1081/1011 \\
\\[-0.25cm]
Ark\,120 & 0.6--53.0 & 1.22 & $2.13\pm0.01$ & $2.7\pm0.6$ & $9^{+2}_{-4}$ & $7.0^{+2.7}_{-2.2}$ & $54^{+6}_{-5}$ & $0.81^{+0.10}_{-0.18}$ & 1316/1177 \\
\\[-0.25cm]
3C\,390.3 & 0.6--49.0 & 1.16 & $1.66\pm0.01$ & $3.1^{+1.2}_{-0.5}$ & $840^{+490}_{-220}$ & (3) & (35) & U & 1302/1259 \\
\\[-0.25cm]
PKS\,0558-504 & 0.6--24.0 & 1.13 & $2.30^{+0.02}_{-0.01}$ & $0.9^{+0.3}_{-0.1}$ & $270^{+70}_{-30}$ & $4.0\pm0.7$ & (45) & $>0.80$ & 990/1022 \\
\\[-0.25cm]
NGC\,7469 & 0.6--52.0 & 1.19 & $1.84^{+0.03}_{-0.02}$ & $2.9^{+1.7}_{-1.0}$ & $200^{+20}_{-50}$ & $>4.6$ & $<54$ & $>0.96$ & 1262/1139 \\
\\[-0.25cm]
Mrk\,110 & 0.6--45.0 & 1.19 & $1.96^{+0.02}_{-0.01}$ & $0.7\pm0.2$ & $310^{+220}_{-80}$ & $>7.4$ & $31^{+4}_{-6}$ & $>0.99$ & 1184/1115 \\
\\[-0.25cm]
Swift\,J0501.9-3239 & 0.6--36.0 & 1.22 & $2.06^{+0.04}_{-0.03}$ & $1.8^{+0.9}_{-0.5}$ & $200^{+10}_{-40}$ & $>5.1$ & $<48$ & $>0.96$ & 1025/1056 \\
\\[-0.25cm]
Mrk\,841 & 0.6--53.0 & 1.19 & $1.85^{+0.03}_{-0.01}$ & $1.0\pm0.2$ & $210^{+20}_{-70}$ & $4.1^{+2.8}_{-1.9}$ & $45^{+7}_{-5}$ & $>0.56$ & 1089/1053 \\
\\[-0.25cm]
Ton\,S180 & 0.6--23.0 & 1.16 & $2.36\pm0.01$ & $0.9^{+0.2}_{-0.1}$ & $280^{+50}_{-20}$ & $>8.1$ & $60^{+3}_{-1}$ & $0.91^{+0.02}_{-0.09}$ & 876/838 \\
\\[-0.25cm]
PDS\,456 & 0.6--17.0 & 1.15 & $2.30^{+0.03}_{-0.01}$ & $>8.4$ & $59^{+17}_{-4}$ & $5.9^{+1.8}_{-1.5}$ & $70^{+3}_{-5}$ & $>0.97$ & 829/826 \\
\\[-0.25cm]
1H\,0323+342 & 0.6--42.0 & 1.25 & $1.91^{+0.03}_{-0.01}$ & $0.8\pm0.2$ & $250^{+40}_{-20}$ & (3) & (45) & $>0.48$ & 864/922 \\
\\[-0.25cm]
UGC\,6728 & 0.6--26.0 & 1.27 & $2.00^{+0.04}_{-0.03}$ & $0.7^{+0.6}_{-0.3}$ & $190^{+80}_{-170}$ & $6.8^{+2.8}_{-1.4}$ & $<55$ & $>0.95$ & 877/885 \\
\\[-0.25cm]
Mrk\,359 & 0.6--21.0 & 1.15 & $1.89^{+0.04}_{-0.03}$ & $1.5^{+0.9}_{-0.6}$ & $21^{+32}_{-16}$ & $>4.1$ & $47\pm6$ & $0.66^{+0.30}_{-0.46}$ & 820/833 \\
\\[-0.25cm]
MCG--2-14-9 & 0.6--37.0 & 1.19 & $1.89\pm0.02$ & (1) & $<10$ & (3) & (45) & U & 802/804 \\
\\[-0.25cm]
ESO\,548-G081 & 0.6--36.0 & 1.23 & $1.70\pm0.03$ & $3.5^{+4.1}_{-1.5}$ & $570^{+560}_{-380}$ & (3) & (45) & U & 853/845 \\
\\[-0.25cm]
Mrk\,1018 & 0.6--41.0 & 1.21 & $1.94^{+0.04}_{-0.03}$ & $2.0^{+1.4}_{-0.7}$ & $5^{+10}_{-4}$ & $>3.9$ & $45^{+14}_{-10}$ & $0.57^{+0.31}_{-0.82}$ & 681/721 \\
\\[-0.25cm]
RBS\,1124 & 0.6--23.0 & 1.22 & $1.86^{+0.04}_{-0.02}$ & $2.9^{+1.5}_{-0.9}$ & $51^{+7}_{-9}$ & $>8.4$ & $66^{+5}_{-15}$ & $>0.98$ & 661/668 \\
\\[-0.25cm]
IRAS\,13224-3809 & 0.6--7.6 & - & (2.7) & (20) & $22\pm3$ & $6.1^{+0.7}_{-0.6}$ & (64) & $>0.995$ & 447/412 \\
\\[-0.25cm]
1H\,0707-495 & 0.6--6.7 & - & (2.7) & (10) & $53^{+1}_{-2}$ & $7.6^{+0.4}_{-0.3}$ & (58) & $>0.994$ & 278/236 \\
\\[-0.25cm]
IRAS\,05262+4432 & 0.6--7.8 & - & $2.18^{+0.13}_{-0.06}$ & (1) & $<51$ & (3) & (45) & U & 234/231 \\
\\[-0.2cm]
\hline
\hline
\end{tabular}
\label{tab_c3_refl}
\end{center}
\end{table*}

\begin{table*}
  \caption[Key Parameters obtained for the reflection based models constructed for
the compiled sample with $C_{\rm PIN/XIS}$ allowed to vary.]
{Key Parameters obtained for the reflection based models constructed for the compiled sample when
$C_{\rm PIN/XIS}$ was free to vary within the indicated range. Again, parameters in parentheses have
not been allowed to vary, and where we were unable to constrain the black hole spin this is indicated
with a `U'.}
\begin{center}
\begin{tabular}{c c c c c c c c c}
\hline
\hline
\\[-0.3cm]
Source & $C_{\rm PIN/XIS}$ (range) & $\Gamma$ & $A_{\rm Fe}$ & $\xi$ & $q$ & $i$ & $a^{*}$ & \rchi \\
\\[-0.25cm]
& & & (solar) & ($\ergcmps$) & & (\deg) & &  ($\chi^{2}$/d.o.f.)\\
\\[-0.3cm]
\hline
\hline
\\[-0.2cm]
Mrk\,509 & $<1.06$ (1.04--1.30) & $1.99\pm0.01$ & $0.5\pm0.1$ & $13^{+8}_{-5}$ & $>4.3$ & $50^{+5}_{-3}$ & $0.36^{+0.20}_{-0.37
}$ & 1302/1332 \\
\\[-0.25cm]
3C\,382 & $<1.03$ (0.99--1.29) & $1.82\pm0.01$ & $1.7^{+0.4}_{-0.3}$ & $200^{+10}_{-70}$ & (3) & (40) & $<0.81$ & 1358/1317 \\
\\[-0.25cm]
Mrk\,335 & $1.19\pm0.11$ (0.79--1.55) & $2.16\pm0.01$ & $1.4\pm0.2$ & $220\pm10$ & $>4.9$ & $50^{+8}_{-9}$ & $0.83^{+0.09}_{-0.13}$ & 1233/1151 \\
\\[-0.25cm]
Fairall 9 & $1.06\pm0.06$ (0.95--1.37) & $2.00\pm0.01$ & $0.9^{+0.2}_{-0.1}$ & $130^{+70}_{-30}$ & $5.2^{+4.5}_{-1.4}$ & $44^{+13}_{-9}$ & $0.82^{+0.09}_{-0.19}$ & 1269/1252 \\
\\[-0.25cm]
1H\,0419-577 & $1.27^{+0.06}_{-0.05}$ (0.93--1.45) & $1.97^{+0.03}_{-0.01}$ & $0.8\pm0.1$ & $120^{+10}_{-20}$ & $5.2^{+0.3}_{-1.3}$ & $49^{+7}_{-4}$ & $>0.89$ & 1380/1293 \\
\\[-0.25cm]
Ark\,564 & $1.40^{+0.17}_{-0.18}$ (0.79--1.61) & $2.53\pm0.01$ & $1.0^{+0.2}_{-0.1}$ & $550^{+40}_{-30}$ & $>6.0$ & $64^{+1}_{-11}$ & $0.96^{+0.01}_{-0.07}$ & 1078/1010 \\
\\[-0.25cm]
Ark\,120 & $<1.04$ (1.00--1.43) & $2.14\pm0.01$ & $1.8^{+0.2}_{-0.3}$ & $6^{+5}_{-2}$ & $5.1^{+2.8}_{-1.1}$ & $47^{+7}_{-2}$ & $0.64^{+0.19}_{-0.11}$ & 1274/1176 \\
\\[-0.25cm]
3C\,390.3 & $1.21^{+0.05}_{-0.04}$ (1.01--1.31) & $1.65\pm0.01$ & $4.5^{+1.7}_{-1.9}$ & $1060^{+360}_{-350}$ & (3) & (35) & U & 1301/1258 \\
\\[-0.25cm]
PKS\,0558-504 & $>1.23$ (0.70--1.56) & $2.30^{+0.02}_{-0.01}$ & $1.0^{+0.3}_{-0.2}$ & $270^{+90}_{-30}$ & $4.1\pm0.7$ & (45) & $>0.95$ & 985/1021 \\
\\[-0.25cm]
NGC\,7469 & $<1.13$ (1.01--1.37) & $1.81^{+0.02}_{-0.01}$ & $1.1^{+0.5}_{-0.2}$ & $200^{+20}_{-100}$ & $>5.5$ & (45) & $0.64^{+0.13}_{-0.20}$ & 1261/1139 \\
\\[-0.25cm]
Mrk\,110 & $<1.07$ (0.96--1.42) & $1.92^{+0.01}_{-0.02}$ & $0.7^{+0.2}_{-0.1}$ & $250^{+230}_{-40}$ & $>5.4$ & $<43$ & $0.96^{+0.03}_{-0.07}$ & 1179/1114 \\
\\[-0.25cm]
Swift\,J0501.9-3239 & $>1.25$ (1.07--1.37) & $2.09\pm0.04$ & $3.1^{+1.4}_{-1.2}$ & $200^{+10}_{-70}$ & $9.0^{+0.4}_{-2.7}$ & $32^{+10}_{-28}$ & $>0.99$ & 1019/1055 \\
\\[-0.25cm]
Mrk\,841 & $1.14^{+0.08}_{-0.10}$ (0.90--1.48) & $1.86\pm0.02$ & $0.9^{+0.3}_{-0.2}$ & $190^{+30}_{-110}$ & $4.1^{+2.8}_{-1.9}$ & $46^{+6}_{-5}$ & $>0.52$ & 1088/1052 \\
\\[-0.25cm]
Ton\,S180 & $>1.46$ (0.42--1.90) & $2.36\pm0.01$ & $0.9^{+0.2}_{-0.1}$ & $280^{+30}_{-20}$ & $>8.3$ & $60^{+3}_{-1}$ & $0.92^{+0.03}_{-0.11}$ & 869/837 \\
\\[-0.25cm]
PDS\,456 & $1.34^{+0.77}_{-0.72}$ (0.00--2.97) & $2.30^{+0.03}_{-0.01}$ & $>8.5$ & $59^{+16}_{-8}$ & $5.9^{+2.2}_{-1.8}$ & $70^{+3}_{-6}$ & $>0.96$ & 829/825 \\
\\[-0.25cm]
1H\,0323+342 & $>1.26$ (0.89--1.61) & $1.92^{+0.01}_{-0.02}$ & $0.8^{+0.3}_{-0.2}$ & $260^{+40}_{-30}$ & (3) & (45) & $>0.37$ & 861/921 \\
\\[-0.25cm]
UGC\,6728 & $<1.41$ (1.02--1.52) & $1.98^{+0.14}_{-0.12}$ & $0.8^{+0.8}_{-0.4}$ & $<500$ & $>4.5$ & (45) & $>0.71$ & 877/884 \\
\\[-0.25cm]
Mrk\,359 & $0.78^{+0.24}_{-0.23}$ (0.33--1.97) & $1.90\pm0.03$ & $1.2^{+0.7}_{-0.4}$ & $21^{+32}_{-17}$ & $>4.1$ & $47^{+10}_{-8}$ & $0.66^{+0.30}_{-0.54}$ & 814/832 \\
\\[-0.25cm]
MCG--2-14-9 & $1.19^{+0.26}_{-0.23}$ (0.54--1.85) & $1.89\pm0.02$ & (1) & $<26$ & (3) & (45) & U & 802/803 \\
\\[-0.25cm]
ESO\,548-G081 & $<1.21$ (0.98--1.48) & $1.71^{+0.05}_{-0.02}$ & $2.2^{+1.8}_{-1.1}$ & $270^{+520}_{-160}$ & (3) & (45) & U & 850/844 \\
\\[-0.25cm]
Mrk\,1018 & $>1.27$ (0.94--1.48) & $1.93^{+0.03}_{-0.04}$ & $2.7^{+1.7}_{-1.2}$ & $7^{+18}_{-5}$ & $>3.4$ & $46\pm9$ & $0.58^{+0.36}_{-0.74}$ & 676/720 \\
\\[-0.25cm]
RBS\,1124 & $>1.46$ (0.84--1.60) & $1.86\pm0.05$ & $2.7^{+1.8}_{-0.9}$ & $50^{+10}_{-40}$ & $>6.9$ & $64^{+7}_{-13}$ & $>0.97$ & 649/667 \\
\\[-0.2cm]
\hline
\hline
\end{tabular}
\label{tab_refl2}
\end{center}
\end{table*}

\begin{figure*}
\begin{center}
\rotatebox{0}{
{\includegraphics[width=159pt]{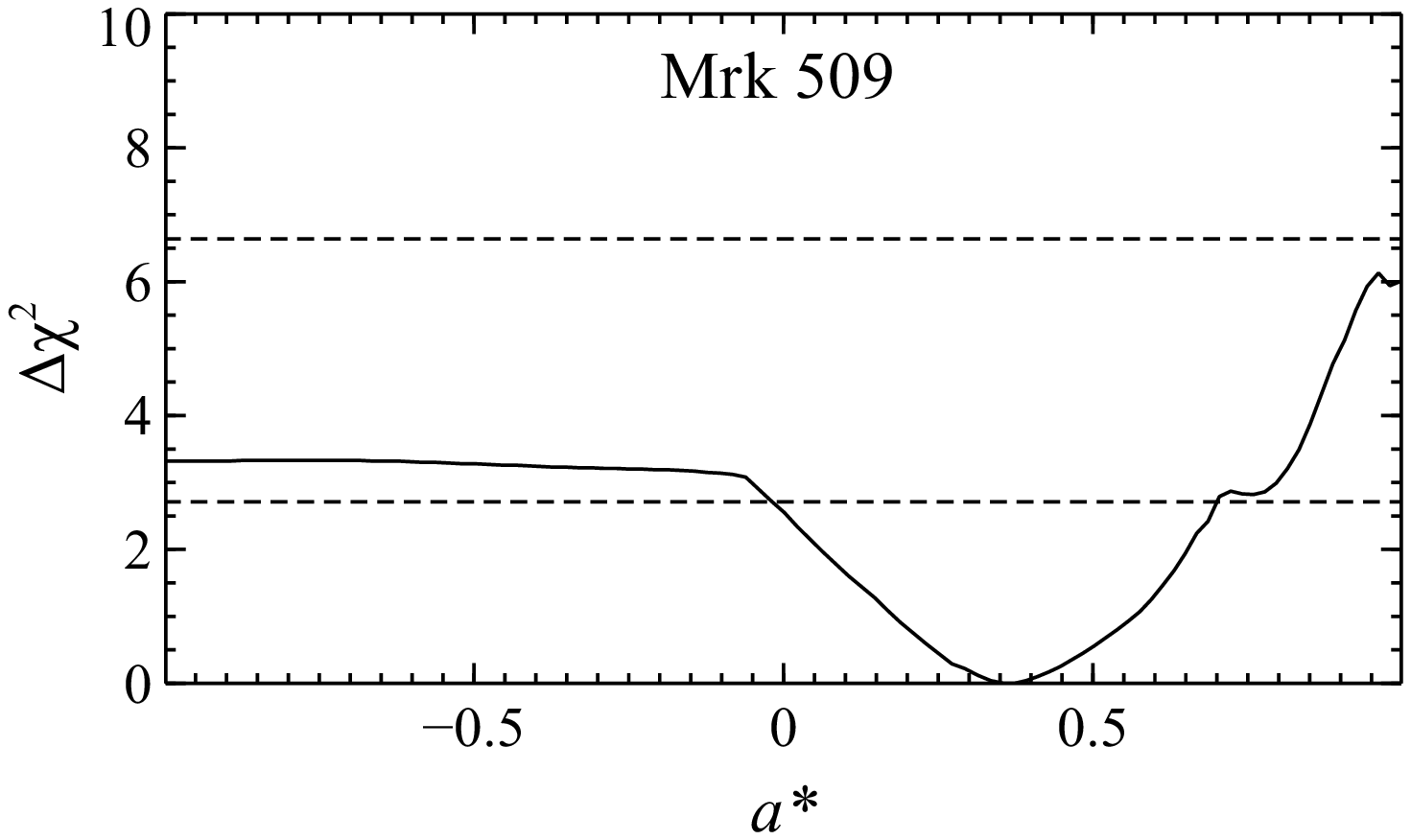}}
}
\rotatebox{0}{
{\includegraphics[width=159pt]{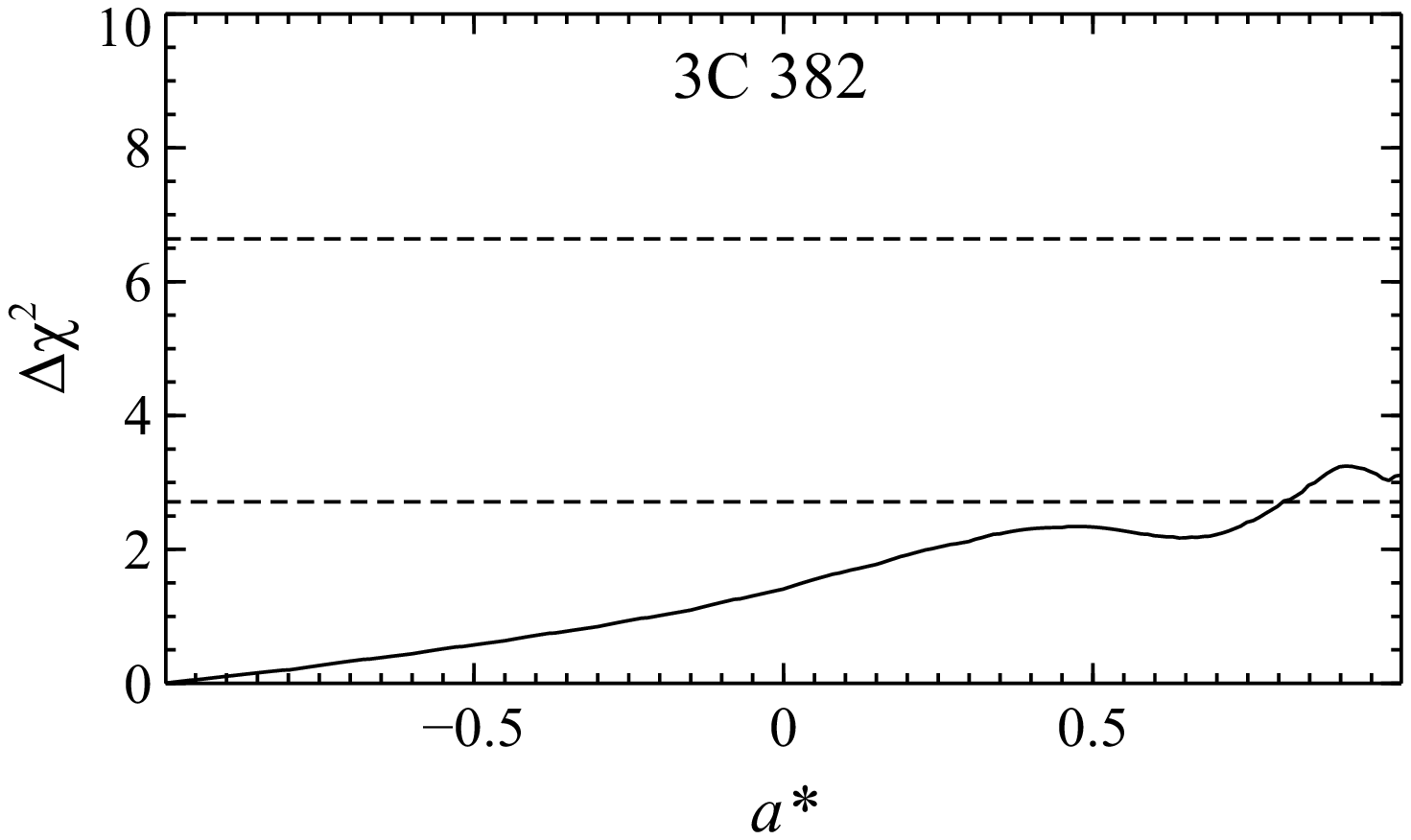}}
}
\rotatebox{0}{
{\includegraphics[width=159pt]{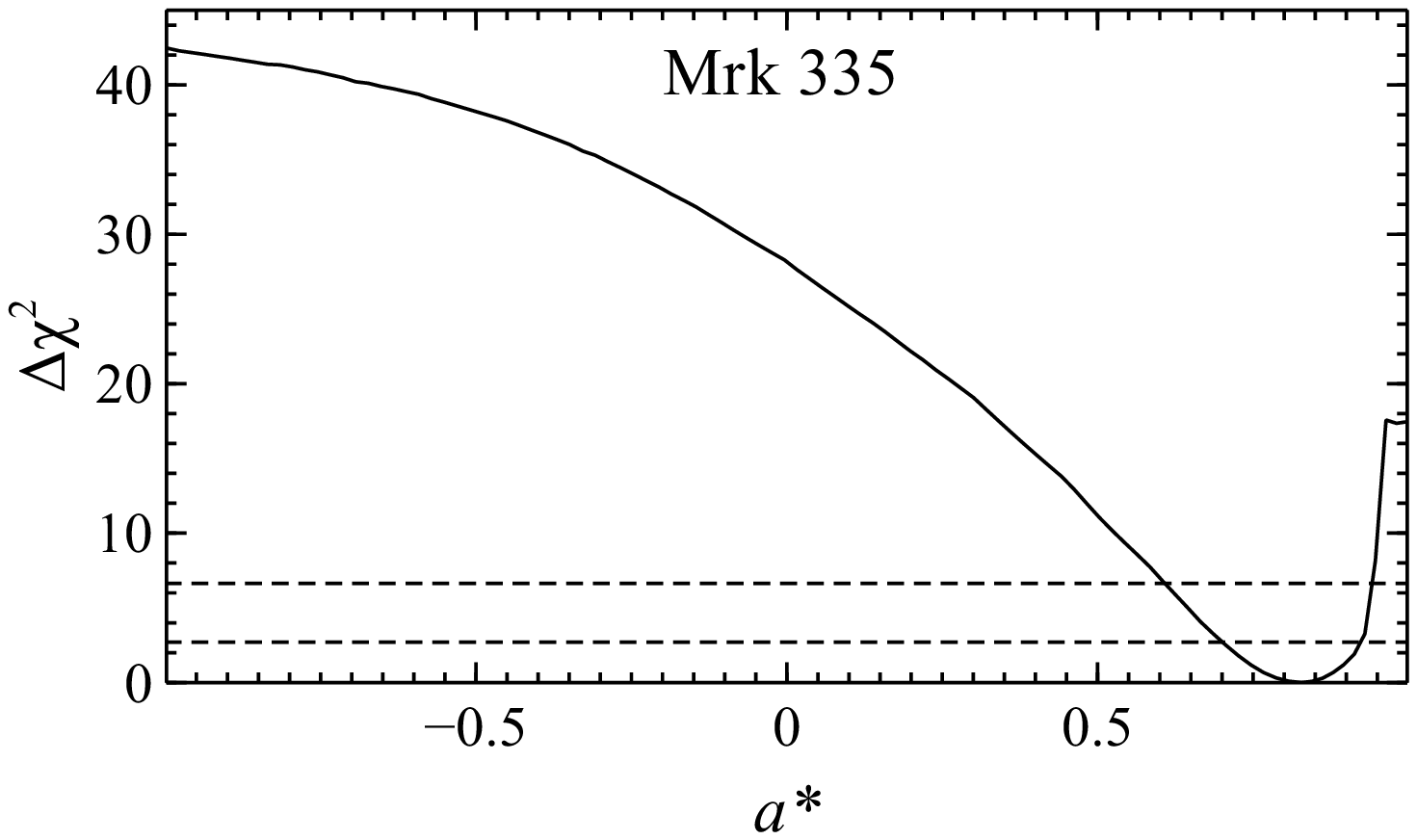}}
}\\
\vspace*{0.4cm}
\rotatebox{0}{
{\includegraphics[width=159pt]{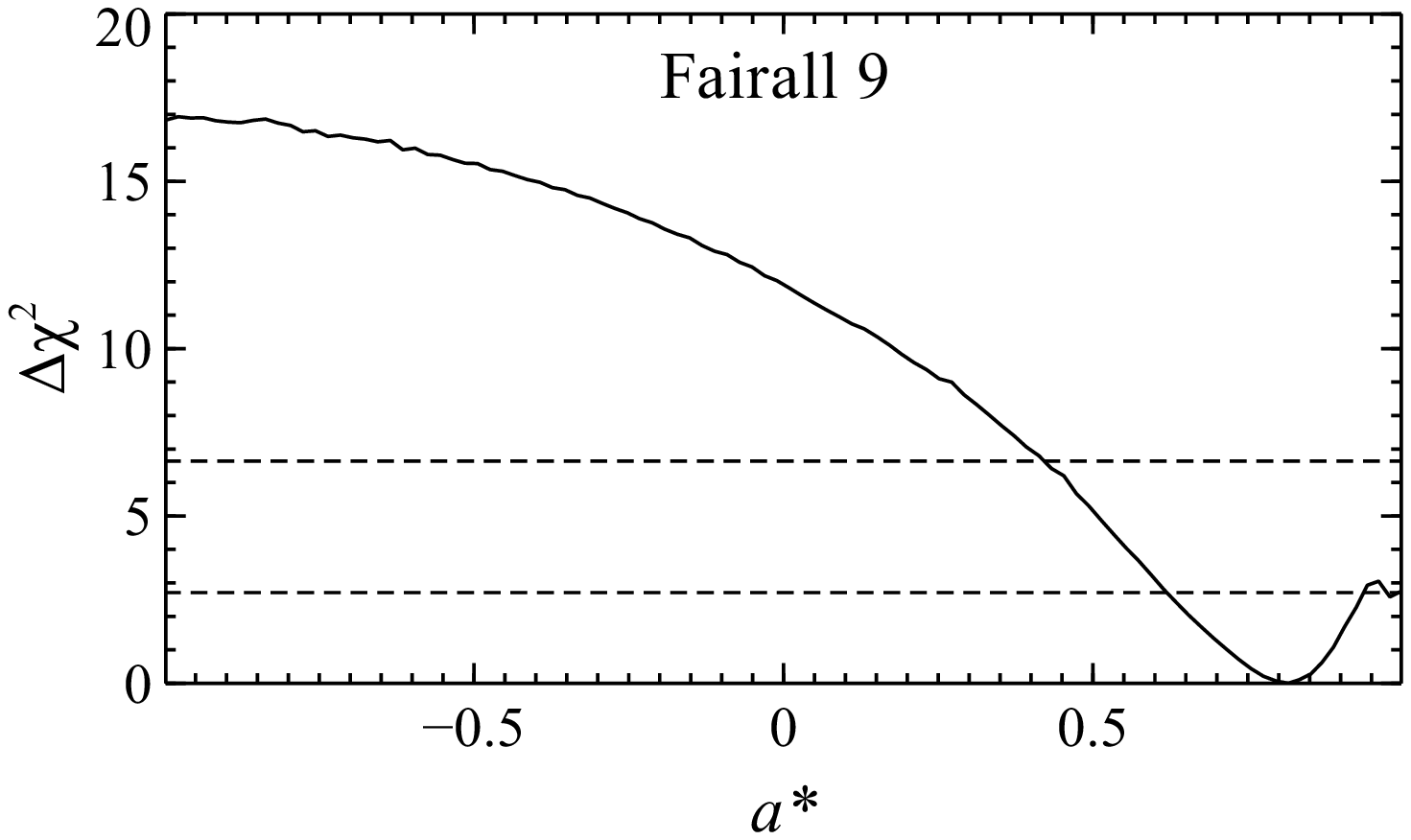}}
}
\rotatebox{0}{
{\includegraphics[width=159pt]{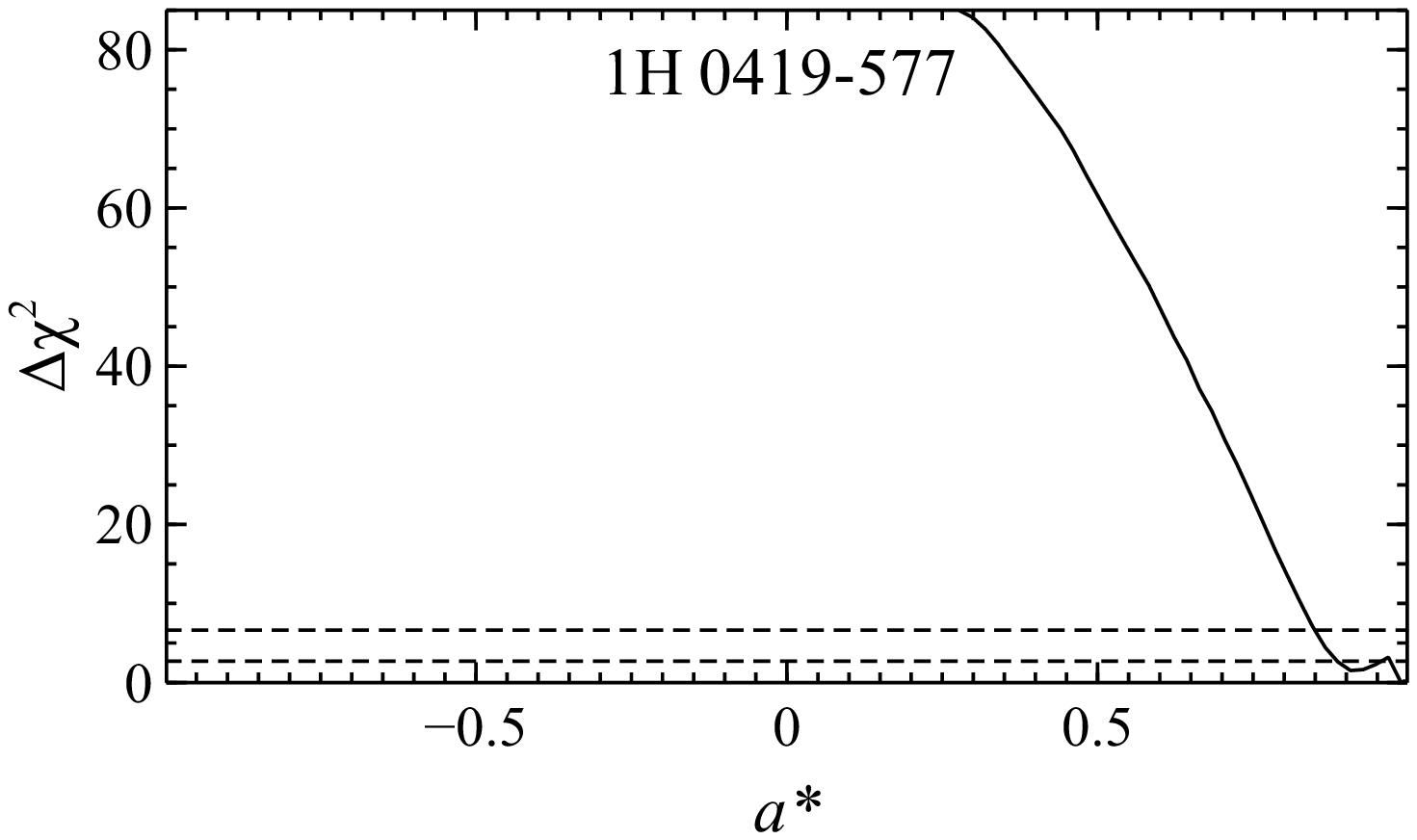}}
}
\rotatebox{0}{
{\includegraphics[width=159pt]{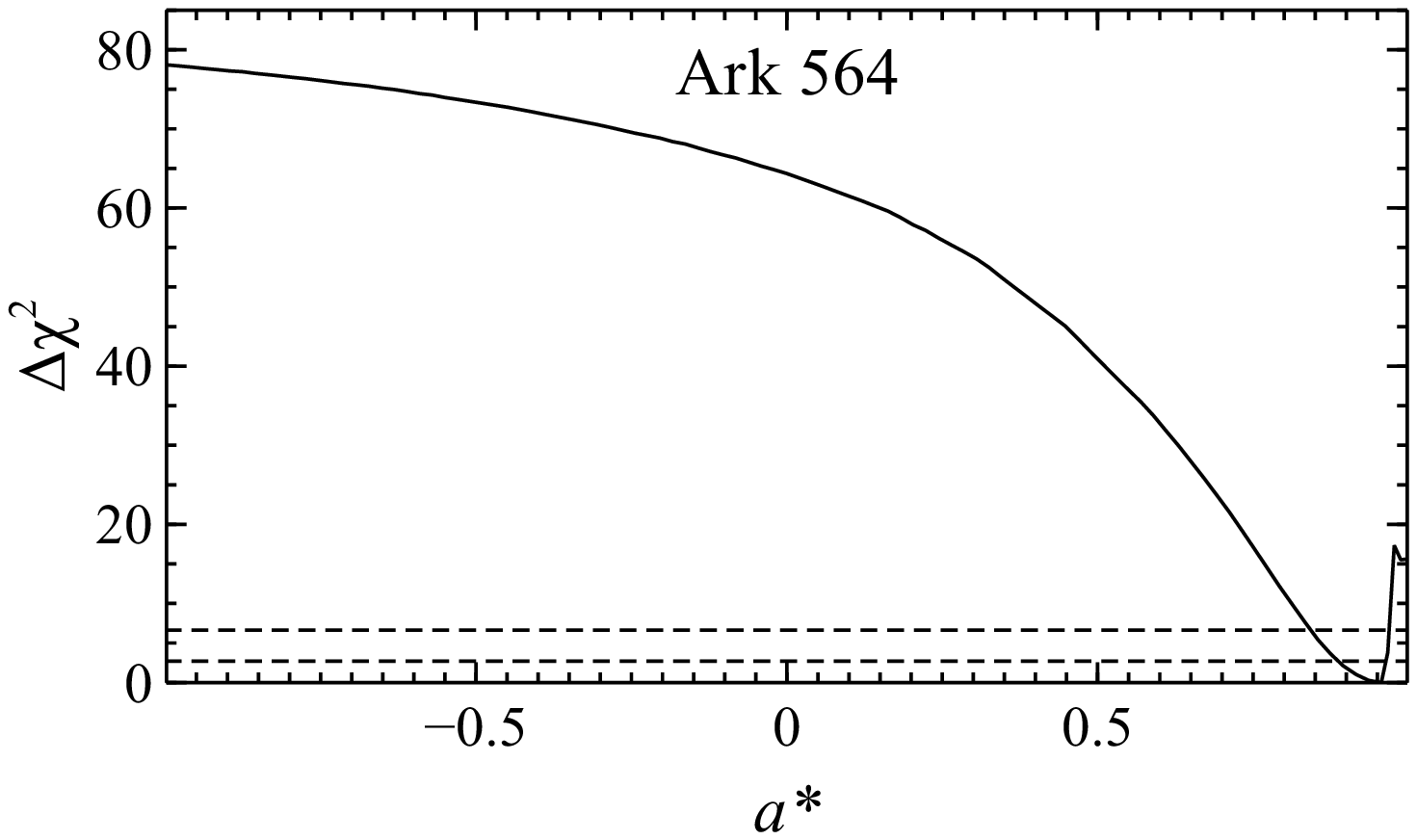}}
}\\
\vspace*{0.4cm}
\rotatebox{0}{
{\includegraphics[width=159pt]{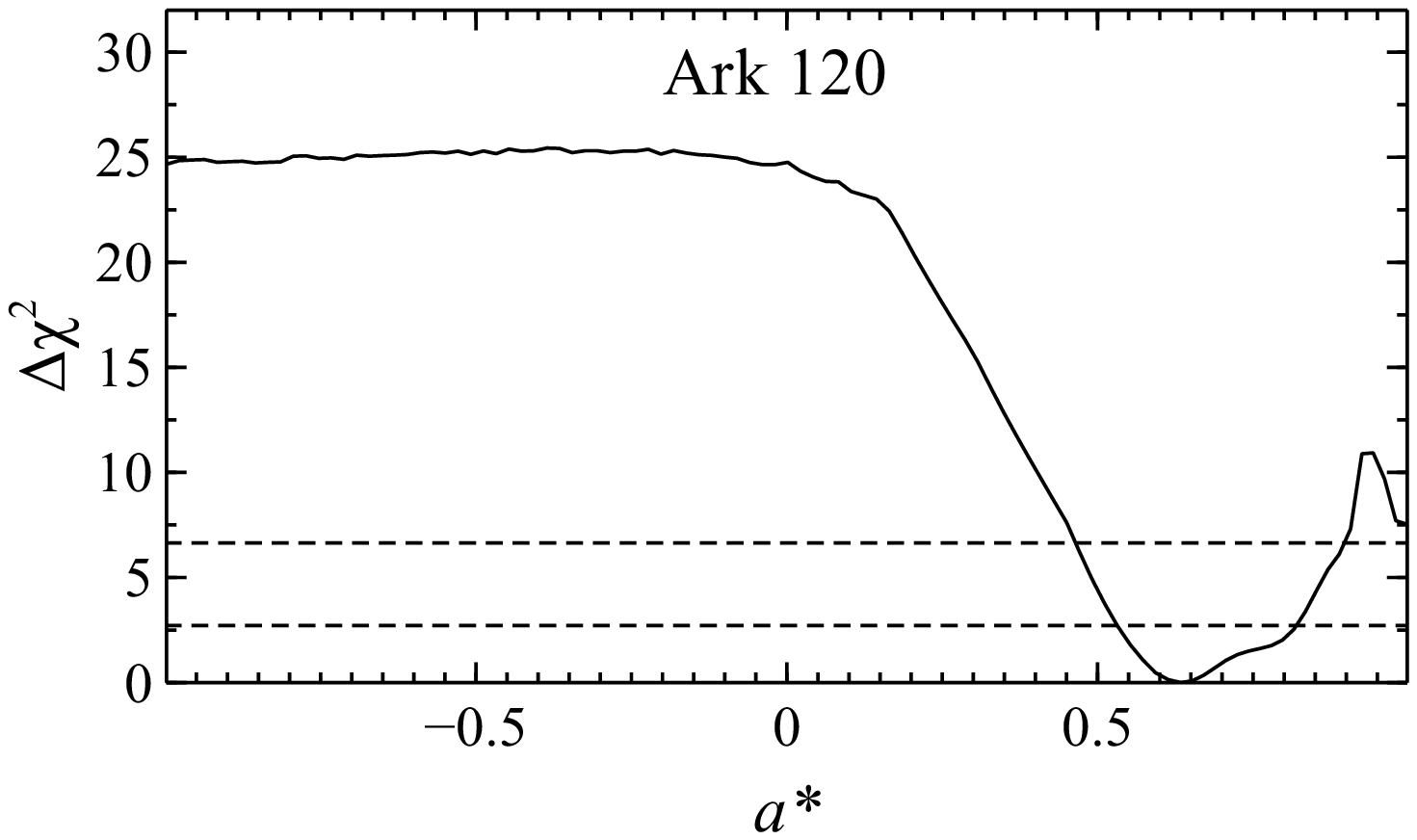}}
}
\rotatebox{0}{
{\includegraphics[width=159pt]{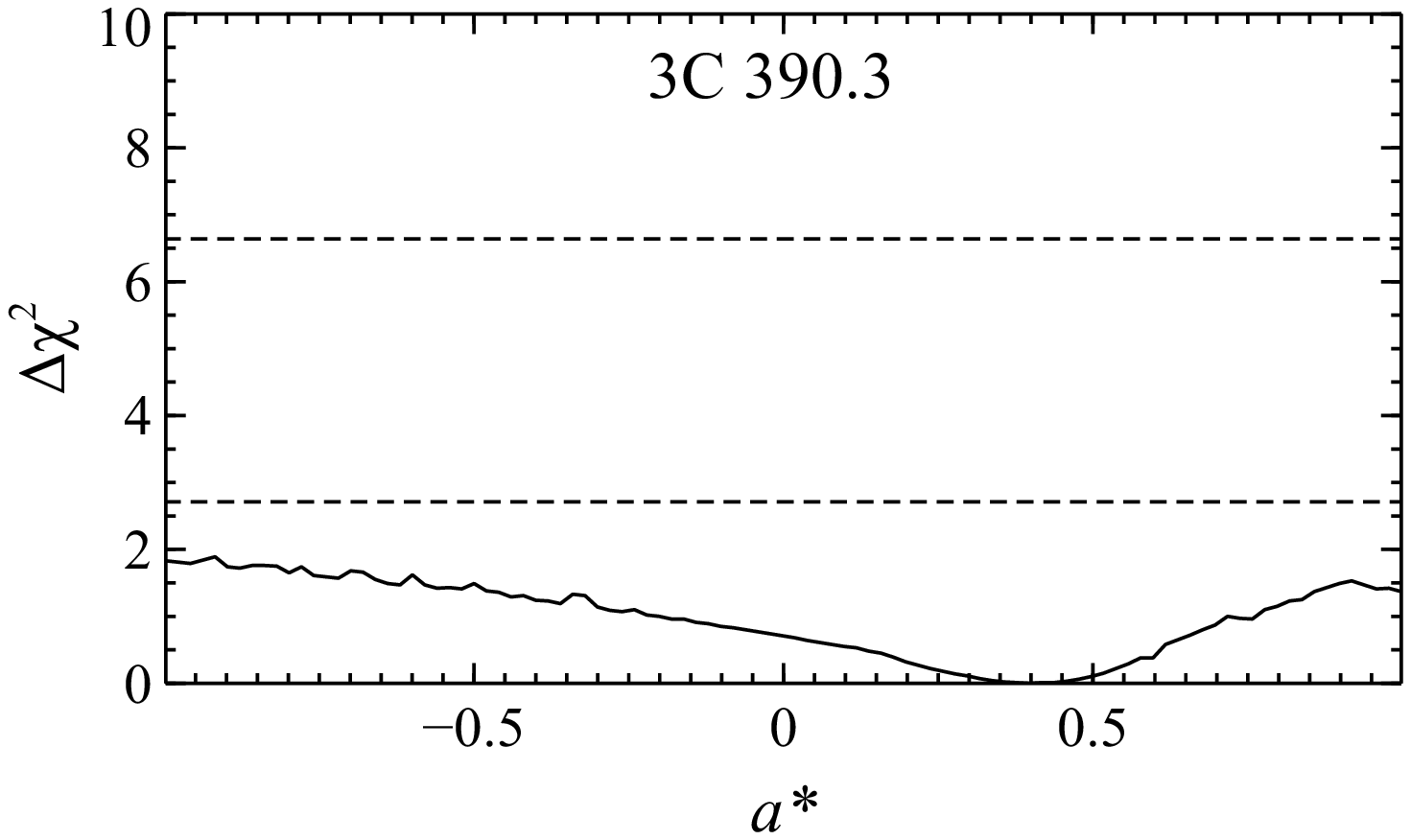}}
}
\rotatebox{0}{
{\includegraphics[width=159pt]{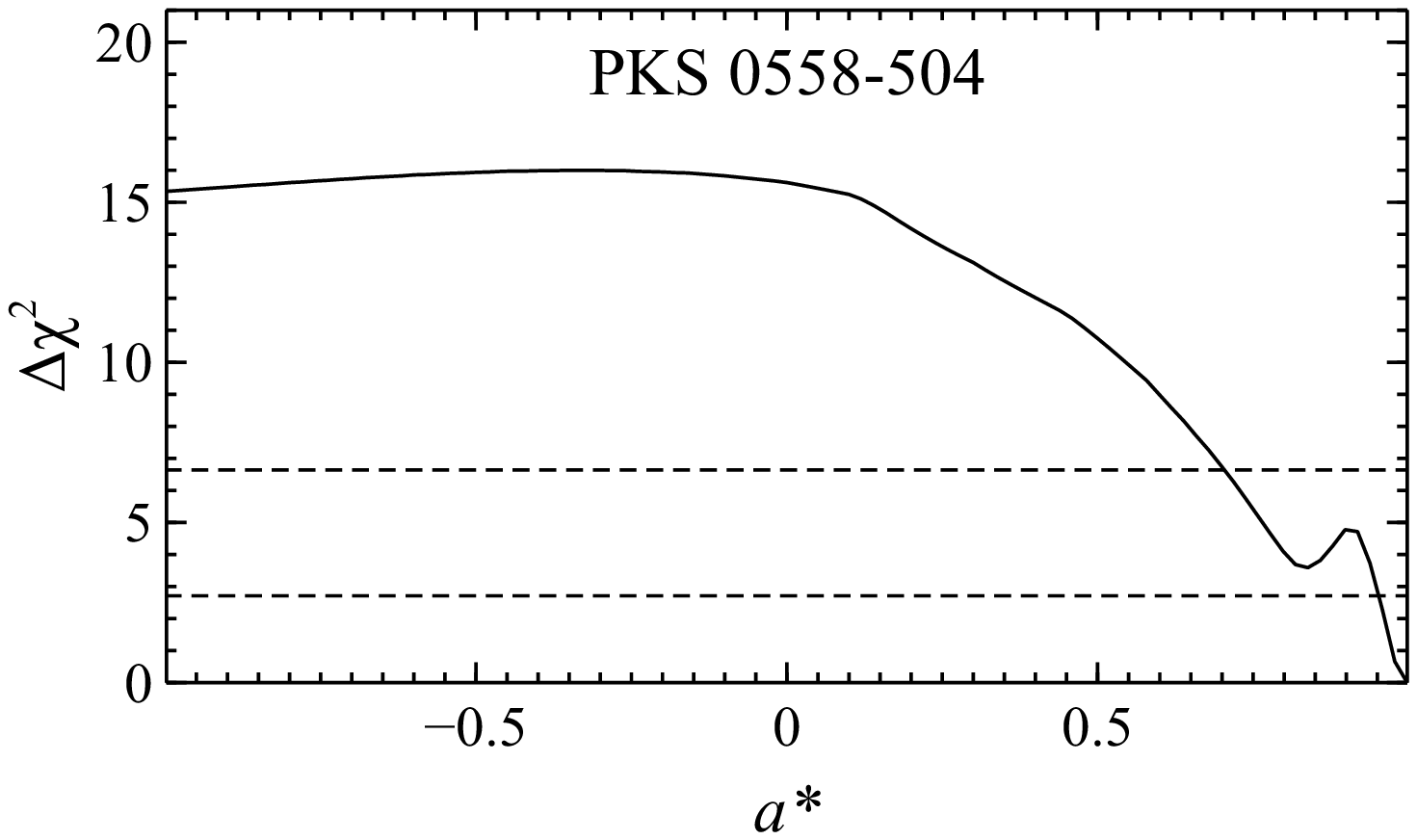}}
}\\
\vspace*{0.4cm}
\rotatebox{0}{
{\includegraphics[width=159pt]{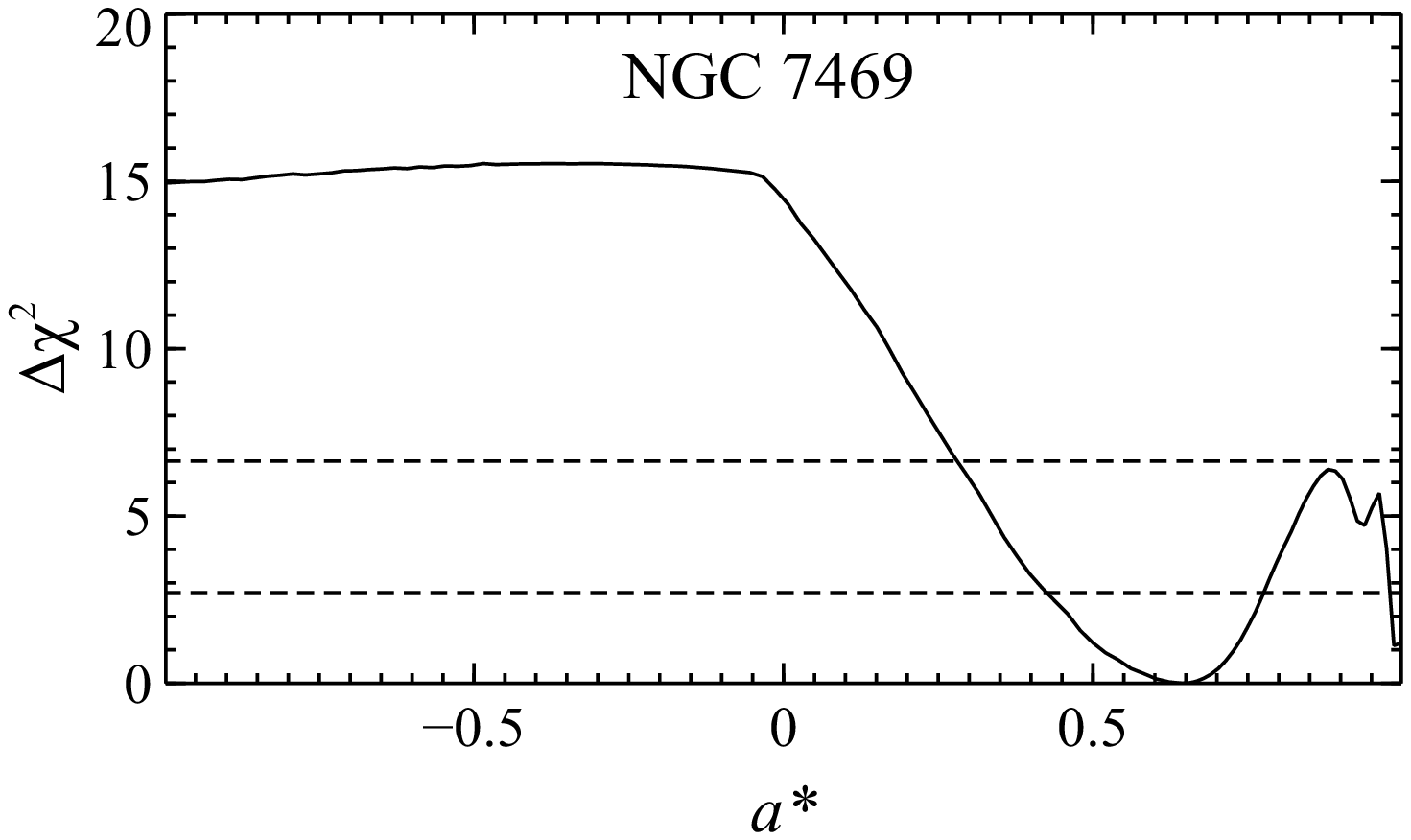}}
}
\rotatebox{0}{
{\includegraphics[width=159pt]{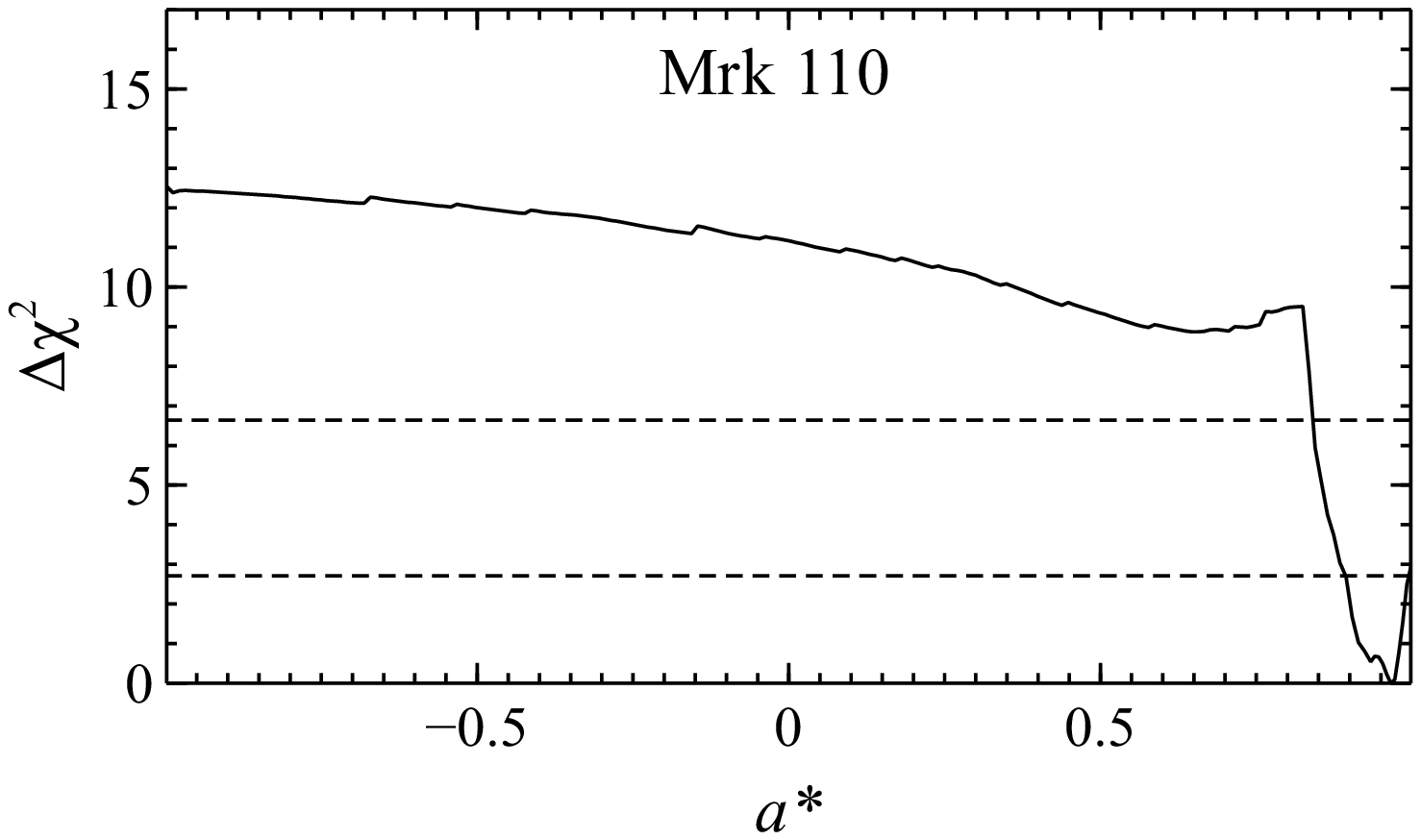}}
}
\rotatebox{0}{
{\includegraphics[width=159pt]{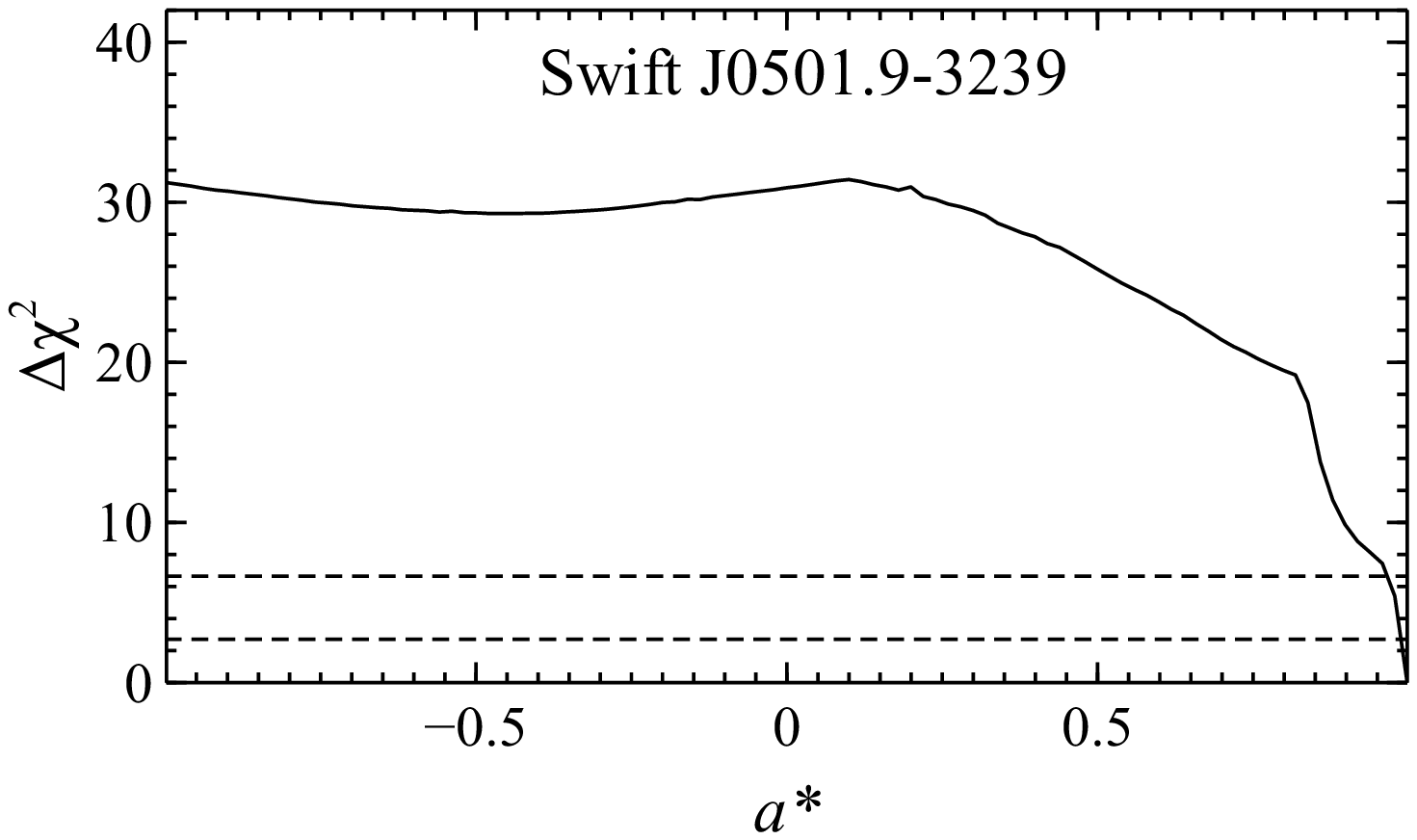}}
}\\
\vspace*{0.4cm}
\rotatebox{0}{
{\includegraphics[width=159pt]{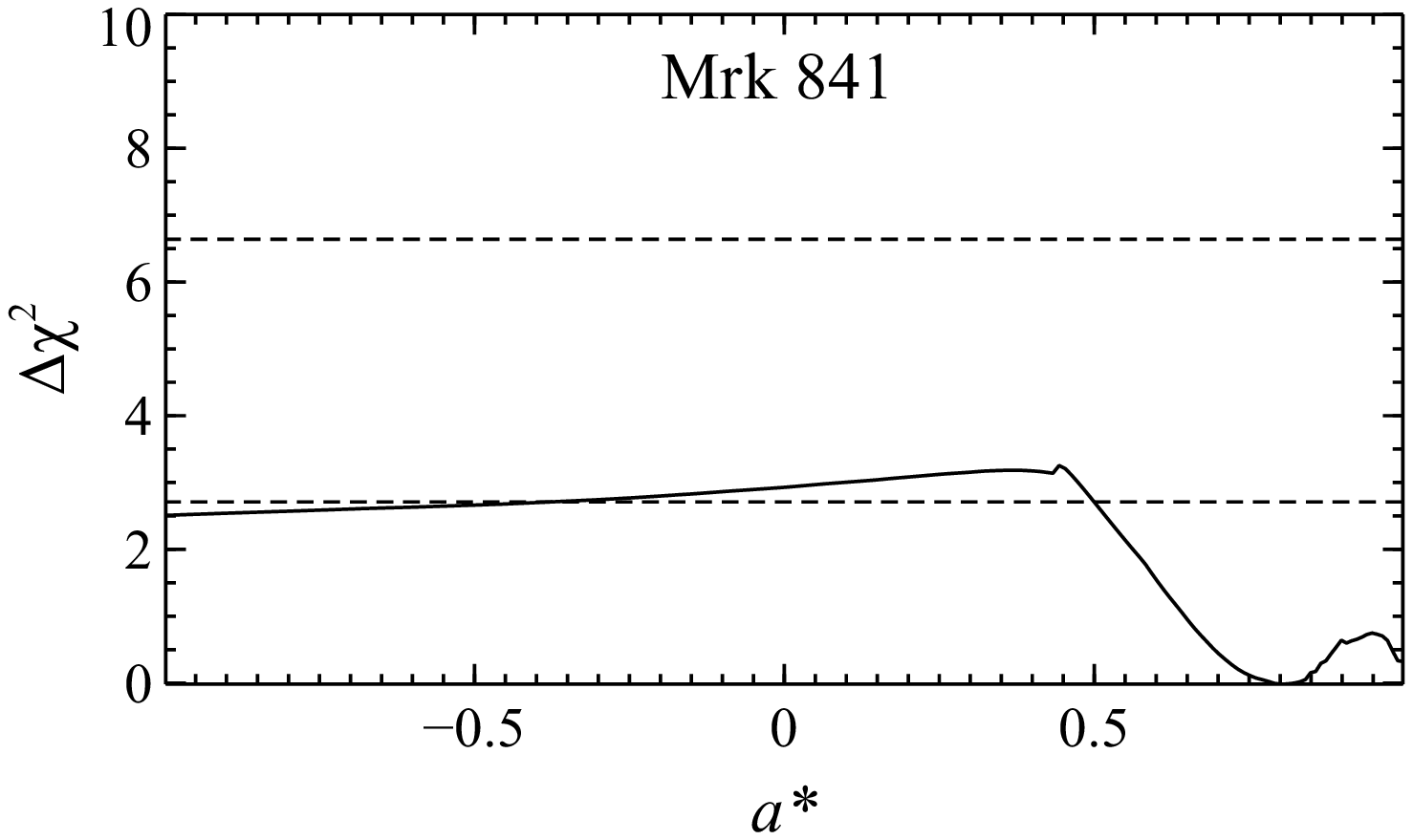}}
}
\rotatebox{0}{
{\includegraphics[width=159pt]{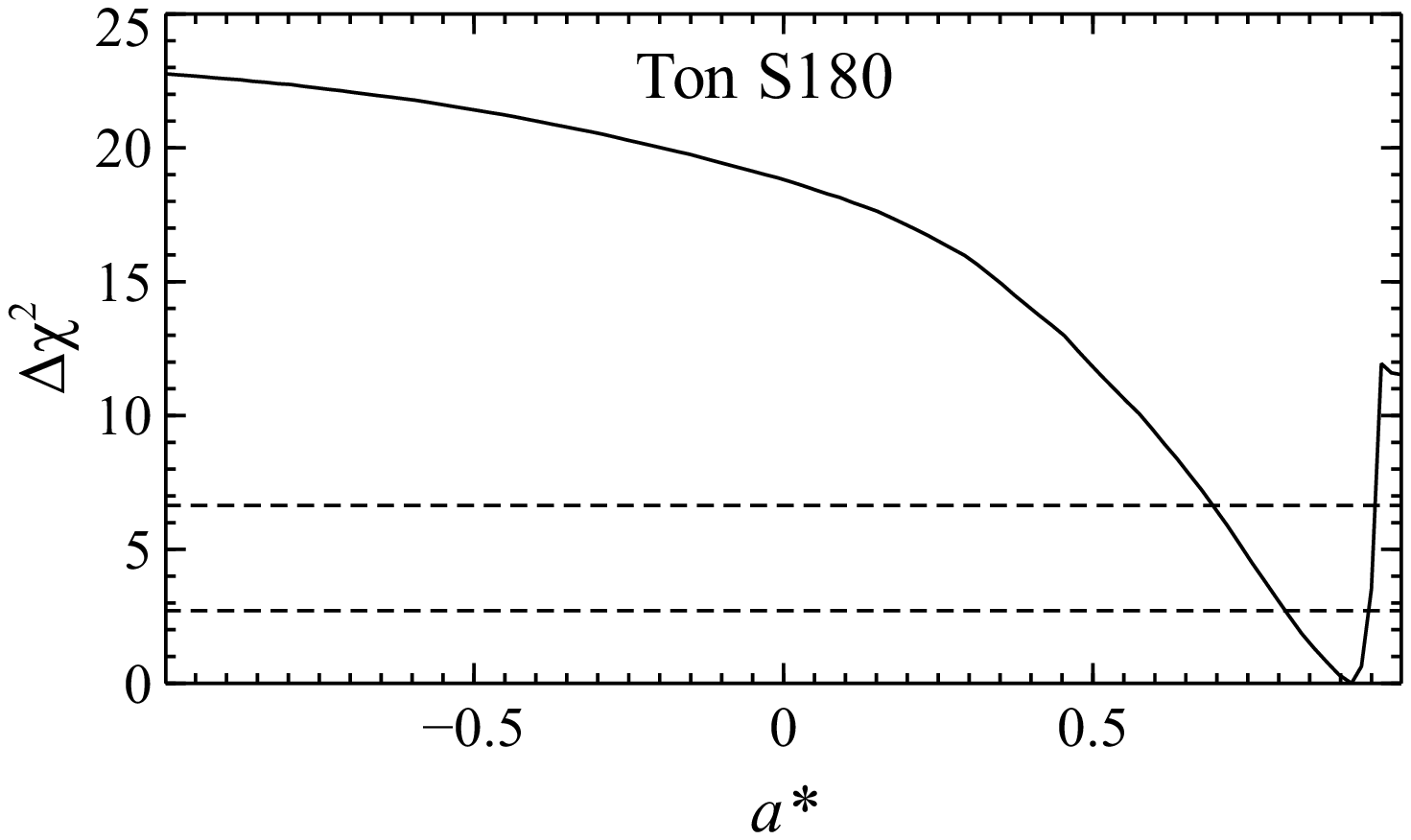}}
}
\rotatebox{0}{
{\includegraphics[width=159pt]{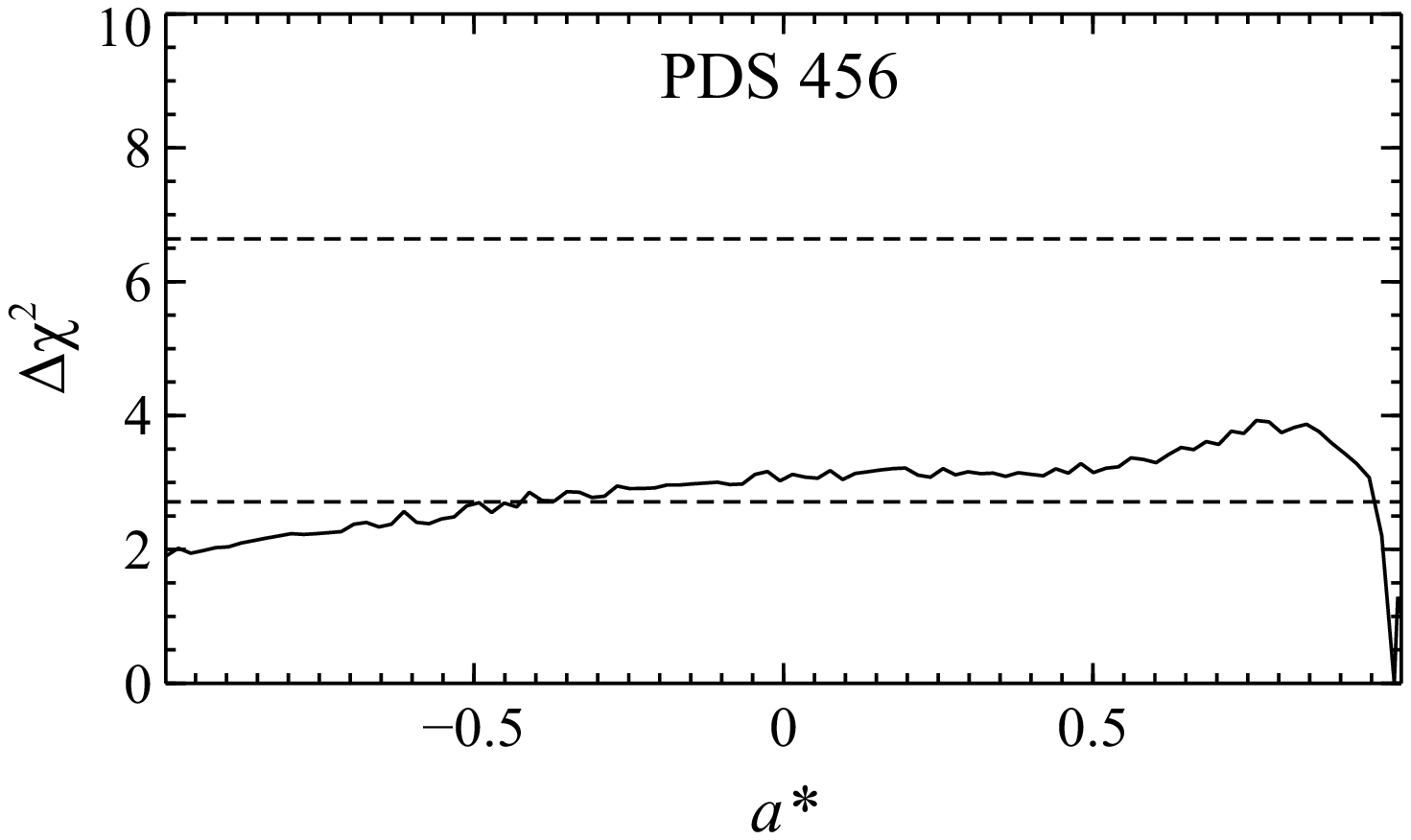}}
}\\
\vspace*{0.4cm}
\rotatebox{0}{
{\includegraphics[width=159pt]{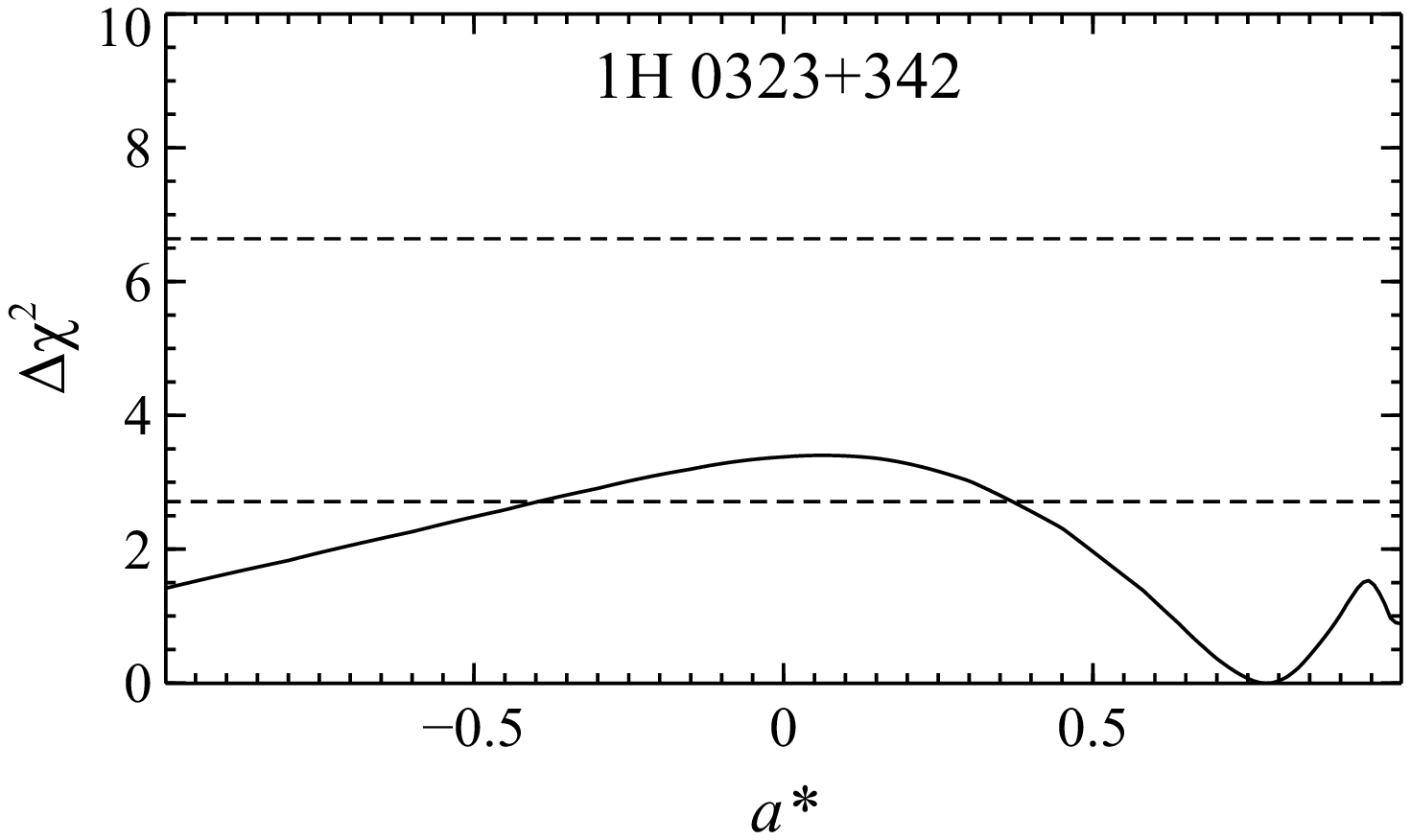}}
}
\rotatebox{0}{
{\includegraphics[width=159pt]{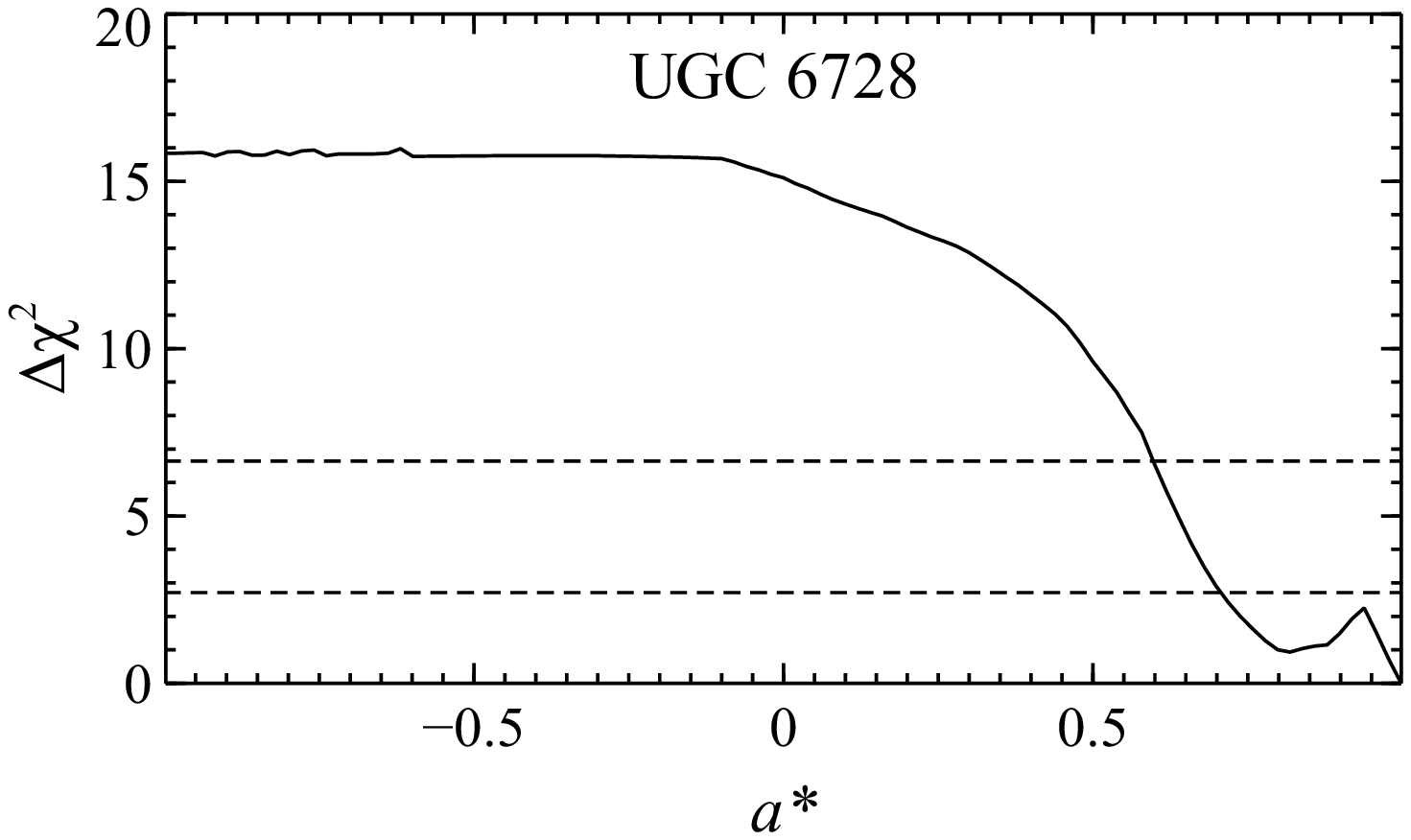}}
}
\rotatebox{0}{
{\includegraphics[width=159pt]{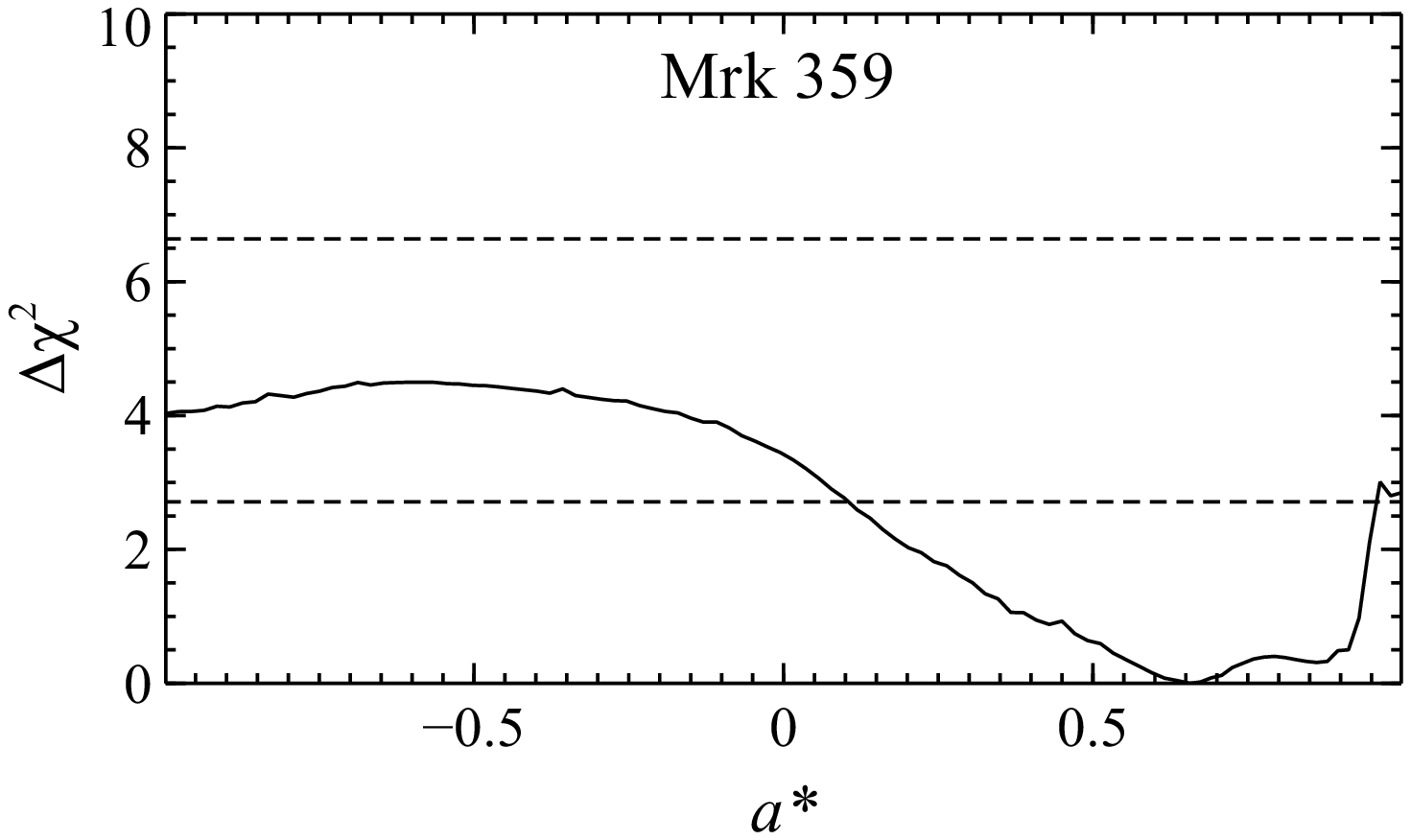}}
}
\end{center}
\caption[\chisq\ confidence contours for the spin measurements obtained for the
compiled sample]
{\chisq\ confidence contours for the spin measurements obtained for the compiled
sample, based (where possible) on the results obtained when allowing $C_{\rm
XIS/PIN}$ to vary (see Table \ref{tab_refl2}). Horizontal dashed lines represent
$\Delta$\chisq\ equivalent to the 90 and 99 confidence intervals.}
\label{fig_c3_spin}
\end{figure*}

Here, we provide further information specific to our individual modelling of each
source, including details on any additional emission or absorption features
included beyond the basic PLC+RDC interpretation. The key reflection parameters
obtained from our spectral modelling are quoted in Table \ref{tab_c3_refl}
($C_{\rm PIN/XIS}$ fixed at the values obtained from equation \ref{eqn_crossnorm})
and in Table \ref{tab_refl2} ($C_{\rm PIN/XIS}$ free to vary within the range
defined by the 3 per cent systematic uncertainty on the modelled PIN background).
We also show 1-D confidence contours of the spin parameter for the full range
considered for each source in Figure \ref{fig_c3_spin}, based where possible on
the results presented in Table \ref{tab_refl2}. Note that these contours have been
generated with the inclination limited to $i \leq 75$\deg\ for sources in which
this parameter is free to vary.

\textbf{\textit{Mrk\,509:}} In addition to the basic PLC+RDC interpretation for the
continuum we also include a moderately ionised absorber at the redshift of the host
galaxy, with $\log\xi = 2.03\pm0.04$\,\ergcmps\ and \nh = $1.9^{+0.4}_{-0.5} \times
10^{21}  $\,\atpcm. The ionisation obtained is consistent with that of the most
prominent absorbing zone found in the detailed RGS analysis presented by
\cite{Detmers11} as part of the substantial multi-wavelength observing campaign
recently carried out on this source (see also \citealt{Kaastra11}), but the column
obtained here is larger. This may be due to our simplistic, single zone treatment of
the absorber, but the inclusion of a second ionised absorbing zone does not offer a
substantial improvement. We also find a narrow emission feature consistent with
neutral iron, so include a distant reflector, and additional emission consistent
with a Fe {\small XXV}. The latter is modelled with a narrow Gaussian emission
component, and we find an equivalent width of $EW_{\rm XXV} = 17^{+9}_{-10}$\,eV;
the inclusion of this feature provides an improvement of $\Delta$\chisq\ = 23 for
an additional degree of freedom. Here, we find that the normalisation between the
XIS and PIN spectra plays an important role in determining the interplay between
the PLC and RDC continuum components, and hence in determining the reflection
parameters. When $C_{\rm PIN/XIS}$ is allowed to vary the inclination obtained is
significantly larger, although still reasonable for an unobscured AGN, and the spin
is lower and less constrained. Furthermore, the ionisation of the absorber increases
to $\log\xi = 2.38^{+0.08}_{-0.13}$\,\ergcmps.

\textbf{\textit{3C\,382:}} As one of the few broad line radio galaxies (BLRGs)
included in this sample, it is possible to obtain an independent constraint on
the inclination at which the source is viewed at from radio observations of its
jet, assuming that the jet axis is aligned perpendicular to the plane of the inner
disc. Although it is difficult to rigorously test this assumption, we note that
the Galactic binary system XTE\,J1550-564, one of the few cases in which it has
been possible to independently constrain the orbit and jet inclinations, appears
to be consistent with this picture (\citealt{Steiner12}). Therefore, when applying
the PLC+RDC continuum, we fixed the inclination at $i = 40$\deg, consistent with
the radio constraint presented by \cite{Giovannini01}.

In addition to this underlying continuum, there is also a narrow feature associated
with neutral iron, and as initially reported by \cite{Sambruna11}, evidence for a
further narrow feature at $\sim$7.5\,\kev. We include a distant reflector to account
for the former, and the latter is modelled with a narrow Gaussian emission line. The
inclusion of this Gaussian results in an improvement of $\Delta\chi^{2}$ = 19 for
2 additional degrees of freedom, and we find a (restframe) line energy of $E_{\rm
G} = 7.53\pm0.04$\,\kev, and an equivalent width of $EW_{\rm G} = 19\pm9$\,eV; the
energy of this feature precludes an association with iron, but is consistent with
moderately ionised nickel. Finally, we also include a moderately ionised absorber
at the redshift of the galaxy, similar to that found by \cite{Torresi10} in their
analysis of the high resolution \xmm\ RGS spectrum of this source, with $\log\xi =
2.5\pm0.1$ \ergcmps\ and \nh\ = $1.4^{+0.7}_{-0.5} \times 10^{21}$\,\atpcm. As with
Mrk\,509, the normalisation between the XIS and PIN spectra plays an important role
in determining the interplay between the PLC and RDC continuum components in this
case, and hence in constraining the reflection parameters. We find that when $C_{\rm
PIN/XIS}$ is allowed to vary the iron abundance obtained is much lower, and the spin
and the emissivity index are essentially unconstrained. Therefore in Table
\ref{tab_refl2} we present the results obtained with the index fixed at the Newtonian
prediction ($q=3$), although the spin is still only very weakly constrained.

\textbf{\textit{Mrk\,335:}} In addition to the basic PLC+RDC model adopted in this
work, we also see evidence for narrow emission features consistent with neutral
iron and Fe {\small XXV}, similar to \cite{Patrick11a}. The former is modelled with
a distant reflector, and the latter is simply modelled with narrow ($\sigma$ =
10\,eV) Gaussian line profile. The inclusion of the latter feature provides a strong
improvement of $\Delta$\chisq\ = 33 for 1 extra degree of freedom, although it does
not significantly effect the parameters obtained for the main PLC and RDC continuum
components, and we find an equivalent width of $EW_{\rm XXV} = 29\pm9$\,eV. The spin
obtained here is very similar to that presented by \cite{Patrick11a} for the same
dataset, and is consistent with both the spin inferred from the inner radius obtained
by \cite{Grupe08} in their analysis of the low flux \xmm\ datasets, and the spin
measured from the recent intermediate flux \xmm\ observations by \cite{Gallo12}.

\textbf{\textit{Fairall 9:}} In addition to the standard PLC+RDC continuum considered
here, we also detect strong, narrow emission features consistent Fe {\small I} and
and Fe {\small XXVI}, and a weaker emission feature consistent with Fe {\small XXV}.
A distant reflector is included to model the former, and the latter two are modelled
with narrow ($\sigma = 10$\,eV) Gaussian line profiles. The equivalent obtained are
$EW_{\rm XXV} = 13^{+7}_{-5}$\,eV and $EW_{\rm XXVI} = 35^{+8}_{-7}$\,eV for the
ionised emission lines, and their inclusion results in respective improvements of
$\Delta$\chisq\ = 14 and 52 for one extra degree of freedom in each case. Although the
inclusion of the Fe {\small XXVI} feature does have a significant effect on some of
the main reflection parameters, the best fit model excluding this line resulted in an
estimate for the inclination of 87\deg, which is unphysically high for an unobscured
AGN. Conversely, the inclusion of the Fe {\small XXV} line does not significantly
modify the key parameters further. The spin obtained here is larger, although still
consistent with those obtained by \cite{Schmoll09} with the same dataset, by
\cite{Patrick11b} with a later \suzaku\ observation not included here (performed after
October 2010), and by \cite{Emman11a} with a long \xmm\ observation, but is lower than
that presented in \cite{Patrick11a} again with the same dataset.

\textbf{\textit{1H\,0419-577:}} In addition to the PLC+RDC components, we also
detect a weak narrow emission feature consistent with neutral iron, and therefore
include a second, distant reflector. The spin constraint obtained here is
consistent with that inferred both in our previous work on the first \suzaku\ 
dataset (\citealt{Walton10hex}), and in the \xmm\ analysis of \cite{Fabian1h0419}.

\textbf{\textit{Ark\,564:}} The basic PLC+RDC interpretation provides a good fit
to the data. No narrow iron emission or absorption lines are detected. The spin
obtained here is consistent with that inferred from the inner radius obtained by
\cite{Dewangan07}, who modelled the long 2005 \xmm\ observation with blurred
reflection.

\textbf{\textit{Ark\,120:}} In addition to the basic PLC+RDC continuum, we also
detect a narrow emission feature consistent with neutral iron, and hence include
a second, distant reflector. The spin obtained here is broadly similar to,
although slightly better constrained than, that obtained by \cite{Nardini11} with
the emissivity index fixed at $q=5$, but is lower than that obtained by
\cite{Patrick11a} with the same dataset.

\textbf{\textit{3C\,390.3:}} Similar to 3C\,382, 3C\,390.3 is a BLRG, and hence
we fix the inclination at $i = 35$\deg\ when applying the PLC+RDC continuum,
consistent with the radio constraint on the inclination presented by
\cite{Giovannini01}. In addition to this underlying continuum, we find evidence
for a strong narrow emission feature consistent with neutral iron, and hence
include a distant reflector. However, owing to the weak features associated with
reflection, we are unable to reliably constrain all the parameters of interest,
and even when fixing the emissivity index at $q=3$ we find the spin to be
unconstrained. Note that similar to \cite{Sambruna09} we also include an
unresolved Gaussian component at $\sim$1.4\,\kev\ to account for a feature in
the background spectrum.

\textbf{\textit{PKS\,0558-504:}} Here the basic PLC+RDC interpretation provides a
good fit to the data; no statistically compelling narrow iron emission lines are
detected. However, when free we find that the inclination obtained is very high
($i \gtrsim 75$\deg), unphysically high for an unobscured active galaxy. This is
most likely due to the complex parameter degeneracies that can arise when using
multi-parameter reflection models (see \eg \citealt{Nardini11}). Therefore,
we instead present the results obtained with the inclination fixed at 45\deg,
similar to the rough estimation of \cite{Gliozzi10} based on the radio morphology
of the source.

\begin{figure}
\begin{center}
\rotatebox{0}{
{\includegraphics[width=235pt]{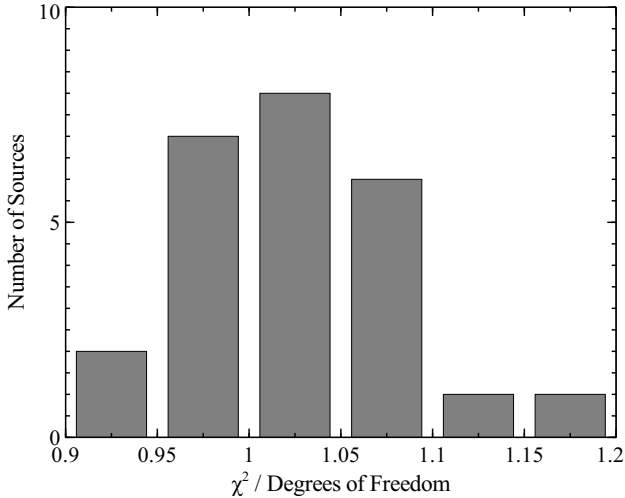}}
}
\end{center}
\vspace{-0.2cm}
\caption{The number distribution of reduced \chisq\ values obtained with our PLC+RDC
interpretations for the presented sample of AGN.}
\label{fig_chi}
\end{figure}

\textbf{\textit{NGC\,7469:}} In addition to the PLC+RDC continuum, there is a
strong narrow emission line consistent with neutral iron, so we again include a
distant cold reflector. We also include a weak, moderately ionised absorber at the
redshift of the galaxy, based on the most prominent absorption component detected
by \cite{Blustin07} when modelling the high resolution \xmm\ RGS spectrum. As the
absorber has an extremely weak effect on the spectrum in this case, we only allow
the column to vary, fixing the ionisation parameter at $\log\xi = 2.73$, and find
\nh\ = $1.7^{+1.2}_{-0.8} \times 10^{21}$\,\atpcm, consistent with that obtained
previously. In this case, when allowing $C_{\rm PIN/XIS}$ to vary it is not
possible to constrain all the reflection parameters of interest, so we present
the results with the inclination fixed at 45\deg. In addition to the golbal minimum
presented in Table \ref{tab_refl2}, there is a further, local minimum with a very
high spin ($a^* \gtrsim 0.98$) which also satisfies $\Delta\chi^{2} < 2.71$ (see
Fig. \ref{fig_chi}), equivalent to the best fit obtained with $C_{\rm PIN/XIS}$
fixed presented in Table \ref{tab_c3_refl}.

\textbf{\textit{Mrk\,110:}} In addition to the PLC+RDC continuum, we also detect a
strong narrow emission feature consistent with neutral iron, and a possible narrow
absorption feature at $\sim7.5$\,\kev. We include a distant reflector to account
for the former, and model the latter with a highly ionised photoionised plasma
using the \xstar\ code. The absorption zone here is assumed to have solar abundances
and a turbulent velocity of $\sim$1000\,\kmps, while the column, ionisation parameter
and outflow velocity are free to vary, and we find \nh\ = $9.6^{+0.7}_{-7.2} \times
10^{22}$\,\atpcm, $\log\xi = 4.2\pm0.3$ and $v_{\rm out} = 0.079\pm0.005c$. The
inclusion of this absorber provides an improvement of $\Delta$\chisq\ = 21 for 3
extra degrees of freedom.

\textbf{\textit{Swift\,J0501.9-3239:}} In addition to the basic PLC+RDC continuum,
we also detect a narrow emission feature consistent with neutral iron, and include
a second, distant reflector.

\begin{figure*}
\begin{center}
\rotatebox{0}{
{\includegraphics[width=159pt]{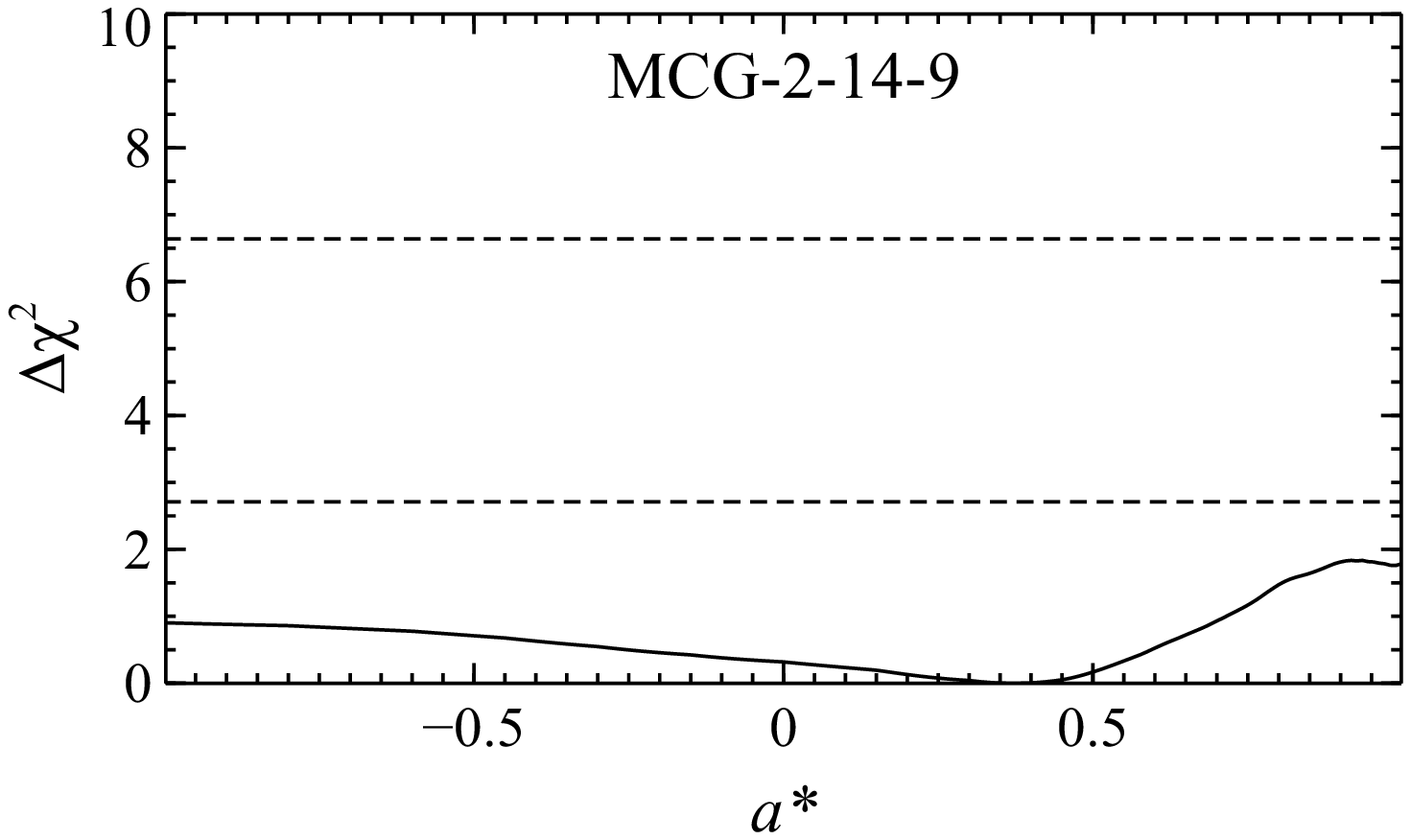}}
}
\rotatebox{0}{
{\includegraphics[width=159pt]{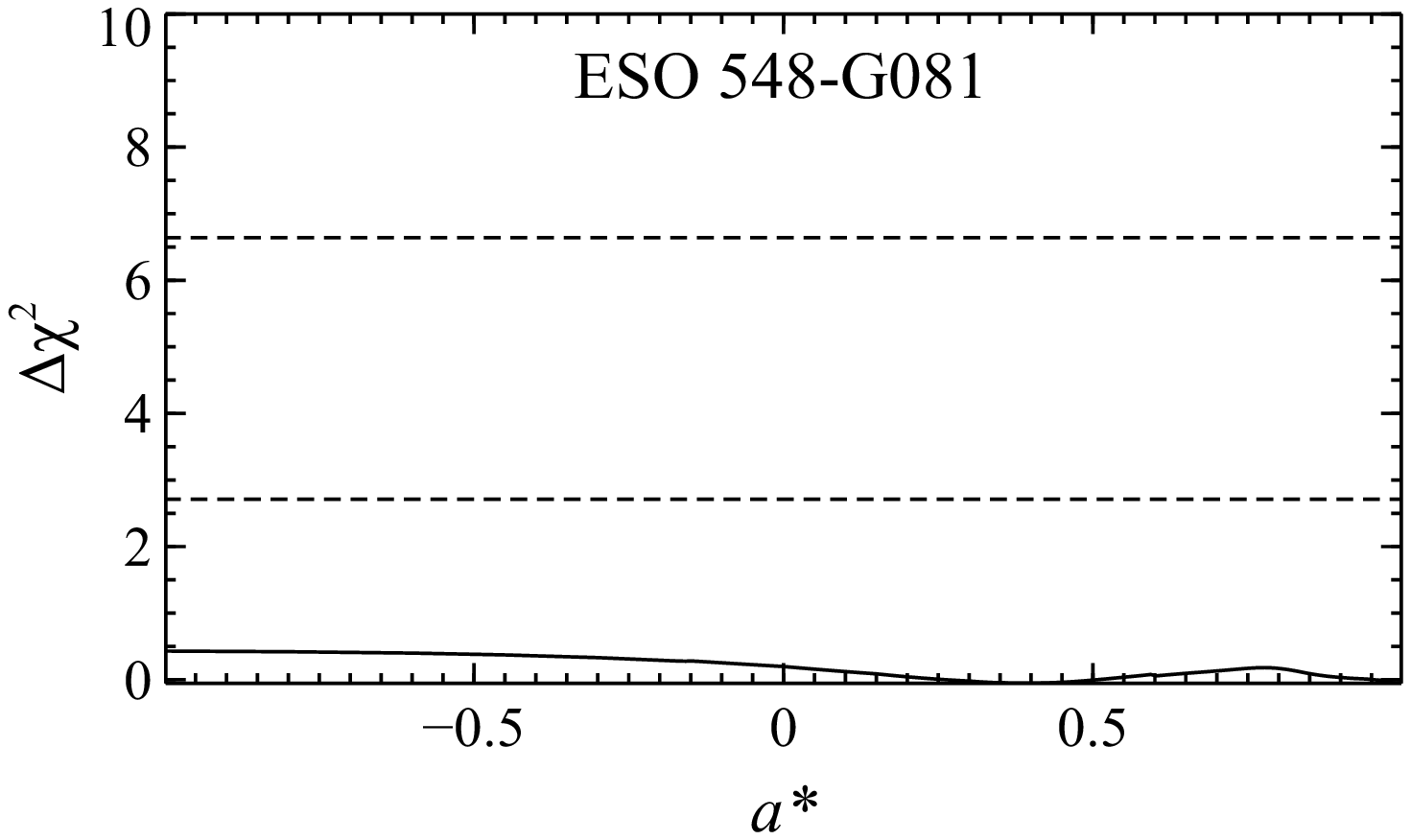}}
}
\rotatebox{0}{
{\includegraphics[width=159pt]{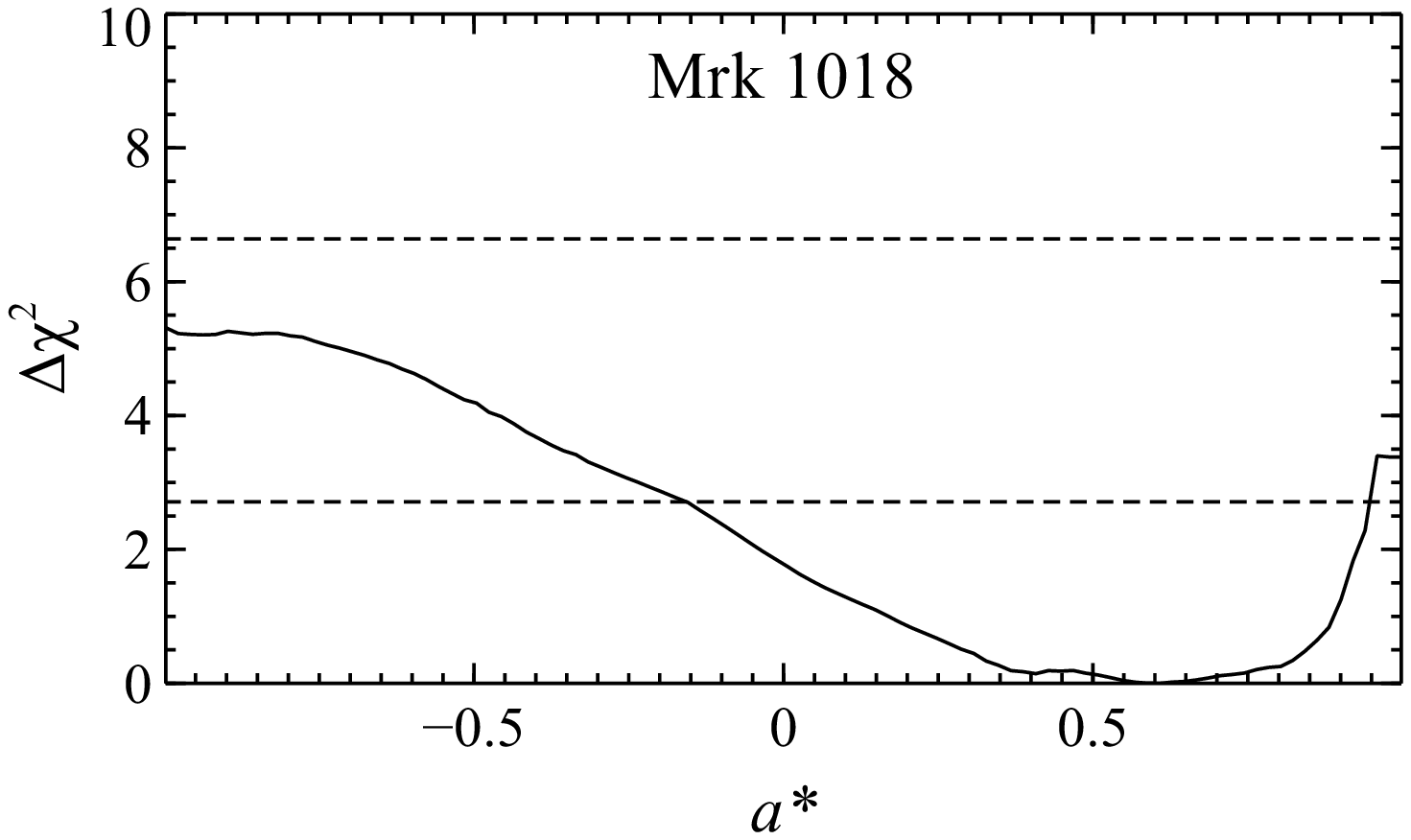}}
}\\
\vspace*{0.4cm}
\rotatebox{0}{
{\includegraphics[width=159pt]{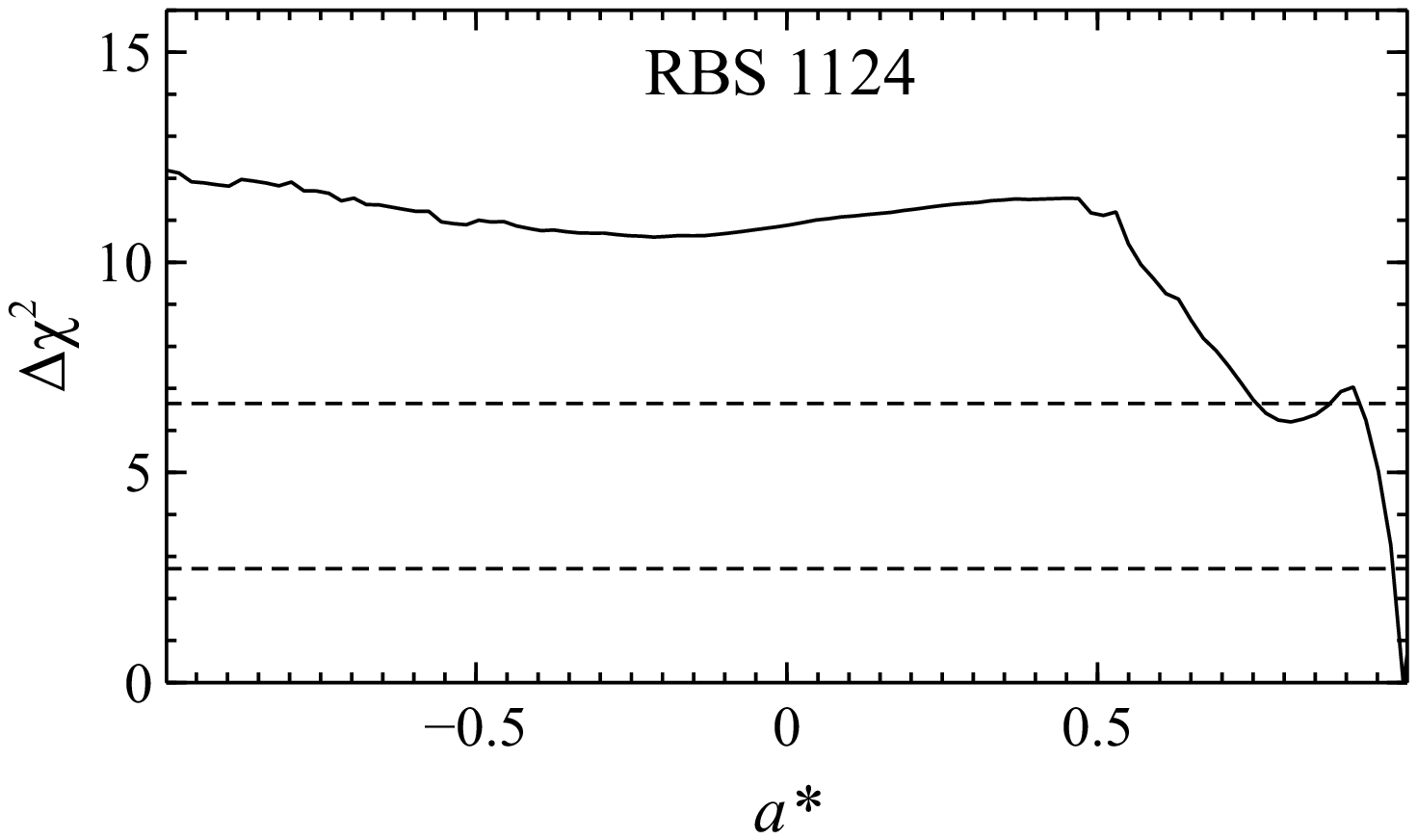}}
}
\rotatebox{0}{
{\includegraphics[width=159pt]{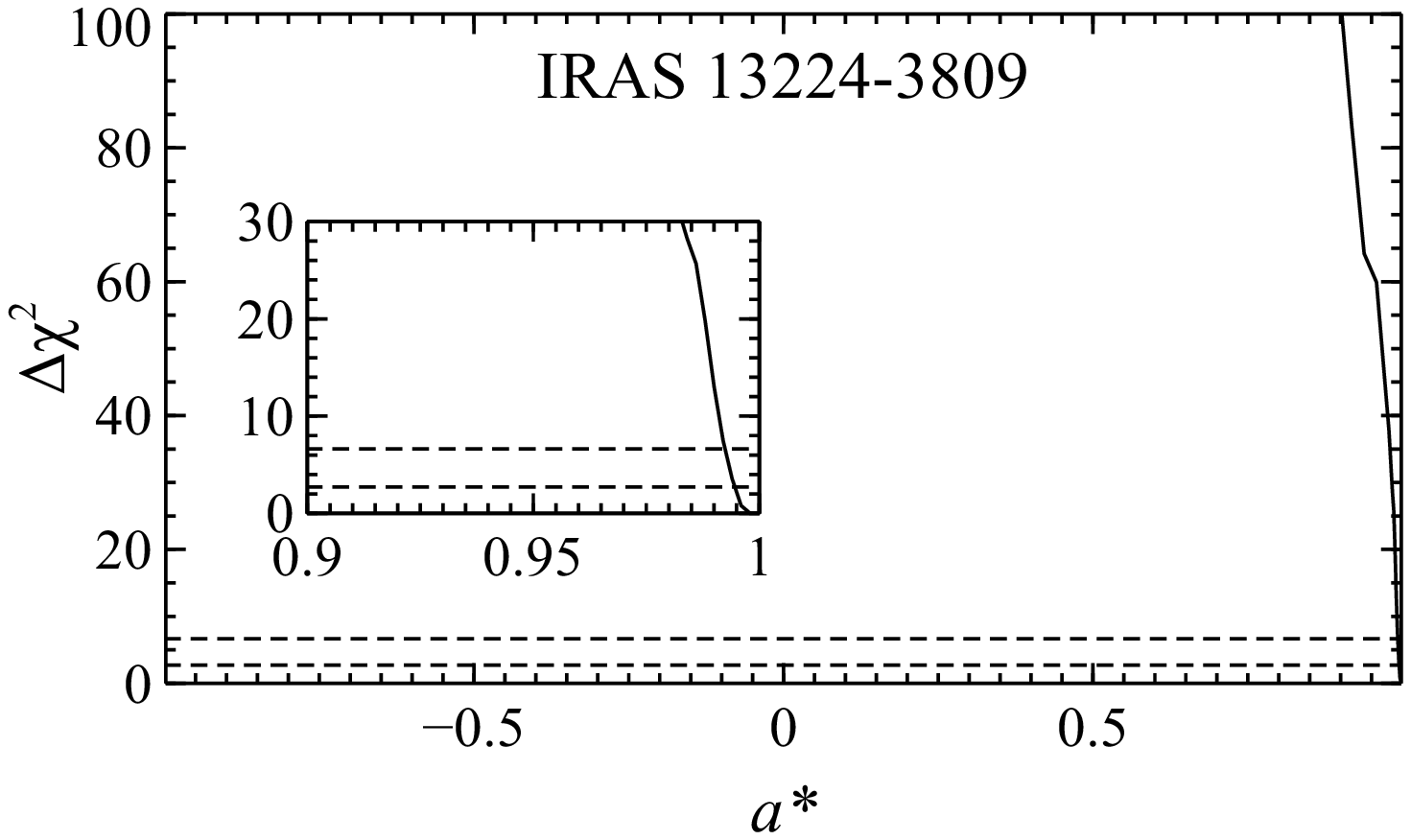}}
}
\rotatebox{0}{
{\includegraphics[width=159pt]{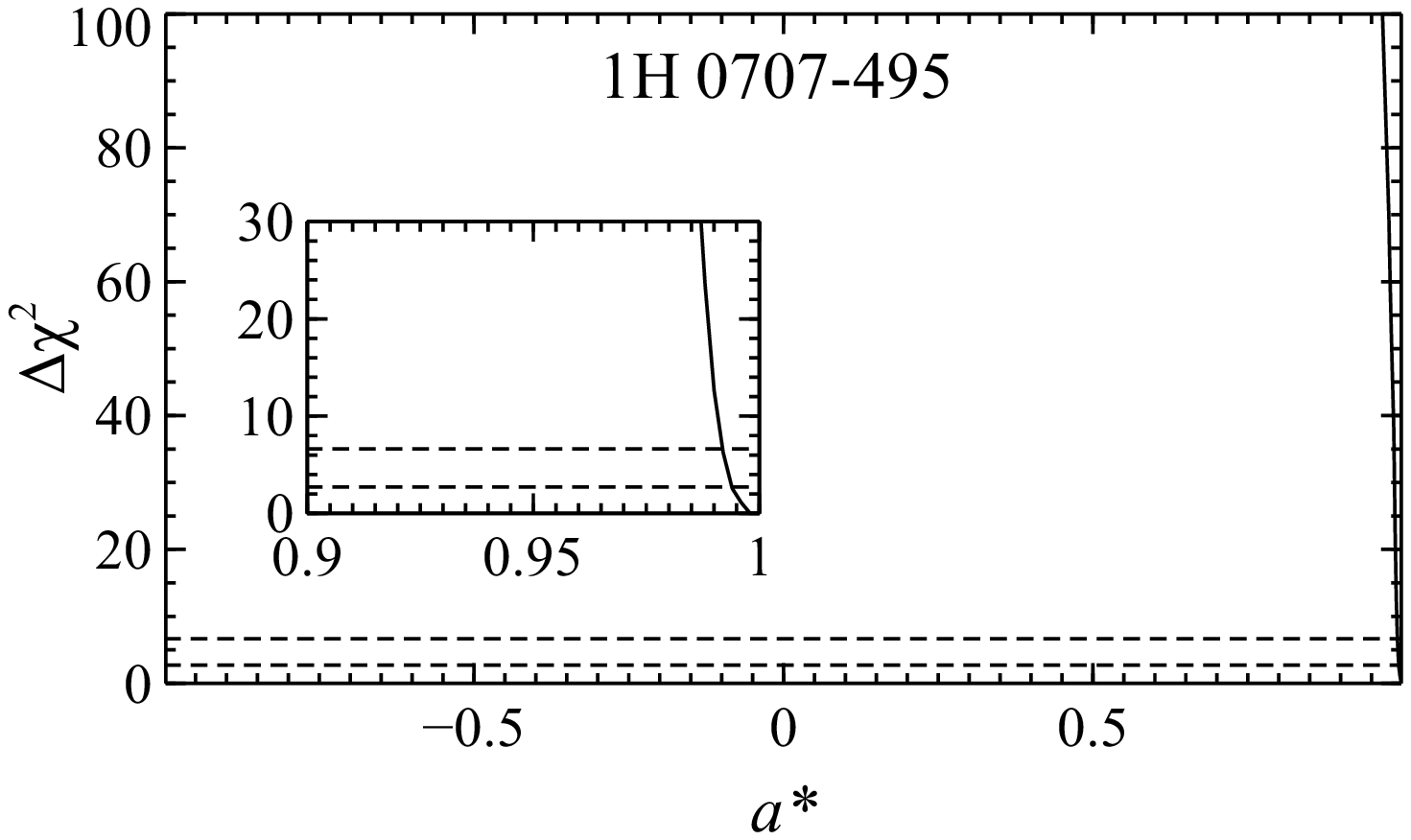}}
}\\
\vspace*{0.4cm}
\rotatebox{0}{
{\includegraphics[width=159pt]{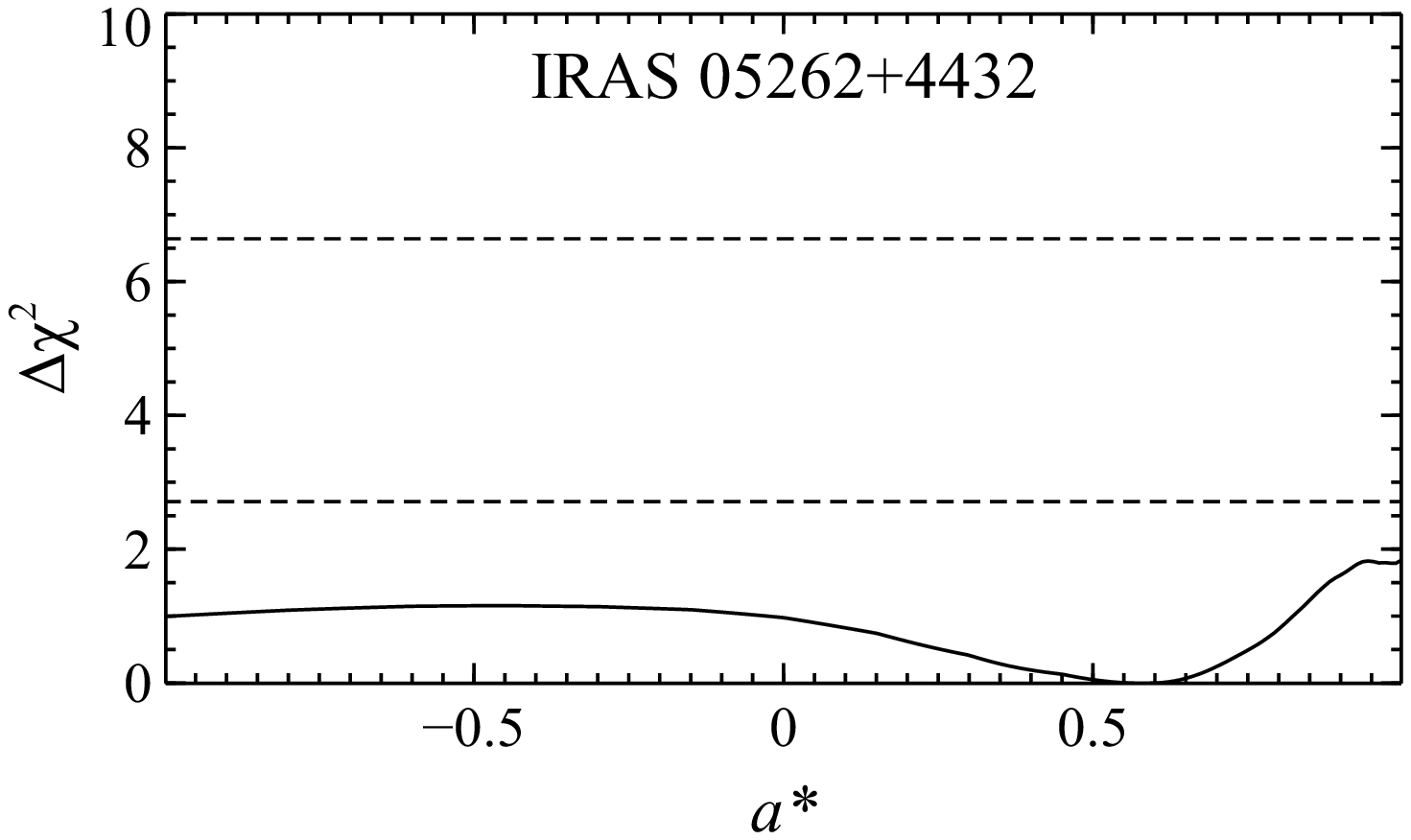}}
}\\
\vspace*{0.25cm}
\textbf{Figure~\ref{fig_c3_spin}.} Continued.
\end{center}
\end{figure*}

\textbf{\textit{Mrk\,841:}} In addition to the PLC+RDC components, we also detect a
narrow emission feature consistent with neutral iron and therefore include a
contribution from a second, distant reflector. We also include a moderately ionised
absorber at the redshift of the galaxy similar to the most prominent absorption
component detected by \cite{Longinotti10} in their analysis of the high resolution
\xmm\ RGS spectrum of this source, with $\log\xi = 2.2\pm0.2$ \ergcmps\ and \nh\ =
$2.1^{+0.9}_{-0.8} \times 10^{21}$\,\atpcm. Although we present the prograde spin
solution in Tables \ref{tab_c3_refl} and \ref{tab_refl2}, there is also a local minimum
with retrograde spin ($a^* \lesssim -0.4$) also just about satisfies $\Delta\chi^{2} <
2.71$ (see Fig. \ref{fig_c3_spin}). The prograde solution is consistent with the inner
radii obtained by \cite{Petrucci07} in their analysis of the available \xmm\ data.

\textbf{\textit{Ton\,S180:}} The basic PLC+RDC continuum provides a good fit to
the data, no narrow emission or absorption lines are detected over the energy range
considered. The spin constraint obtained here is very similar to that presented by
\cite{Nardini12}, who kept the inclination fixed at 45\deg.

\textbf{\textit{PDS\,456:}} In addition to the PLC+RDC continuum, there is also
well documented evidence for absorption due to an extremely high velocity outflow
(\citealt{Reeves09}). Owing to the high iron abundance obtained, we follow our
original approach and model the outflow with a photoionised plasma with a variable
iron abundance, fixed to that of the reflector for consistency. The other
abundances are assumed to be solar, and the turbulent velocity is assumed to be
$\sim$3000\,\kmps\ following \cite{Reeves09}. The column, ionisation parameter and
outflow velocity are allowed to vary, and we find \nh\ = $7^{+2}_{-3} \times
10^{21}$\,\atpcm\ and $\log\xi = 3.52^{+0.35}_{-0.03}$, consistent with our
previous work, and $v_{\rm out} = 0.31\pm0.01c$, consistent with \cite{Reeves09}.
As discussed in \cite{Walton10hex}, the high iron abundance and in particular
the high inclination obtained here are driven by the association of the emission
feature at $\sim$0.8\,\kev\ with reflected iron-L emission, as opposed to arising
due to complex parameter degeneracies as is likely to be the case for PKS\,0558-504,
so in this instance we consider the values obtained to be robust. As with Mrk\,841,
there is also a local minimum with a retrograde spin ($a^* \lesssim -0.5$) that
satisfies $\Delta\chi^{2} < 2.71$ (see Fig. \ref{fig_chi}). However, the prograde
solution obtained is consistent with our previous work on this dataset. Note finally
that in this case, in order to include a PIN detection it was necessary to relax our
rebinning requirement to a S/N of 2 instead of 3.

\textbf{\textit{1H\,0323+342:}} In this case the basic PLC+RDC interpretation
provides a good fit to the data; no statistically compelling narrow iron emission
lines are detected. However, similar to the earlier case of PKS\,0558-504, when
free we find that the inclination obtained is unphysically high for an unobscured
active galaxy ($i = 82\pm3$\deg). Therefore, we again present the results obtained
with the inclination fixed at 45\deg. However, we are still not able to constrain
all the reflection parameters of interest, so we present the results obtained
assuming $q=3$. Note that when $C_{\rm PIN/XIS}$ is free to vary, in addition to the
prograde solution presented in Table \ref{tab_refl2}, there is also a local minimum
with a retrograde spin ($a^* \lesssim -0.4$) which also satisfies $\Delta\chi^{2} <
2.71$ (see Fig. \ref{fig_c3_spin}), similar to Mrk\,841.

\textbf{\textit{UGC\,6728:}} In addition to the underlying PLC+RDC continuum, we
also detect a narrow emission feature consistent with neutral iron, and hence include
a second, distant reflector. Here, as for NGC\,7469, when allowing $C_{\rm PIN/XIS}$
to vary we were not able to reliably constrain all the reflection parameters of
interest, so we present the results with the inclination fixed at 45\deg.

\textbf{\textit{Mrk\,359:}} In addition to the basic PLC+RDC continuum, we also detect
a narrow emission feature consistent with neutral iron, and hence include a second,
distant reflector. We also include an unresolved Gaussian component at $\sim$5.9\,\kev\
to account for a feature in the background spectrum. The formal global minimum gives a
solution in $i > 85$\deg and the spin is unconstrained. However, as with PKS\,0558-504
and 1H\,0323+342, we consider this to be unphysically high for an unobscured source,
and there is a local minimum of essentially equal statistical quality with $i \sim
45$\deg, so we instead present this solution. The spin is only weakly constrained.

\textbf{\textit{MCG--2-14-9:}} In addition to the PLC+RDC continuum, we also detect a
narrow emission feature consistent with neutral iron, and hence include a second, distant
reflector. A very good fit is obtained, however in this case, owing to the moderate data
quality and the relatively weak reflection features, even when fixing the inclination at
45\deg, the iron abundance at the solar value, and the emissivity index at the Newtonian
prediction ($q=3$) we are unable to provide a preliminary spin constraint.

\textbf{\textit{ESO\,548-G081:}} In addition to the basic PLC+RDC continuum, we also
detect a strong narrow emission feature consistent with neutral iron, and include a
second, distant reflector. As with MCG--2-14-9 a very good fit is obtained, but owing to
the moderate data quality and the relatively weak reflection features, even fixing the
inclination at 45\deg and the emissivity index at the Newtonian prediction we are unable
to provide a preliminary spin constraint.

\textbf{\textit{Mrk\,1018:}} In this case, the PLC+RDC continuum provides an excellent
fit, no statistically compelling narrow lines are detected. However, the spin is only
very weakly constrained.

\textbf{\textit{RBS\,1124:}} Here, the basic PLC+RDC continuum also provides a good fit
to the available data. The spin constraint for this dataset initially presented by
\cite{Miniutti10} is broadly consistent with that obtained here.

\textbf{\textit{IRAS\,13224-3809:}} In this case, the XIS data quality is fairly poor,
only extending up to $\sim$7.6\,\kev, and even relaxing our binning constraint to a S/N
of 2 we do not obtain a robust PIN detection. However, we have recently undertaken a
long observing campaign on this source with \xmm, in which the source is found to
display a sharp, deep drop above $\sim$7.5\,\kev\ which is not well detected here owing
to the poor data quality (see \citealt{Fabian12iras}, and also \citealt{Ponti10},
\citealt{Boller03}). This feature, and the spectrum at higher energies, is extremely
important in determining the iron abundance, inclination and also the photon index, so
in this case we take these values from our \xmm\ analysis. Furthermore, to remain
consistent with this more detailed analysis, we also include a low temperature ($kT =
0.106 \pm 0.005$\,\kev) blackbody component, which may also be associated with
irradiation of the accretion disc. In addition, we also include an unresolved Gaussian
at $\sim$5.9\,\kev\ to account for a feature in the background spectrum.

Owing to the extreme nature of this source, by fixing the above parameters we still
obtain a very strong spin constraint despite the relatively poor data quality, which
is consistent with the inner radius originally obtained by \cite{Ponti10}, and the spin
obtained with our long \xmm\ exposure (\citealt{Fabian12iras}). This remains the case
even when the key parameters informed by our \xmm\ analysis are allowed to vary within
their quoted uncertainties. If we allow instead for a dual-ionisation reflector ($\xi_{1}
= 20^{+1}_{-7}$, $\xi_{2} = 490^{+50}_{-200}$\,\ergcmps), as proposed in
\cite{Fabian12iras}, the fit improves to \rchi\ = 417/410, but the spin constraint
obtained does not change. Attempts to fit lower spins, while still imposing the same
limits on the other key parameters, result in the model significantly underpredicting the
data at high energies. Removing the constraint on the photon index results in $\Gamma =
2.22^{+0.06}_{-0.05}$ and a substantially weaker constraint on the spin, $a^* > 0.9$,
although this is still constrained to be high. If the constraints on the iron abundance
and the disc inclination are also removed, the model becomes too degenerate to reliably
constrain the spin solely using the \suzaku\ data, even if the blackbody component is
also removed, and the values obtained for these quantities are substantially different
to those obtained with the significantly higher quality \xmm\ data ($A_{\rm Fe}$/solar =
$5^{+2}_{-1}$, $i = 30\pm4$\deg), despite not being expected to evolve with time.
Furthermore, we stress that neither of these cases with relaxed parameter constraints
correctly include the sharp high energy drop known to be present in this source.
Therefore, we will proceed with the model in which the photon index, iron abundance and
disc inclination are fixed at the values obtained with \xmm.

\textbf{\textit{1H\,0707-495:}} Again, the XIS data quality here is rather poor, the
source only being detected up to $\sim$6.7\,\kev, and there is no robust PIN detection.
As with IRAS\,13224-3809, this source is well documented as having a deep, sharp drop
above $\sim$7\,\kev\ (\citealt{Boller02, Gallo04, Fabian04, FabZog09, Zoghbi10,
Fabian12}), which is not detected owing to the poor data quality, leading to the same
problems in determining the iron abundance, disc inclination and photon index. The
spectrum obtained with this \suzaku\ observation is fairly reminiscent of the most
recent \xmm\ observation, presented in \cite{Fabian12}, so we adopt the photon index
of 2.7 obtained in that work. We also fix the inclination and the iron abundance values
at those obtained in the more detailed studies presented by \cite{Zoghbi10},
\cite{Fabian12} and most recently \cite{Kara12}.

As with IRAS\,13324-3809, by fixing these key parameters to their previously obtained
values we are still able to obtain a strong spin constraint despite the poor data
quality, owing again to the extreme nature of this source. The spin obtained here is
consistent with that originally presented in \cite{FabZog09}, and also that found in
subsequent reflection-based analyses (see also \citealt{Dauser12}). Again, allowing
the key parameters informed by the higher quality \xmm\ data to vary within their
quoted uncertainties does not significantly effect the spin constraint obtained, nor
does adopting a dual-ionisation reflector ($\xi_{1} < 41$, $\xi_{2} =
130^{+110}_{-60}$\,\ergcmps) similar to that proposed in \cite{Fabian12}, despite the
latter modification providing a moderate improvement in the quality of fit (\rchi\ =
264/234). Attempts to force lower spin values with the same constraints on the other
key parameters result in severe under-estimation of the data at higher energies. In
fact, this source was caught in such a reflection dominated state (see section
\ref{sec_R}) that, even removing the constraints on the photon index, iron abundance
and disc inclination, the spin is still constrained to be high ($a^* > 0.88$).
However, the values obtained for the iron abundance and disc inclination ($A_{\rm
Fe}$/solar = $3.4^{+1.2}_{-0.8}$, $i = 40^{+3}_{-10}$\deg), which should not vary with
time, are highly discrepant with those obtained from any high quality \xmm\ observation,
and as with IRAS\,13224-3809 the sharp high energy drop known to be present disappears
from the model. We therefore again proceed with the model in which the parameters are
fixed based on previous \xmm\ analysis.

\textbf{\textit{IRAS\,05262+4432:}} Here, the XIS data quality is extremely poor, only
extending up to $\sim$7.8\,\kev, and as with IRAS\,13224-3809 and 1H\,0707-495 there is
no robust PIN detection. A good fit is obtained with the basic PLC+RDC interpretation,
but even when fixing the inclination at 45\deg, the iron abundance at the solar value,
and the emissivity index at the Newtonian prediction we are unable to provide a
preliminary spin constraint.

\section{Discussion}
\label{sec_c3_disc}

We have presented a spectral analysis of a sample of 25 `bare' active galaxies observed
with the \suzaku\ satellite. These are sources that display either no intrinsic
absorption, or only very weak intrinsic absorption, and have been selected on the basis
of the presence of a relatively smooth soft excess below $\sim$2\,\kev\ after
extrapolation of the $\sim$2--10\,\kev\ continuum to lower energies. Owing to this lack
of absorption, such sources offer the best opportunity to study the intrinsic emission
spectra of active galaxies. In addition to the soft excesses required for selection, a
number of the sources also show evidence for broad iron emission components (\eg
Mrk\,335, Fairall\,9, Ark\,564, see Fig. \ref{fig_ratio_po}), and/or for `hard' excesses
above $\sim$10\,\kev\ in the PIN data, although these features are not ubiquitous within
the presented sample. 

Our analysis is based around interpreting these spectral complexities as reflected
emission from the accretion disc. For each source we have constructed a continuum model
consisting of a powerlaw-like component (PLC), associated with Compton up-scattering in
an X-ray emitting corona, and a reflection dominated component (RDC) arising through
irradiation of the surface of the accretion disc by the Comptonised emission. The
reflected emission has been modelled with the physically self consistent reflection code
\reflionx\ (\citealt{reflion}), which treats both the backscattered continuum emission
and the atomic features produced when optically thick material is irradiated by a
powerlaw-like X-ray continuum, and the relativistic effects relevant for emission from
an accretion disc in the strong gravity regime close to a black hole are included with
the \relconv\ convolution kernel (\citealt{relconv}). In addition to the their continuum
emission, many of the AGN analysed also display narrow emission lines consistent with
neutral iron, which we associate with reflection from cold, distant material, and treat
with a second, unmodified \reflionx\ component. A small minority also show evidence for
discrete emission lines associated with highly ionised iron (Fe {\small XXV} and/or
{\small XXVI}), which we model with simple Gaussian emission line profiles. All the
emission components are modified by neutral, Galactic absorption. Finally, a further
minority also show evidence for some minor absorption, either at low energies from
moderately ionised material or at high energies from highly ionised material. Where
present, we have modelled this absorption with the \xstar\ photoionisation code
(\citealt{xstar}). In this work, we have attempted to simultaneously combine a
systematic approach to analysing the sample as a whole, with a detailed physical
treatment of each individual source (see section \ref{sec_spec_details}).

In all the cases analysed, we have been able to successfully model the available
broadband spectra with a reflection based interpretation; Fig. \ref{fig_chi} shows the
distribution of the values obtained for the reduced chi-squared ($\chi^{2}_{\nu}$ =
\chisq/D.o.F.) with our PLC+RDC based interpretations for the analysed sample. This
distribution clearly displays a fairly narrow peak around $\chi^{2}_{\nu} = 1$. This
demonstration of the flexibility of disc reflection is important, as we have previously
argued in \cite{Walton12xrbAGN} that this process is a fundamental consequence of the
widely accepted disc--corona accretion geometry.

\subsection{Sample Properties}

Having successfully modelled the observed data, we now consider the results obtained for
some of the physical properties of the accretion flow for the compiled sample, focusing
conservatively on the results obtained with $C_{\rm XIS/PIN}$ allowed to vary (Table
\ref{tab_refl2}). In general, when they could be constrained, the disc inclinations
obtained from our analysis are either moderate or low, as would be expected for a sample
selected on the basis of a lack of obscuration\footnote{Recall that the probability
distribution for the inclination of randomly orientated disc-like sources follows $P(i)
\propto \sin(i)$}. For a couple of the AGN considered the formal best fit does return
unphysically high inclinations when this parameter is left free to vary given the nature
of the selected sources. However, we attribute this to the complex dependencies and
degeneracies that can arise between the various parameters in such a complex,
multi-parameter model, which can be especially problematic when modelling data of only
moderate quality and/or a relatively featureless spectrum. Crucially, we are still able
to obtain acceptable solutions in these cases even when fixing the inclination at a
physically more reasonable value (45\deg).

It may at first seem surprising that the majority of inclinations obtained are moderate
($\sim$50\deg) rather than low ($\lesssim$30\deg). However, there are two factors that
result in a natural bias towards higher inclinations (up to the point that sources are
excluded due to obscuration by the torus, above $\sim$70\deg). We have already mentioned
that for disc-like structures, $P(i) \propto \sin(i)$, so we naturally expect there to
be more AGN with high inclinations. In addition, the majority of sources analysed appear
to require high spin (see section \ref{sec_spin}). For such sources, the reflected flux
observed in the soft X-ray bandpass is greater at high inclinations than low inclinations,
owing to the larger line-of-sight velocities of the material in the disc at high
inclinations. This results in an enhancement of the reflected flux observed owing to the
additional special relativistic beaming along angles close to the plane of the disc, and
also in additional reflected flux being blueshifted into the soft X-ray bandpass. The
combined strength of these effects depends on both the spin of the black hole and the
emissivity profile; for a maximally rotating black hole with a steep ($q = 6$) emissivity,
the reflected flux at soft energies for a source observed at $i$ = 80\deg\ can be more
than a factor of $\sim$2 larger than that for the same source observed at $i$ = 10\deg,
while for a non-rotating black hole with an emissivity of $q = 3$ the reflected flux in
the soft band appears practically isotropic. Therefore, by basing our sample selection
on the presence of a soft excess we are probably introducing an additional bias towards
higher inclination sources, since rapid spins are generally inferred.

\begin{figure}
\begin{center}
\rotatebox{0}{
{\includegraphics[width=235pt]{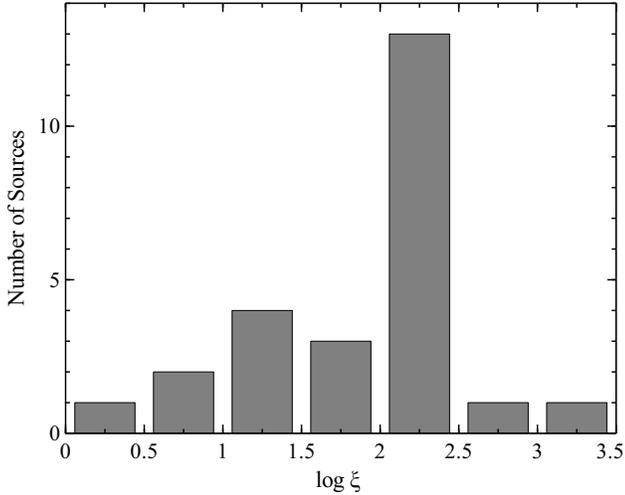}}
}
\end{center}
\vspace{-0.2cm}
\caption{The number distribution obtained for the logarithm of the ionisation parameter
of the reflecting medium with our PLC+RDC based interpretations.}
\label{fig_ion}
\end{figure}

In terms of the ionisation state of the reflecting medium (\ie the surface of the disc),
as shown in Fig. \ref{fig_ion}, we find a fairly broad range of ionisation parameters,
ranging from $\sim$10--1000\,\ergcmps. The vast majority of sources are found to have
$\xi \lesssim 400$, suggesting that, as expected, iron is very rarely highly ionised in
AGN. There is a fairly strong peak in the obtained distribution around $\xi \sim 300$,
corresponding to iron being primarily found in its Fe {\small XIX--XXI} ions
(\citealt{Kallman04}). At these ionisation states, resonant trapping and Auger
destruction of the K$\alpha$ photons is expected to be very efficient, and the emitted
iron line should only be very weak (\citealt{Matt93, Ross96}). This offers a natural
explanation as to why a number of the sources do not show any strong evidence for
relativistically broadened iron emission (see Fig. \ref{fig_ratio_po}).

\begin{figure}
\begin{center}
\rotatebox{0}{
{\includegraphics[width=235pt]{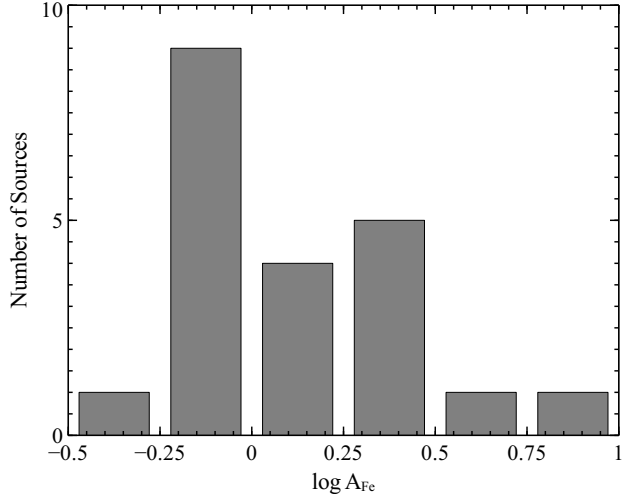}}
}
\end{center}
\vspace{-0.2cm}
\caption{The number distribution obtained for the logarithm of the iron abundance of the
reflecting medium with our PLC+RDC based analysis.}
\label{fig_feabund}
\end{figure}

In Fig. \ref{fig_feabund} we also plot the number distribution of the best fit iron
abundances obtained; only sources for which we were able to constrain this quantity are
included. The distribution clearly peaks around a solar iron abundance, with the vast
majority of sources consistent with having an abundance within a factor of $\sim$2 of
this value. This is similar to the preference for a roughly solar abundance reported by
\cite{Crummy06} in their earlier application of disc reflection to a general sample of
AGN, despite the two samples only sharing 5 sources in common. Our analysis results in a
substantially super-solar iron abundance for only one source, PDS\,456 (although highly
super-solar iron abundances are also adopted for 1H\,0707-495 and IRAS\,13224-3809, this
is based on prior analyses rather than being constrained by the data analysed here). As
discussed previously, this is driven by the association of the emission feature observed
at $\sim$0.9\,\kev\ with iron L shell emission in the reflected contribution.

\subsection{Reflection Strength}
\label{sec_R}

A further quantity of interest is the implied strength of the reflected emission from
the disc relative to the intrinsic PLC emission, $R_{\rm disc}$, which can provide
information on the geometry of the accretion flow. In order to estimate $R_{\rm disc}$,
in a manner consistent with the literature, we make brief use of the \pexriv\ Compton
reflection code (\citealt{PEXRAV}), in which $R$ is directly included as a free
parameter. $R$ is normalised such that a reflecting medium subtending a solid angle
of $\Omega = 2\pi$, as seen by the illuminating source, gives a value of $R$ = 1. The
other key parameters of this model are the photon index and the high energy cut-off of
the ionising continuum, and the inclination, abundances and ionisation of the reflecting
medium. However, as \pexriv\ does not self consistently include any flourescent atomic
emission, which can be of significant importance in determining the reflected flux at
soft energies, we estimate $R_{\rm disc}$ from the strength of the Compton reflection
hump in our best fit models, rather than applying \pexriv\ to the data directly, as
follows.

To remain consistent with our detailed modelling, the photon index, disc inclination,
iron abundance and ionisation parameter are all set to the best fit values presented
previously, and to remain consistent with the manner in which the \reflionx\ model was
computed the high energy cut-off was set to 300\,\kev\ and the other elemental
abundances were set to their solar values. The normalisation of the \pexriv\ component
was then set such that, when $R_{\rm disc}$ is set to zero the 2--10\,\kev\ flux
matches that of the PLC component. Finally, $R_{\rm disc}$ is determined such that the
15--50\,\kev\ flux matches that of the combined PLC+RDC components in the rest frame,
\ie once the relativistic effects applied by \relconv\ have been removed. We repeat
this process for the best fit models obtained both when $C_{\rm PIN/XIS}$ was fixed
and when it was free to vary, and the results obtained are given in Table \ref{tab_R}.
As we are not directly applying \pexriv\ to the data, we do not estimate any
statistical uncertainties on the values obtained for $R_{\rm disc}$, and stress that
these should only be considered as approximate estimates given the nature in which
they are determined.

\begin{table}
  \caption{Approximate estimates for the relative strength of the reflected
emission from the accretion disc, based on the strength of the best fit Compton
reflection hump (see text).}
\begin{center}
\begin{tabular}{c c c}
\hline
\hline
\\[-0.3cm]
Source & \multicolumn{2}{c}{$R_{\rm disc}$} \\
\\[-0.25cm]
& $C_{\rm PIN/XIS}$ fixed & $C_{\rm PIN/XIS}$ free \\
\\[-0.3cm]
\hline
\hline
\\[-0.2cm]
Mrk\,509 & 0.7 & 0.5 \\
\\[-0.25cm]
3C\,382 & 0.05 & 0.1 \\
\\[-0.25cm]
Mrk\,335 & 0.8 & 0.8 \\
\\[-0.25cm]
Fairall 9 & 0.4 & 0.4 \\
\\[-0.25cm]
1H\,0419-577 & 1.5 & 1.3 \\
\\[-0.25cm]
Ark\,564 & 1.5 & 1.5 \\
\\[-0.25cm]
Ark\,120 & 1.6 & 1.5 \\
\\[-0.25cm]
3C\,390.3 & 0.03 & 0.03 \\
\\[-0.25cm]
PKS\,0558-504 & 1.0 & 1.0 \\
\\[-0.25cm]
NGC\,7469 & 0.8 & 0.2 \\
\\[-0.25cm]
Mrk\,110 & 2.6 & 1.2 \\
\\[-0.25cm]
Swift\,J0501.9-3239 & 2.1 & 3.7 \\
\\[-0.25cm]
Mrk\,841 & 0.3 & 0.3 \\
\\[-0.25cm]
Ton\,S180 & 1.4 & 1.4 \\
\\[-0.25cm]
PDS\,456 & 1.2 & 1.2 \\
\\[-0.25cm]
1H\,0323+342 & 0.5 & 0.5 \\
\\[-0.25cm]
UGC\,6728 & 1.5 & 1.4 \\
\\[-0.25cm]
Mrk\,359 & 1.0 & 1.0 \\
\\[-0.25cm]
MCG--2-14-9 & 0.3 & 0.3 \\
\\[-0.25cm]
ESO\,548-G081 & 0.1 & 0.1 \\
\\[-0.25cm]
Mrk\,1018 & 1.3 & 1.2 \\
\\[-0.25cm]
RBS\,1124 & 3.3 & 2.9 \\
\\[-0.25cm]
IRAS\,13224-3809 & 16 & - \\
\\[-0.25cm]
1H\,0707-495 & 275 & - \\
\\[-0.25cm]
IRAS\,05262+4432 & 1.3 & - \\
\\[-0.2cm]
\hline
\hline
\end{tabular}
\label{tab_R}
\end{center}
\end{table}

\begin{figure}
\begin{center}
\rotatebox{0}{
{\includegraphics[width=235pt]{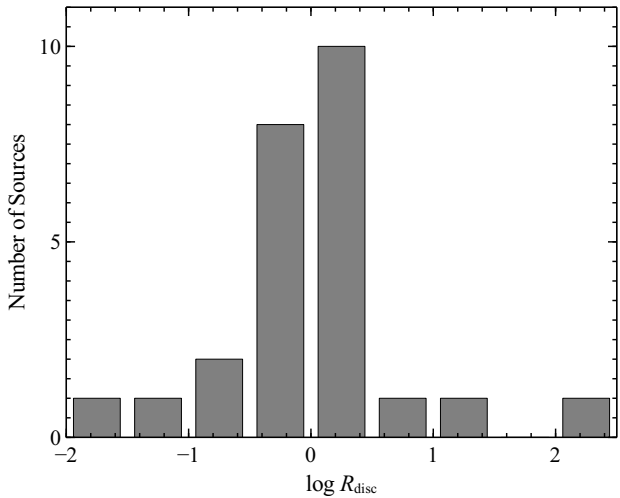}}
}
\end{center}
\vspace{-0.2cm}
\caption{The number distribution obtained for $\log R_{\rm disc}$ estimated
from our best fit PLC+RDC continuum models.}
\label{fig_R}
\end{figure}

In Fig. \ref{fig_R} we show the distribution of the values for $R_{\rm disc}$
obtained, where possible, when $C_{\rm PIN/XIS}$ was free to vary. This distribution
shows a clear peak around $R_{\rm disc} \sim 1$ (the median value is $\sim$1.2), as
broadly expected for a central isotropic emission source illuminating a geometrically
thin accretion disc. However, there are also some specific individual results worth
drawing attention to. First of all, the reflection fraction inferred for 1H\,0707-495
is {\it extremely} large. This is consistent with our previous assertion that the
\suzaku\ data is reminiscent of the \xmm\ observation obtained recently in 2011,
which caught 1H\,0707-495 in an almost completely reflection dominated state (see
\citealt{Fabian12}). The reflection fraction obtained for IRAS\,13224-3809 is also
large, consistent with that estimated from our recent long \xmm\ observation
(\citealt{Fabian12iras}). At the other end of the scale, it is interesting to note
that both of the radio-loud AGN included in the compiled sample display extremely weak
reflection features. This would be expected if the illuminating X-ray corona were 
associated with the base of the jet, as suggested by \cite{Markoff05} (see also
\citealt{Beloborodov99}). It would be of great interest to further test this scenario
by investigating whether systematically weak reflection features are a general trait
of radio loud AGN with a much larger sample of such sources.

\subsection{Black Hole Spin}
\label{sec_spin}

In addition to generally testing the robustness of the disc reflection interpretation,
during the course of this analysis we have endeavoured to use the inferred contribution
from the reflected emission to place initial constraints on the black hole spin, although
this has not always been possible, and it should be noted that the constraints presented
are model dependent. Where the spin could be well constrained, rapid prograde spins are
generally obtained, although the black holes are certainly not always inferred to be
maximally rotating. Even when the spin is only relatively loosely constrained, some kind
of rotation is generally implied. There are no sources that are strongly constrained to
be Schwarschild black holes, although there are a few sources that remain consistent with
this interpretation. There are also no sources that are strongly constrained to have a
retrograde spin, although again there are a few sources that remain consistent with this
scenario. Therefore, if relativistic disc reflection is responsible for the intrinsic
spectral complexities displayed by AGN, and as discussed previously there is very good
evidence to suggest this is the case, the implication is that the majority of AGN may be
rotating fairly rapidly. Where the spin has been successfully measured via reflection for
other sources not included in this work, the results obtained also seem to be consistent
with this picture (\citealt{kerrconv, Brenneman11, Reis3783, Cowperthwaite12}; note that
in the latter case, although the favoured spin implies a retrograde rotation with respect
to the material in the disc, the black hole is still inferred to be rapidly rotating). For
the majority of the sources analysed, where either a measurement of the spin or of the
inner radius of the disc has been made previously, either using one of the same datasets
analysed here or a different observation obtained at some other epoch, the spin obtained
here is consistent with the previous work, although there are some notable exceptions
highlighted in section \ref{sec_spec_details}. Where it has not been possible to reliably
constrain the spin, this has largely been due to either a data quality issue, and/or a
lack of strong reflection features.

\begin{figure}
\begin{center}
\rotatebox{0}{
{\includegraphics[width=235pt]{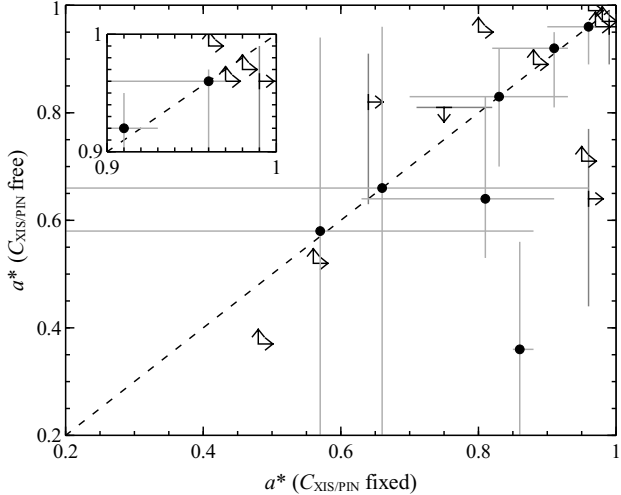}}
}
\end{center}
\vspace{-0.2cm}
\caption{A comparison of the spin constraints obtained with $C_{\rm XIS/PIN}$ fixed
and with $C_{\rm XIS/PIN}$ free to vary. Where limits are obtained, these are
indicated with arrows. In many cases, allowing $C_{\rm XIS/PIN}$ to vary does not
significantly change the spin constraint obtained, although as discussed in section
\ref{sec_spec_details} there are some notable exceptions.}
\label{fig_comp}
\end{figure}

In this work we have carefully attempted to determine the correct cross-normalisation
constant to use between the combined XIS and PIN spectra based on the recommendations of
the \suzaku\ HXD calibration team. To do so, we have first compared the combined XIS
spectrum with the single XIS0 spectrum, and used this comparison to calculate the
normalisation constant $C_{\rm XIS/PIN}$ from the normalisation constants obtained by
the \suzaku\ team between the XIS0 and PIN detectors using equation \ref{eqn_crossnorm}.
In general we find a good agreement between the combined XIS and XIS0 spectra, as would
be expected, although in rare cases the normalisations of these spectra can be discrepant
by up to $\sim$8 per cent (see UGC\,6728). We have also investigated what effect allowing
this value to vary has on the results obtained, as it does not account for any systematic
uncertainties on the level of the modelled instrument background. These are estimated to
be $\sim$3 per cent, and so we have allowed $C_{\rm XIS/PIN}$ to vary within an appropriate
range based on this uncertainty. The results obtained allowing $C_{\rm XIS/PIN}$ to vary
in this manner are quoted in Table \ref{tab_refl2}. Focusing on the spin parameters, doing
so does not appear to have any single systematic effect on the results obtained, either in
terms of the preferred value or its statistical uncertainty. As demonstrated in Fig.
\ref{fig_comp}, allowing $C_{\rm XIS/PIN}$ to vary has little or no effect on the spin
obtained in many cases, even when the PIN spectrum carries some significant statistical
weight (\eg Mrk\,841), while in some the spin becomes more tightly constrained
(\eg PKS\,0558-504), and in others it becomes substantially more difficult to constrain
(\eg Mrk\,509). However, the general conclusions drawn above still seem to be robust.

\begin{figure}
\begin{center}
\rotatebox{0}{
{\includegraphics[width=235pt]{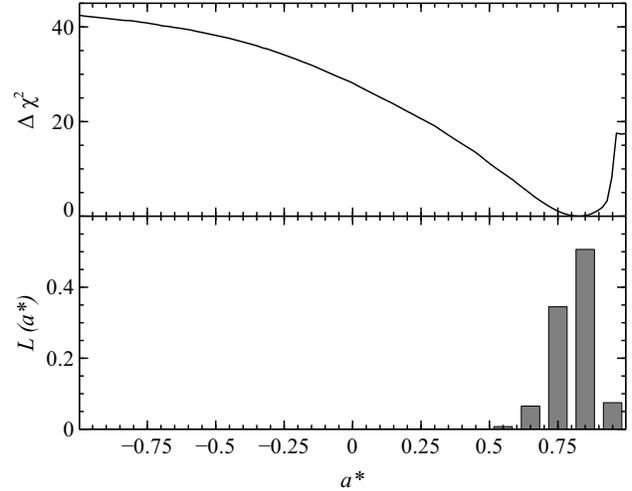}}
}
\end{center}
\vspace{-0.2cm}
\caption{The likelihood distribution obtained for the spin parameter for Mrk\,335
(\textit{bottom panel}) from the \chisq\ confidence contour (\textit{top panel});
each bin spans a range of $\Delta a^* = 0.1$.}
\label{fig_Ldist}
\end{figure}

\begin{figure*}
\begin{center}
\rotatebox{0}{
{\includegraphics[width=235pt]{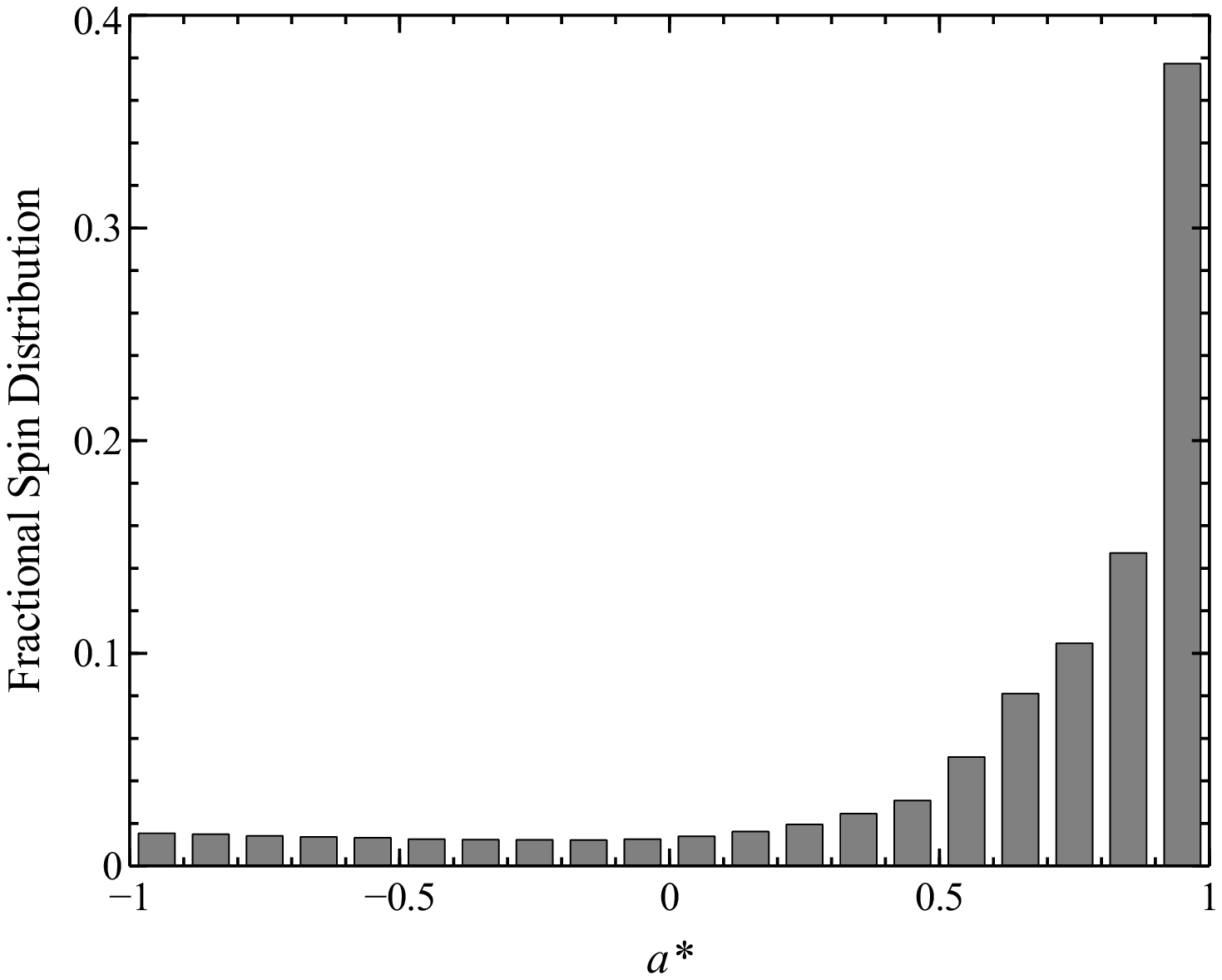}}
}
\hspace*{0.75cm}
\rotatebox{0}{
{\includegraphics[width=235pt]{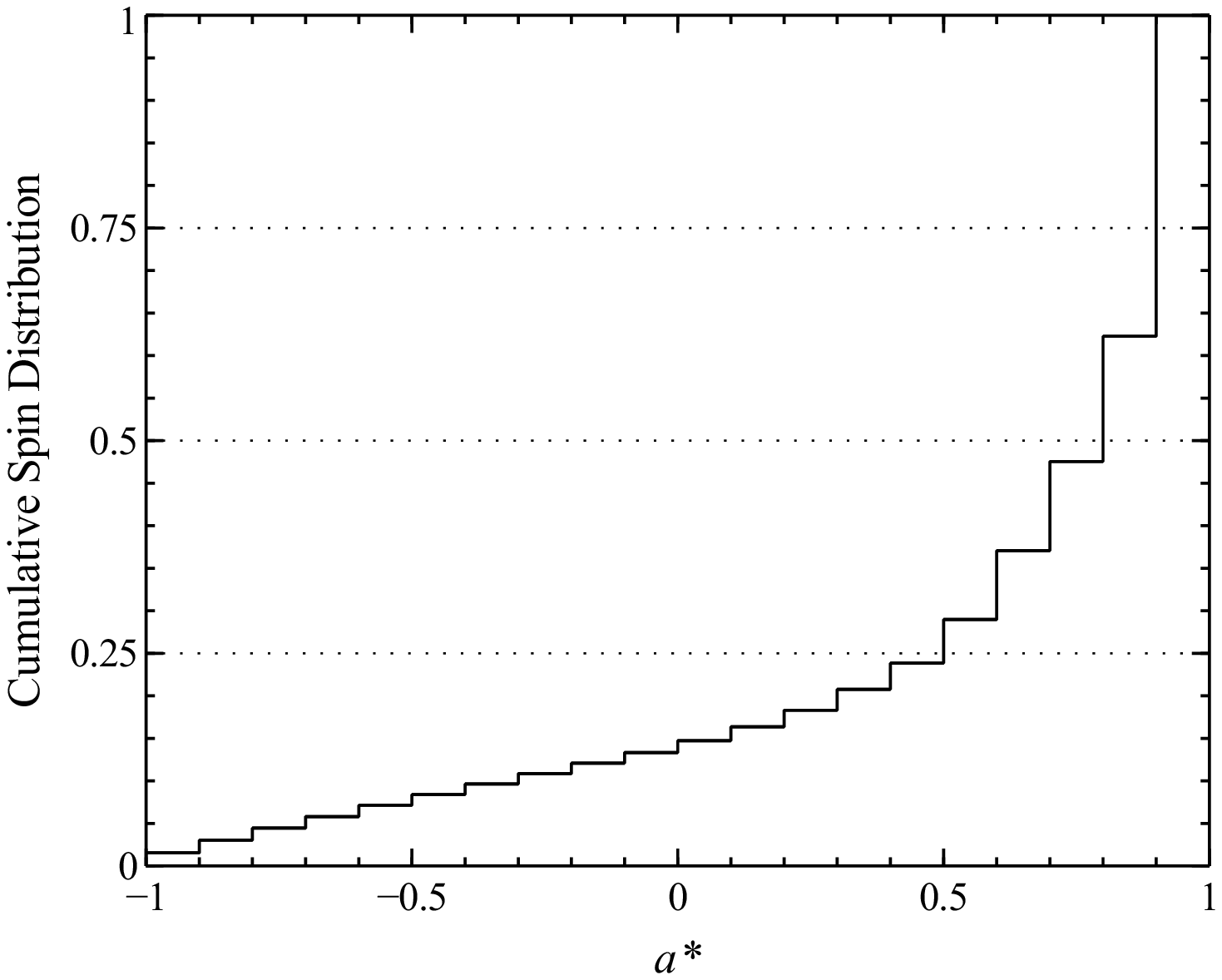}}
}
\end{center}
\vspace{-0.2cm}
\caption{Fractional (\textit{left panel}) and cumulative (\textit{right panel}) distributions
for the black hole spins obtained for the compiled sample of AGN (see text); each bin spans a
range of $\Delta a^* = 0.1$. The obtained distribution is clearly peaked at higher spin values.}
\label{fig_c3_dist}
\end{figure*}

In Fig. \ref{fig_c3_dist} we plot a fractional distribution of the results shown in Fig.
\ref{fig_c3_spin} for the black hole spins of the compiled sample. Here, we again only
consider sources for which the data quality and/or the strength of the reflected features
were sufficient to allow at least some constraint to be placed on the spin at the 90 per
cent confidence level. However, in many cases the statistical uncertainties obtained are
non-Gaussian and cover a broad range of spin values, which can be significantly larger
than any desirable bin size for such a distribution. As a compilation of AGN spin is one
of the primary goals of this work, we take a more careful approach and attempt to account
for these uncertainties when plotting the distribution of spins we obtain. We do so by
combining the likelihood distributions\footnote{These distributions are designed to
represent the underlying probability distributions for the spin parameters given the
available data, but as this is not a fully Bayesian analysis they cannot formally be
classed as such.} for the spin parameters, $L(a^*)$, of the sample rather than simply
the best fit values. These distributions are derived from the \chisq\ confidence contours
displayed in Fig. \ref{fig_c3_spin}, assuming that $L(a^*) \propto \exp(-\Delta\chi^{2}/2)$
and normalised such that the sum of the likelihoods over the $-0.998 \leq a^* \leq 0.998$
range considered here is unity. As an example, in Fig. \ref{fig_Ldist} we show the
likelihood distribution obtained from the \chisq\ contour for Mrk\,335. Finally, the
individual likelihood distributions are then combined and renormalised again to form a
composite likelihood distribution for the compiled sample. The distribution obtained
clearly displays a strong peak at higher spin values. This is consistent with the
typically small inner radii obtained by \cite{Crummy06}.

As discussed previously, the spin of the supermassive black holes powering AGN is widely
expected to be determined by their growth history. Over the lifetime of these nuclear
black holes, galaxy mergers and prolonged accretion can impart enough angular momentum
to determine the spin distribution of AGN in the current epoch. \cite{Berti08} calculated
the local spin distributions expected for various different SMBH growth scenarios under
the assumption that the SMBH seeds were not rotating. When SMBH growth is dominated
solely by mergers, \ie galaxy mergers resulting in the coalescence of the two central
SMBHs, the expected spins appear to be fairly randomly distributed, with a slight peak
at low spin owing to their assumption regarding the initial seeds. However, prolonged
accretion is found to be the dominant mechanism in determining the expected distribution
of AGN spins. When the accretion is ordered, the central black holes are generally spun
up, leading to a strong preference for rapidly rotating black holes. If instead the
accretion is chaotic, \ie the angular momenta of the accreted material and the AGN are
not typically aligned, the AGN will experience the opposite effect and be spun down,
leading to a strong preference for slowly rotating black holes. Therefore, measuring the
spin for a large sample of AGN has important implications regarding the nature in which
their black holes grow. Our observational analysis suggests a preference for at least
some rotation, and probably for rapid rotation, which if taken at face value is most
consistent with the scenario in which SMBH growth is dominated by ordered accretion.

However, before drawing any firm conclusions, we must first consider whether the general
high spin preference obtained here is likely to be representative of the true underlying
spin distribution. \cite{Brenneman11} suggest there may actually be an observational bias
towards AGN with high spin, based on the widely held expectation that the radiative
efficiency of accretion onto a black hole increases with black hole spin, which should
result in AGN with high spin being brighter on average than their low spin counterparts.
Based on this, \cite{Reynolds12} estimate that for a flux limited sample drawn from an AGN
population with a flat intrinsic spin distribution, only $\sim$30 per cent of the sources
in the sample would have a spin of $a^* < 0.5$, despite the true value being 50 per cent.
Furthermore, as demonstrated in Appendix \ref{app_param}, we find that it is easier to
statistically constrain higher spins, which is probably a natural consequence of the form of
the relation between black hole spin and the radius of the ISCO (see \citealt{Bardeen72}).
Given our method for compiling the sample likelihood distribution, this may also introduce
some slight bias towards higher spins. Nevertheless, where we have been able to constrain
the spin well at the 90 per cent confidence level, rapid rotation (\ie $a^* > 0.5$) is
almost always required, and this subset of the sample compiled is large enough that even
given this probable observational bias we might have expected to find at least one source
that is constrained to have low spin if the intrinsic distribution does not truly peak at
higher spins. However, we stress that the sample presented is certainly not a complete,
flux limited sample, so the picture that appears to be growing could yet change.

A further important consequence of our analysis, independent of the presence of any potential
biases, regards the radio loud/radio quiet AGN dichotomy. It has previously been suggested
that the observational differences between these classes of AGN are due to differences in
the spins of the black holes powering these sources (\citealt{Blandford90, Wilson95,
Moderski98, Garofalo10}). In this picture, radio loud AGN host black holes with high spin
that can launch powerful jets via magnetic extraction of the rotational energy of the black
hole, as proposed by \cite{BZ77}, while radio quiet AGN have low spin, and cannot launch jets
via this method. However, our analysis implies that, if the PLC+RDC interpretation adopted
here is correct, then this cannot be the case, as the majority of AGN included in this work
are radio quiet, yet are found to have high spin. This is not to say that the process
outlined by \cite{BZ77} is not present at all; when jets are present they may well carry more
power for AGN with high spin than for AGN with low spin. However, black hole spin would not
be sufficient by itself to solely determine jet production in AGN. This is similar to the
observed behaviour of Galactic black hole binaries (BHBs), where the presence of jets appears
to be determined by the accretion state of the black hole (see \eg \citealt{Fender04}); the
same BHB can be observed both with and without jets at different epochs, despite the fact
that its spin cannot have evolved substantially. Black hole spin may yet prove to be important
for certain jet properties, but it does not appear to play a substantial role in determining
whether jets are produced in the first place.

Although we have presented the largest compilation of AGN spins to date, in order to build
a more robust picture the number of sources with reliable spin measurements must still be
increased. There are two important aspects to addressing this issue. First and foremost, as
we have stressed previously, in many of the cases in which we have been unable to
provide an initial spin constraint, this has been due to the data currently available
having a relatively poor S/N. This deficiency can naturally be addressed with further long
observations with existing instrumentation. The other major uncertainty on some of the
individual constraints obtained in this work is the cross-normalisation of the XIS and PIN
instruments. In order to address this issue, it will be vital to obtain observations of
nearby active galaxies with the upcoming \nustar\ observatory (\citealt{NUSTAR}), which is
due to begin operation imminently. \nustar\ will offer continuous spectral coverage over
the 5--80\,\kev\ energy range, which includes a key region of overlap with the soft X-ray
spectrometers currently available, and being an imaging spectrometer will allow simultaneous
measurement of the background emission. Simultaneous \nustar\ and either \xmm\ or \suzaku\
observations will be able to provide spin measurements robust to the current
cross-calibrational uncertainties. In addition, the microcalorimeter due to fly aboard
\textit{Astro-H} (\citealt{ASTROH_TMP}; 2014) will enable the profiles of the soft excesses
and the iron lines to be resolved in much greater detail, enabling us to further test the
robustness of the reflection interpretation and improve the spin measurements obtained.
Finally, the effect of any high spin bias will ultimately need to be tested in the future
with large, volume limited samples of AGN.

\section{Conclusions}
\label{sec_c3_conc}

We have investigated the relativistic disc reflection interpretation for a large sample of
25 `bare' active galactic nuclei, sources with little or no complicating intrinsic
absorption, observed with \suzaku. In constructing reflection based models for this sample,
we have taken the unique approach of attempting to simultaneously undertake a systematic
analysis of the whole sample as well as a detailed treatment of each individual source,
and find that disc reflection has the required flexibility to successfully reproduce the
broadband spectra of all of the sources considered. Where possible, we also use the
reflected emission to place constraints on the black hole spin for this sample of sources.
Our analysis suggests a general preference for rapidly rotating black holes, which if taken
at face value is most consistent with the scenario in which SMBH growth is dominated by
prolonged, ordered accretion. However, there may be strong observational biases towards
AGN with high spin in the compiled sample, which limits our ability to draw strong
conclusions for the overall population at this stage. Our analysis also implies that the
radio loud/radio quiet AGN dichotomy is not related to black hole spin.

\section*{ACKNOWLEDGEMENTS}

This research has made use of data obtained from the \suzaku\ observatory, a collaborative
mission between the space agencies of Japan (JAXA) and the USA (NASA). DJW acknowledges
the financial support provided by STFC in the form of a PhD scholarship, EN is supported
by NASA grants GO2-13124X and NNX11AG99G, and ACF thanks the Royal Society. RCR thanks the
Michigan Society of Fellows, and is supported by NASA through the Einstein Fellowship
Program, grant number PF1-120087. Special thanks go to both Jeremy Sanders and Jack Steiner
for their computational assistance, which greatly aided in the publication of this research.
The figures included in this work have been produced with the
Veusz\footnote{http://home.gna.org/veusz/} plotting package, written by Jeremy Sanders.
Finally, the authors would like to thank the referee for their feedback, which helped
improve the depth of this work.

\bibliographystyle{mnras}

\bibliography{/home/dwalton/papers/references}

\appendix

\section{Multi-Parameter Reflection Models}
\label{app_param}

\begin{figure*}
\begin{center}
\rotatebox{0}{
{\includegraphics[width=159pt]{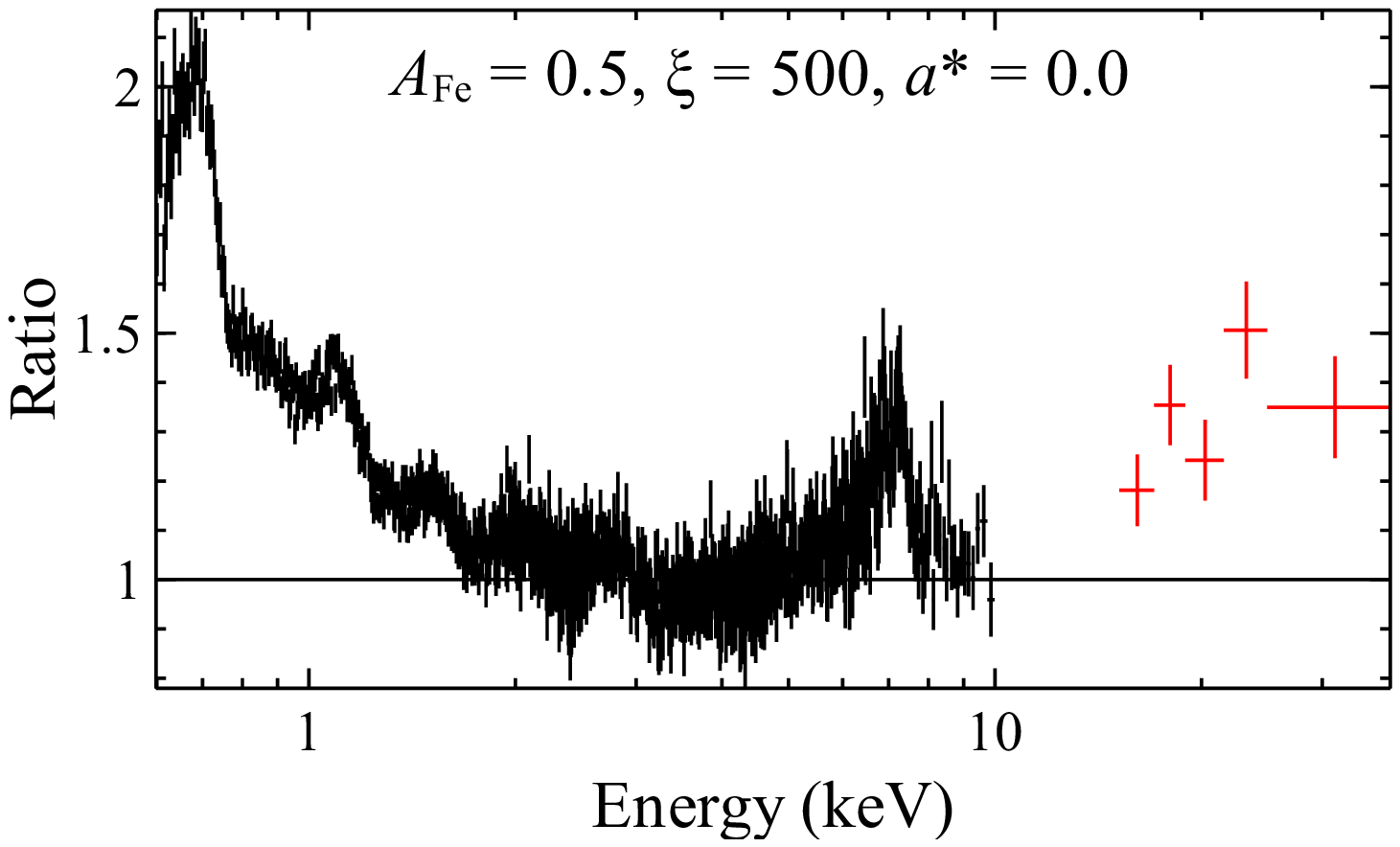}}
}
\rotatebox{0}{
{\includegraphics[width=159pt]{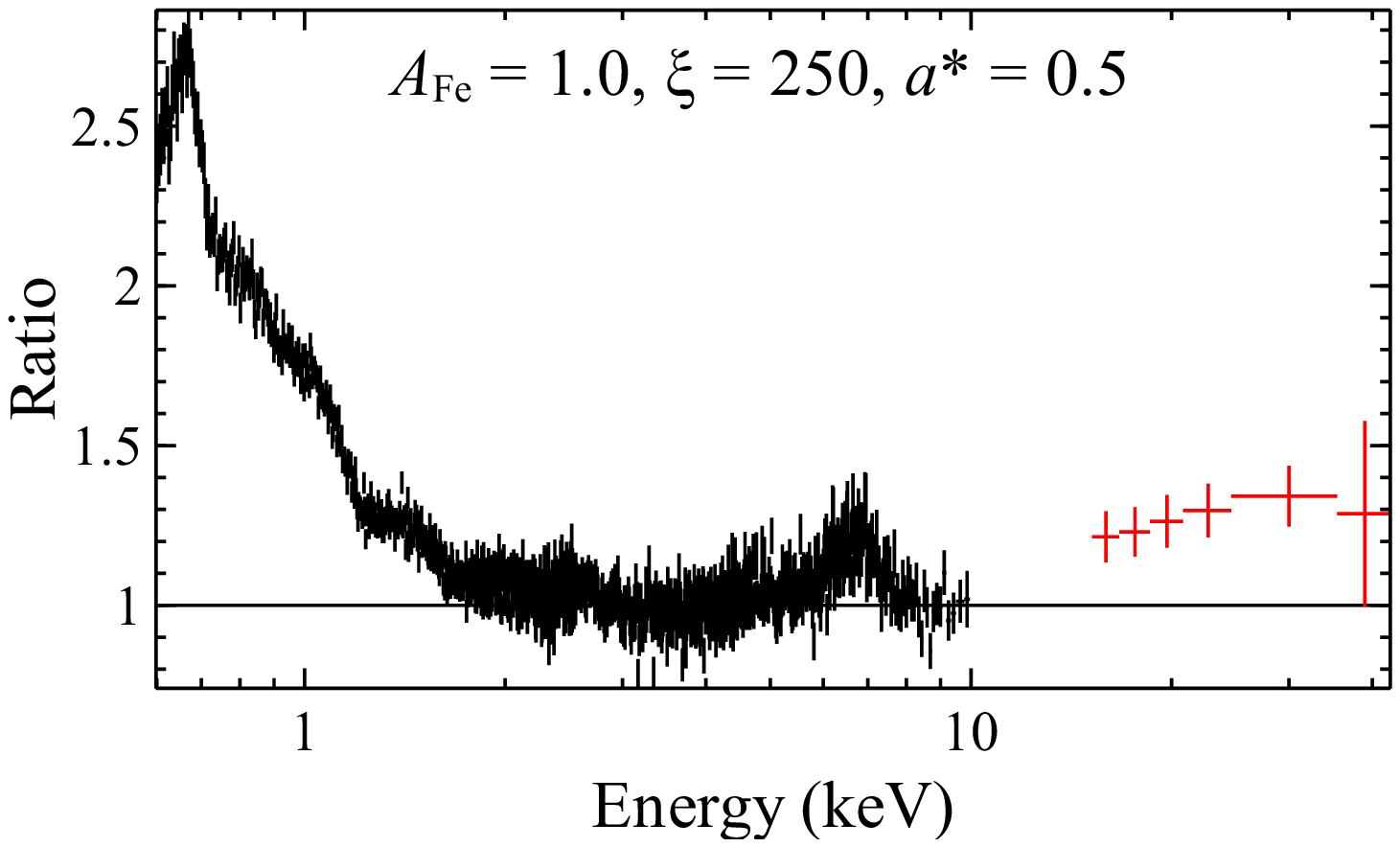}}
}
\rotatebox{0}{
{\includegraphics[width=159pt]{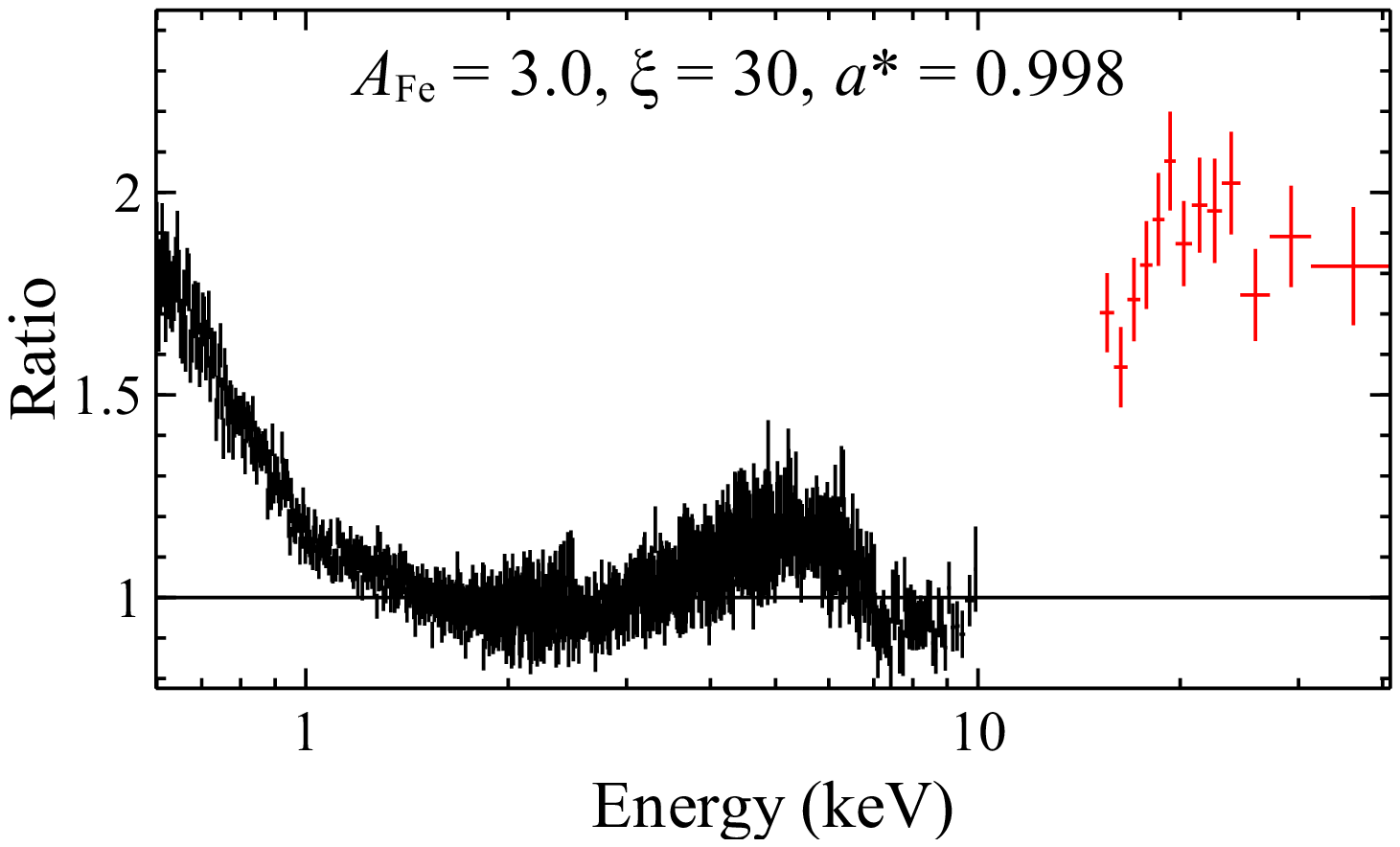}}
}
\end{center}
\caption{Data/model ratio plots for a selection of the simulated \suzaku\ \xis\
(front illuminated; black) and \pin\ (red) spectra to a simple powerlaw continuum
model. The plots for the two higher spin values shown display reasonable qualiatative
similarities with many of the real datasets shown in Fig. \ref{fig_ratio_po}. The
simulated data shown have been re-binned for plotting purposes only.}
\label{fig_sims}
\end{figure*}

\begin{table*}
  \caption{The various input parameter combinations simulated, and the values obtained
modelling these simulated spectra with our PLC+RDC continuum; all the spectra are
simulated with $\Gamma$ = 2.0, $q$ = 6 and $i$ = 45\deg. The input parameters are
generally successfully reproduced.}
\begin{center}
\begin{tabular}{c c c c c c c c c c}
\hline
\hline
\\[-0.3cm]
\multicolumn{3}{c}{Input Parameters} & & \multicolumn{6}{c}{Values Obtained} \\
\\[-0.3cm]
$A_{\rm Fe}$ & $\xi$ & $a^{*}$ & & $\Gamma$ & $i$ & $q$ & $A_{\rm Fe}$ & $\xi$ & $a^{*}$ \\
(solar) & (\ergcmps) & & & & (\deg) & & (solar) & (\ergcmps) & \\
\\[-0.25cm]
\hline
\hline
\\[-0.2cm]
0.5 & 30 & 0 & & $2.00\pm0.01$ & $45^{+2}_{-1}$ & $>3.4$ & $0.50^{+0.05}_{-0.04}$ & $33^{+4}_{-3}$ & $0.02^{+0.18}_{-0.25}$ \\
\\[-0.3cm]
0.5 & 30 & 0.5 & & $2.00\pm0.01$ & $44^{+3}_{-2}$ & $5.3^{+1.4}_{-0.8}$ & $0.5\pm0.1$ & $32^{+5}_{-3}$ & $0.46^{+0.11}_{-0.10}$ \\
\\[-0.3cm]
0.5 & 30 & 0.998 & & $2.00^{+0.02}_{-0.04}$ & $41^{+9}_{-4}$ & $5.5^{+1.2}_{-0.6}$ & $0.5^{+0.3}_{-0.1}$ & $34^{+19}_{-9}$ & $>0.97$ \\
\\[-0.3cm]
0.5 & 250 & 0 & & $2.01\pm0.01$ & $45^{+2}_{-1}$ & $4.2^{+2.4}_{-1.0}$ & $0.48\pm0.02$ & $246\pm6$ & $0.24^{+0.16}_{-0.28}$ \\
\\[-0.3cm]
0.5 & 250 & 0.5 & & $2.01\pm0.01$ & $42^{+7}_{-3}$ & $>4.3$ & $0.54\pm0.04$ & $243\pm6$ & $0.28^{+0.31}_{-0.22}$ \\
\\[-0.3cm]
0.5 & 250 & 0.998 & & $2.01^{+0.02}_{-0.01}$ & $45^{+3}_{-9}$ & $6.3^{+1.1}_{-0.9}$ & $0.47^{+0.05}_{-0.03}$ & $250\pm20$ & $>0.98$ \\
\\[-0.3cm]
0.5 & 500 & 0 & & $2.00\pm0.01$ & $46^{+2}_{-1}$ & $>5.7$ & $0.52\pm0.02$ & $500^{+10}_{-50}$ & $0.05^{+0.25}_{-0.14}$ \\
\\[-0.3cm]
0.5 & 500 & 0.5 & & $2.01\pm0.01$ & $46^{+4}_{-3.}$ & $>5.8$ & $0.51\pm0.02$ & $450^{+50}_{-40}$ & $0.52^{+0.23}_{-0.18}$ \\
\\[-0.3cm]
0.5 & 500 & 0.998 & & $2.01\pm0.01$ & $48\pm7$ & $6.8^{+1.5}_{-1.0}$ & $0.48^{+0.05}_{-0.04}$ & $460^{+60}_{-80}$ & $>0.98$ \\
\\[-0.3cm]
1 & 30 & 0 & & $2.00\pm0.01$ & $46^{+1}_{-2}$ & $>4.8$ & $1.00^{+0.09}_{-0.045}$ & $30\pm2$ & $0.04^{+0.15}_{-0.24}$ \\
\\[-0.3cm]
1 & 30 & 0.5 & & $2.00\pm0.01$ & $44\pm2$ & $5.8^{+1.0}_{-0.8}$ & $1.0\pm0.1$ & $30^{+4}_{-3}$ & $0.46\pm0.08$ \\
\\[-0.3cm]
1 & 30 & 0.998 & & $2.01\pm0.03$ & $41^{+6}_{-4}$ & $5.4^{+1.0}_{-0.5}$ & $1.1^{+0.2}_{-0.1}$ & $26^{+8}_{-4}$ & $>0.98$ \\
\\[-0.3cm]
1 & 250 & 0 & & $2.006^{+0.003}_{-0.005}$ & $46\pm1$ & $>6.5$ & $0.98\pm0.03$ & $252^{+2}_{-5}$ & $-0.11^{+0.18}_{-0.12}$ \\
\\[-0.3cm]
1 & 250 & 0.5 & & $2.01\pm0.01$ & $44^{+6}_{-3}$ & $>4.7$ & $1.1\pm0.1$ & $251\pm6$ & $0.42^{+0.26}_{-0.19}$ \\
\\[-0.3cm]
1 & 250 & 0.998 & & $2.02\pm0.01$ & $39^{+9}_{-8}$ & $5.7^{+1.3}_{-0.9}$ & $1.0^{+0.2}_{-0.1}$ & $240\pm10$ & $>0.97$ \\
\\[-0.3cm]
1 & 500 & 0 & & $2.00\pm0.01$ & $46^{+2}_{-1}$ & $>4.6$ & $1.1\pm0.1$ & $510\pm10$ & $0.11^{+0.22}_{-0.30}$ \\
\\[-0.3cm]
1 & 500 & 0.5 & & $2.01\pm0.01$ & $43^{+7}_{-2}$ & $5.4^{+4.1}_{-1.2}$ & $1.0\pm0.1$ & $490^{+20}_{-40}$ & $0.41^{+0.27}_{-0.16}$ \\
\\[-0.3cm]
1 & 500 & 0.998 & & $2.00\pm0.01$ & $45^{+4}_{-5}$ & $6.0^{+0.7}_{-0.6}$ & $1.0^{+0.2}_{-0.1}$ & $510\pm10$ & $>0.98$ \\
\\[-0.3cm]
3 & 30 & 0 & & $2.00\pm0.01$ & $44\pm1$ & $6.4^{+2.0}_{-1.3}$ & $2.9\pm0.2$ & $32^{+3}_{-2}$ & $-0.13^{+0.13}_{-0.18}$ \\
\\[-0.3cm]
3 & 30 & 0.5 & & $2.00\pm0.01$ & $47\pm2$ & $7.0\pm1.0$ & $3.3\pm0.2$ & $31^{+3}_{-2}$ & $0.57^{+0.07}_{-0.10}$ \\
\\[-0.3cm]
3 & 30 & 0.998 & & $1.99\pm0.03$ & $42\pm3$ & $5.5\pm0.4$ & $3.6\pm0.4$ & $31^{+17}_{-5}$ & $>0.98$ \\
\\[-0.3cm]
3 & 250 & 0 & & $2.01\pm0.01$ & $45^{+3}_{-1}$ & $>4.1$ & $3.2\pm0.2$ & $250\pm5$ & $0.01^{+0.41}_{-0.31}$ \\
\\[-0.3cm]
3 & 250 & 0.5 & & $2.00\pm0.01$ & $47^{+3}_{-2}$ & $>6.1$ & $2.9\pm0.2$ & $244^{+5}_{-4}$ & $0.53^{+0.14}_{-0.13}$ \\
\\[-0.3cm]
3 & 250 & 0.998 & & $2.00\pm0.01$ & $48\pm6$ & $6.7^{+1.3}_{-0.9}$ & $3.3^{+0.2}_{-0.4}$ & $280\pm20$ & $>0.97$ \\
\\[-0.3cm]
3 & 500 & 0 & & $2.01\pm0.01$ & $45\pm1$ & $6.1^{+3.5}_{-1.9}$ & $3.0\pm0.1$ & $500^{+10}_{-20}$ & $-0.06^{+0.25}_{-0.24}$ \\
\\[-0.3cm]
3 & 500 & 0.5 & & $2.01\pm0.01$ & $47\pm2$ & $7.7^{+1.6}_{-1.3}$ & $3.0\pm0.1$ & $500^{+10}_{-30}$ & $0.53\pm0.10$ \\
\\[-0.3cm]
3 & 500 & 0.998 & & $2.01\pm0.01$ & $49^{+3}_{-4}$ & $6.8\pm0.7$ & $3.5\pm0.5$ & $510^{+10}_{-90}$ & $>0.99$ \\
\\[-0.2cm]
\hline
\hline
\end{tabular}
\label{tab_sims}
\end{center}
\end{table*}

One of the major aspects of this work is attempting to constrain the black
hole spin from the full contribution of the reflected emission from the
accretion disc observed in AGN, \ie from the combination of the soft excess,
any broadened iron K emission and the high energy reflected continuum. In
order to do so, we make use of a physically self consistent reflection model,
\reflionx\ (\citealt{reflion}), combined with the latest relativistic
convolution kernal, \relconv\ (\citealt{relconv}). This combination results
in a complex, multi-parameter model for the reflected emission observed,
including the ionisation of the surface of the disc, its iron abundance,
the the radial emissivity profile, the disc inclination and of course the
black hole spin. The combination of the ionisation parameter and the iron
abundance in particular can have a strong effect on the broadband spectral
shape. In order to simultaneously constrain all of these parameters, some
balance between the strength of the reflection features and the quality of
data available is naturally required. In this appendix, we present a series
of simulations demonstrating that reasonable quality \suzaku\ data can
provide reliable constraints for all of these parameters.

Using the {\small FAKEIT} command in \xspec, we simulate \suzaku\ data based
on our underlying PLC+RDC continuum model (modified by neutral absorption),
using standard response and background files for the (front illuminated) XIS
and PIN detectors. These spectra are rebinned to the same S/N requirements
as the real data presented in the main body of the paper, \ie a minimum S/N
per energy bin of 5 for the combined FI XIS spectra, and 3 for the PIN
spectra. The 2--10\,\kev\ flux for the PLC component in these models is set
to $2\times10^{-11}$\ergpcmsqps, and for simplicity both the PLC and RDC
components are required to contribute the same flux over the much broader
0.1--1000.0\,\kev\ energy range (\ie the energy range over which the version
of the \reflionx\ model utilised here is calculated). The exposure time for
the simulations is set to 125\,ks, such that we are simulating a fairly good
\suzaku\ observation of a fairly bright AGN, relative to the sample compiled.
We repeat this process for a variety of input parameter combinations, varying
the ionistion of the disc, the iron abundance and the black hole spin. The
other key parameters adopted are \nh\ = $3\times10^{20}$\,\atpcm, $\Gamma$ =
2, $q$ = 6, $i$ = 45\deg, relatively typical of the results obtained in the
main body of the paper. For illustration, Fig. \ref{fig_sims} shows some
examples of the ratio plots obtained with the simulated spectra when modelled
with a simple absorbed powerlaw model, applied as described in section
\ref{sec_sel}. The plots obtained for the two higher spin values, although
obviously showing idealised scenarios in which just the PLC and disc
reflection components are present, are reasonably similar to many of those
obtained with the real data (see Fig. \ref{fig_ratio_po}). Each of the
simulated spectra are then modelled with our PLC+RDC continuum model, and the
parameters obtained are presented in Table \ref{tab_sims}.

In general, the input parameters are successfully reproduced, particularly
the parameters that significantly help determine the broadband spectral shape,
\ie\ the iron abundance, disc ionisation and photon index. The relativistic
blurring parameters are also generally well reproduced, although there is one
interesting trend worth briefly discussing. It is clear from Table
\ref{tab_sims} that it is easier to constrain the spin when the spin is high.
This can be understood in terms of the evolution of the theoretical relation
between the radius of the ISCO with black hole spin (see \citealt{Bardeen72}).
At higher spins, a given deviation in spin results in larger deviations in
the radius of the ISCO. As the model utilised here ultimitely infers the spin
by determining the inner radius of the disc from the observed spectrum, and
assumes this is coincident with the ISCO, it is therefore naturally easier to
statistically constrain higher spins via this method. Nevertheless, these
simulations demonstrate that \suzaku\ data with reasonably good S/N (as well
as reasonably strong reflection features) should be sufficient to determine
even moderate spins, as well as the other key parameters of interest in this
work, from the observed broadband X-ray spectrum.

\label{lastpage}

\end{document}